\documentclass[traditabstract,longauth]{aa}

\usepackage{graphicx}
\usepackage{txfonts}
\usepackage{txfonts}
\usepackage{natbib}
\usepackage[colorlinks=true,citecolor=blue]{hyperref}
\usepackage{xspace}
\usepackage{longtable}
\usepackage{array}
\usepackage{lscape}
\usepackage{rotating}
\usepackage{subcaption} 
\usepackage[labelformat=simple]{caption,subcaption}
\usepackage{gensymb}

\usepackage{enumitem}
\usepackage{amsmath}  

\bibpunct{(}{)}{;}{a}{}{,}

\newcommand{\MJup}{\ensuremath{M_{\mathrm{Jup}}}\xspace}

\newcommand{\degre}{\degree\xspace}

\newcommand{\as}{\hbox{$^{\prime\prime}$}\xspace}

\defcitealias{Desidera2020}{Paper~I}

\begin{document}

\title{The SPHERE infrared survey for exoplanets (SHINE)}
\subtitle{II. Observations, Data reduction and analysis,
   Detection performances and early-results}
\titlerunning{The SPHERE infrared survey for exoplanets (SHINE). II.}
\author{    
    M. Langlois \inst{\ref{CRAL},\ref{lam}} \and
    R. Gratton\inst{\ref{padova}} \and
    A.-M. Lagrange\inst{\ref{ipag},\ref{lesia}} \and
    P. Delorme\inst{\ref{ipag}} \and
    A. Boccaletti\inst{\ref{lesia}} \and
    M. Bonnefoy\inst{\ref{ipag}} \and
    A.-L. Maire\inst{\ref{star},\ref{mpia}} \and
    D. Mesa\inst{\ref{padova}} \and
    G. Chauvin\inst{\ref{ipag},\ref{umi}} \and 
    S. Desidera\inst{\ref{padova}} \and
    A. Vigan\inst{\ref{lam}} \and
    A. Cheetham\inst{\ref{geneva}} \and
    J. Hagelberg\inst{\ref{geneva}} \and
    M. Feldt\inst{\ref{mpia}} \and
    M. Meyer\inst{\ref{umich},\ref{eth}}  \and
    P. Rubini\inst{} \and
    H. Le Coroller\inst{\ref{lam}} \and
    F. Cantalloube\inst{\ref{mpia}} \and
    B. Biller \inst{\ref{mpia},12,13} \and
    M. Bonavita\inst{\ref{ifa_uoe}} \and
    T. Bhowmik\inst{\ref{lesia}} \and
    W. Brandner\inst{\ref{mpia}} \and
    S. Daemgen\inst{\ref{eth}} \and
    V. D'Orazi\inst{\ref{padova}} \and
    O. Flasseur\inst{\ref{CRAL}}  \and
    C. Fontanive\inst{\ref{bern}},\inst{\ref{padova}} \and
    R. Galicher \inst{\ref{lesia}} \and
    J. Girard\inst{\ref{ipag}} \and
    P. Janin-Potiron\inst{\ref{lesia}} \and
    M. Janson\inst{\ref{Stockholm}},\inst{\ref{mpia}} \and
    M. Keppler\inst{\ref{mpia}} \and
    T. Kopytova\inst{\ref{mpia}},\inst{\ref{russia}},\inst{\ref{DMIC}} \and
    E. Lagadec \inst{\ref{oca}} \and
    J. Lannier\inst{\ref{ipag}} \and
    C. Lazzoni\inst{\ref{padova2}} \and
    R. Ligi \inst{\ref{oca}} \and
    N. Meunier\inst{\ref{ipag}} \and
    A. Perreti\inst{\ref{geneva}} \and
    C. Perrot \inst{\ref{lesia}},\inst{\ref{Valparaiso}},\inst{\ref{NPF}} \and
    L. Rodet\inst{\ref{ipag}} \and
    C. Romero \inst{\ref{ipag}},\inst{\ref{eso_chili}} \and
    D. Rouan\inst{\ref{lesia}} \and
    M. Samland\inst{\ref{Stockholm}},\inst{\ref{mpia}} \and
    G. Salter\inst{\ref{lam}} \and
    E. Sissa\inst{\ref{padova}} \and
    T. Schmidt\inst{\ref{lesia}} \and
    A. Zurlo \inst{\ref{diegoportales1},\ref{diegoportales2},\ref{lam}} \and
    D. Mouillet\inst{\ref{ipag}} \and
    L. Denis\inst{\ref{LHC}} \and
    E. Thi\'ebaut\inst{\ref{CRAL}} \and
    J. Milli\inst{\ref{ipag}} \and
    Z.~Wahhaj\inst{\ref{eso_chili}} \and
    J.-L. Beuzit \inst{\ref{lam}} \and
    C. Dominik \inst{\ref{ams}} \and
    Th. Henning\inst{\ref{mpia}} \and
    F. M\'enard \inst{\ref{ipag}} \and
    A. M\"uller\inst{\ref{mpia}} \and
    H.M. Schmid \inst{\ref{eth}} \and
    M. Turatto \inst{\ref{padova}} \and
    S. Udry \inst{\ref{geneva}} \and
    L.~Abe\inst{\ref{oca}} \and
    J.~Antichi\inst{\ref{padova}} \and
    F. Allard\inst{\ref{CRAL}}
    A.~Baruffolo\inst{\ref{padova}} \and
    P.~Baudoz\inst{\ref{lesia}} \and
    J.~Baudrand\inst{\ref{lesia}} \and
    A.~Bazzon\inst{\ref{eth}} \and
    P.~Blanchard\inst{\ref{lam}} \and
    M.~Carbillet\inst{\ref{oca}} \and
    M.~Carle\inst{\ref{lam}} \and
    E.~Cascone\inst{\ref{padova}} \and
    J.~Charton\inst{\ref{ipag}} \and
    R.~Claudi\inst{\ref{padova}} \and
    A.~Costille\inst{\ref{lam}} \and
    V.~De Caprio\inst{\ref{capodimonte}} \and
    A.~Delboulb\'e\inst{\ref{ipag}} \and
    K.~Dohlen\inst{\ref{lam}} \and
    D.~Fantinel\inst{\ref{padova}} \and
    P.~Feautrier\inst{\ref{ipag}} \and
    T.~Fusco\inst{\ref{onera},\ref{lam}} \and
    P.~Gigan\inst{\ref{lesia}} \and
    E.~Giro\inst{\ref{padova}} \and
    D.~Gisler\inst{\ref{eth}} \and
    L.~Gluck\inst{\ref{ipag}} \and
    C.~Gry\inst{\ref{lam}} \and
    N.~Hubin\inst{\ref{eso_garching}} \and
    E.~Hugot\inst{\ref{lam}} \and
    M.~Jaquet\inst{\ref{lam}} \and
    M.~Kasper\inst{\ref{eso_garching},\ref{ipag}} \and
    D.~Le Mignant\inst{\ref{lam}} \and
    M.~Llored\inst{\ref{lam}} \and
    F.~Madec\inst{\ref{lam}} \and
    Y.~Magnard\inst{\ref{ipag}} \and
    P.~Martinez\inst{\ref{oca}} \and
    D.~Maurel\inst{\ref{ipag}} \and
    S. Messina\inst{\ref{catania}} \and
    O.~M\"oller-Nilsson\inst{\ref{onera}} \and
   L.~Mugnier\inst{\ref{onera}} \and 
    T.~Moulin\inst{\ref{ipag}} \and
    A.~Orign\'e\inst{\ref{lam}} \and
    A.~Pavlov\inst{\ref{mpia}} \and
    D.~Perret\inst{\ref{lesia}} \and
    C.~Petit\inst{\ref{onera}} \and
    J.~Pragt\inst{\ref{ipag}} \and
    P.~Puget\inst{\ref{ipag}} \and
    P.~Rabou\inst{\ref{ipag}} \and
    J.~Ramos\inst{\ref{ipag}} \and
    F.~Rigal\inst{\ref{ipag}} \and
    S.~Rochat\inst{\ref{ipag}} \and
    R.~Roelfsema\inst{\ref{nova}} \and
    G.~Rousset\inst{\ref{lesia}} \and
    A.~Roux\inst{\ref{ipag}} \and
    B.~Salasnich\inst{\ref{padova}} \and
    J.-F.~Sauvage\inst{\ref{onera},\ref{lam}} \and
    A.~Sevin\inst{\ref{lesia}} \and
    C.~Soenke\inst{\ref{eso_garching}} \and
    E.~Stadler\inst{\ref{ipag}} \and
    M.~Suarez\inst{\ref{eso_garching}} \and
    L.~Weber\inst{\ref{geneva}} \and
    F.~Wildi\inst{\ref{geneva}} 
	E.~Rickman\inst{\ref{geneva}}
}

\institute{
    CRAL, UMR 5574, CNRS, Universit\'e de Lyon, ENS, 9 avenue Charles Andr\'e, 69561 Saint Genis Laval Cedex, France \label{CRAL} 
    \\ \email{\href{mailto:maud.langlois@univ-lyon1.fr}{maud.langlois@univ-lyon1.fr}} 
    \and
  Aix Marseille Univ, CNRS, CNES, LAM, Marseille, France \label{lam}
    \and 
    INAF - Osservatorio Astronomico di Padova, Vicolo della Osservatorio 5, 35122, Padova, Italy \label{padova}
      \and
    Univ. Grenoble Alpes, CNRS, IPAG, F-38000 Grenoble, France \label{ipag}
       \and
    LESIA, Observatoire de Paris, Universit\'e PSL, CNRS, Sorbonne Universit\'e, Univ. Paris Diderot, Sorbonne Paris Cit\'e, 5 place Jules Janssen, 92195 Meudon, France \label{lesia} 
     \and 
    Max Planck Institute for Astronomy, K\"onigstuhl 17, D-69117 Heidelberg, Germany \label{mpia}
    \and 
    STAR Institute, Universit\'e de Li\`ege, All\'ee du Six Ao\^ut 19c, B-4000 Li\`ege, Belgium \label{star}
    \and
    Geneva Observatory, University of Geneva, Chemin des Mailettes 51, 1290 Versoix, Switzerland \label{geneva}
    \and
    Department of Astronomy, University of Michigan, Ann Arbor, MI 48109, USA \label{umich}
    \and
    Institute for Particle Physics and Astrophysics, ETH Zurich, Wolfgang-Pauli-Strasse 27, 8093 Zurich, Switzerland \label{eth}
    \and 
    Institute for Astronomy, University of Edinburgh, EH9 3HJ, Edinburgh, UK \label{ifa_uoe}
    \and
    Scottish Universities Physics Alliance (SUPA), Institute for Astronomy, University of Edinburgh, Blackford Hill, Edinburgh EH9 3HJ, UK \label{supa}
    \and
    Center for Space and Habitability, University of Bern, 3012 Bern, Switzerland \label{bern}
    \and
    Universit\'e C\^ote d’Azur, OCA, CNRS, Lagrange, France \label{oca}
    \and
    European Southern Observatory, Karl-Schwarzschild-Str. 2, 85748 Garching, Germany \label{eso_garching}
    \and
    European Southern Observatory, Alonso de Cordova 3107, Casilla 19001 Vitacura, Santiago 19, Chile \label{eso_chili}
    \and
    N\'ucleo de Astronom\'ia, Facultad de Ingenier\'ia y Ciencias, Universidad Diego Portales, Av. Ejercito 441, Santiago, Chile 
    \label{diegoportales1}
    \and
    Escuela de Ingenier\'ia Industrial, Facultad de Ingenier\'ia y Ciencias, Universidad Diego Portales, Av. Ejercito 441, Santiago, Chile \label{diegoportales2}
    \and
    Anton Pannekoek Institute for Astronomy, Science Park 9, NL-1098 XH Amsterdam, The Netherlands \label{ams}
    \and
    Unidad Mixta Internacional Franco-Chilena de Astronom\'{i}a, CNRS/INSU UMI 3386 and Departamento de Astronom\'{i}a, Universidad de Chile, Casilla 36-D, Santiago, Chile \label{umi}
    \and
    DOTA, ONERA (Office National dEtudes et de Recherches Arospatiales), Université Paris Saclay, F-92322 Chatillon, France
     \label{onera}
    \and
    NOVA Optical Infrared Instrumentation Group, Oude Hoogeveensedijk 4, 7991 PD Dwingeloo, The Netherlands \label{nova}
    \and
    INAF - Osservatorio Astronomico di Capodimonte, Salita Moiariello 16, 80131 Napoli, Italy \label{capodimonte}
    \and
    Instituto de F\'isica y Astronom\'ia, Facultad de Ciencias, Universidad de Valpara\'iso, Av. Gran Breta\~na 1111, Valpara\'iso, Chile \label{Valparaiso}
    \and
    N\'ucleo Milenio Formaci\'on Planetaria - NPF, Universidad de Valpara\'iso, Av. Gran Breta\~na 1111, Valpara\'iso, Chile
    \label{NPF}
    \and
    Univ Lyon, UJM-Saint-Etienne, CNRS, Institut d Optique Graduate School, Laboratoire Hubert Curien UMR 5516, F-42023, SAINT-ETIENNE, France \label{LHC}
    \and 
    Dipartimento di Fisica a Astronomia "G. Galilei", Universita' di Padova, Via Marzolo, 8, 35121 Padova, Italy \label{padova2}
    \and
    Department of Astronomy, Stockholm University, AlbaNova University Center, SE-10691 Stockholm, Sweden
    \label{Stockholm}
    \and
    Division of Medical Image Computing, German Cancer Research Center (DKFZ), 69120 Heidelberg, Germany \label{DMIC}
    \and
    Ural Federal University, Yekaterinburg, 620002, Russia \label{russia}
    \and
    INAF- Catania Astrophysical Observatory, via S. Sofia 78, I-95123 Catania, Italy \label{catania}
}

\date{Received ???; accepted ???}

\abstract{
Over the past decades, direct imaging has confirmed the existence of substellar companions (exoplanets or brown dwarfs) on wide orbits (>10 au) from their host stars. To understand their formation and evolution mechanisms, we have initiated in 2015 the SPHERE infrared survey for exoplanets (SHINE), a systematic direct imaging survey of young, nearby stars to explore their demographics.}
{We aim to detect and characterize the population of giant planets and brown dwarfs beyond the snow line around young, nearby stars. Combined with the survey completeness, our observations offer the opportunity to constrain the statistical properties (occurrence, mass and orbital distributions, dependency on the stellar mass) of these young giant planets.}
{In this study, we present the observing and data analysis strategy, the ranking process of the detected candidates, and the survey performances for a subsample of 150 stars, which are representative of the full SHINE sample. The observations were conducted in an homogeneous way from February 2015 to February 2017 with the dedicated ground-based VLT/SPHERE instrument equipped with the IFS integral field spectrograph and the IRDIS dual-band imager covering a spectral range between 0.9 and 2.3 $\mu$m. We used coronographic, angular and spectral differential imaging techniques to reach the best detection performances for this study down to the planetary mass regime.}
{
We have processed in a uniform manner more than 300 SHINE observations and datasets to assess the survey typical sensitivity as a function of the host star, and of the observing conditions. The median detection performance reaches $5\sigma$-contrasts of 13\,mag at 200\,mas and 14.2\,mag at 800\,mas with the IFS (YJ and YJH bands), and of 11.8\,mag at 200\,mas, 13.1\,mag at 800\,mas and 15.8\,mag at 3\,as with IRDIS in H  band, delivering one of the deepest sensitivity surveys so far for young nearby stars.
A total of  sixteen   substellar   companions   were   imaged  in this first part of  SHINE: seven   brown   dwarf   companions, and ten planetary-mass companions. They include the two new discoveries HIP\,65426\,b and  HIP\,64892\,B, but not the planets around PDS70 not originally select in the SHINE core sample. A total of 1483 candidates were detected, mainly in the large field-of-view of IRDIS. Color-magnitude diagrams, low-resolution spectrum when available with IFS, and follow-up observations, enabled to identify the nature (background contaminant or comoving companion) of about 86\,\% of them. The remaining cases are often connected to crowded field missing follow-up observations. Finally, although SHINE was not designed for disk searches, twelve circumstellar disks were imaged including three new detections around the HIP\,73145, HIP\,86598 and HD\,106906 systems. }
{Nowadays, direct imaging brings a unique opportunity to probe the outer part of exoplanetary systems beyond 10\,au to explore planetary architectures as highlighted by the discoveries of one new exoplanet, one new brown dwarf companion, and three new debris disks during this early phase of SHINE. It also offers the opportunity to explore and revisit the physical and orbital properties of these young giant planets and brown dwarf companions (relative position, photometry and low-resolution spectrum in near-infrared, predicted masses, contrast to search for additional companions). Finally, these results highlight the importance to finalize the SHINE systematic observation of about 500 young, nearby stars, for a full exploration of their outer part to explore the demographics of young giant planets beyond 10\,au, and to nail down the most interesting systems for the next generation of high-contrast imagers on very large and extremely large telescopes.
}
\maketitle
\section{Introduction and context}
\label{sec:introduction}
The discovery of the first brown dwarf companion Gl\,229\,B benefited from the combined technological innovation of infrared detectors and high contrast techniques \citep{Nakajima1995}. Following this discovery, the first generation of dedicated planet imagers on 10-m class telescopes (NaCo at VLT, NIRC2 at Keck, NICI at Gemini) conducted systematic surveys of young and nearby stars. They led the first direct detections of planetary mass companions in the early 2000's. These companions were detected at distances larger than several hundreds astronomical units (au) and/or with a mass ratio not much smaller than a tenth with respect to their host primaries (except for brown dwarf primaries), giving hints of formation via gravo-turbulent fragmentation \citep{2011ApJ...743L..29H} or gravitational disk instability \citep{1997LPI....28..137B}. Thanks to the improvement of direct imaging observation and data analysis techniques with ground-based adaptive optics systems (AO) or space telescopes, a few planetary mass objects and low-mass brown dwarfs have been detected since the first detection by \cite{Chauvin2004}. Moreover, these developments, enabled the discoveries of giant planets within 100\,au around young, nearby, and dusty early type stars like HR8799 b,c,d,e \citep{Marois2008,2010Natur.468.1080M}, $\beta$ Pictoris b \citep{Lagrange2009}, and more recently HD\,95086\,b \citep{2013ApJ...772L..15R}, and GJ\,504\,b \citep{2013ApJ...774...11K}.

Direct imaging is the only viable technique to probe for planets at large separations with single epoch observations, but detecting them requires to overcome the difficulties caused by the angular proximity and the high contrast involved. With improved instruments and data reduction techniques, we are currently initiating the characterization of the giant planet population at wide orbits, between typically 10-100\,au. More than a decade of direct imaging surveys targeting several hundred young, nearby stars have lead to the discovery of approximately a dozen sub-stellar companions excluding the companions to brown dwarfs located in the star vicinity (within 100\,au). Despite the relatively small number of discoveries compared with other techniques, such as radial velocity and transit, each new imaged giant planet has provided unique clues on the formation, evolution and physics of young massive Jupiters. 

Early surveys performed using the first generation of planet imagers enabled to conduct systematic surveys relatively modest in size sampling each less than hundreds of young nearby stars \citep{Chauvin2018}. Various strategies were followed for the target selection of these surveys: i/ complete census of given associations \citep{2003A&A...404..157C,2005ApJ...625.1004M,2007A&A...472..321K,Chauvin2010}, ii/ selection of young, intermediate-mass stars \citep{Janson2011,2013A&A...553A..60R,2013ApJ...776....4N}, iii/ or very low-mass stars \citep{Delorme2012,2012A&A...548A..33C,2015ApJS..216....7B,2016A&A...596A..83L}, iv/ application of figures of merit considering detection rate with toy models of planet population \citep{2013ApJ...777..160B}. Numerous direct imaging surveys to detect giant planet companions have reported null-detections \citep{2005ApJ...625.1004M,2007A&A...472..321K,Chauvin2010,Biller2007, Lafreniere2007, Ehrenreich2010, Chauvin2010, Janson2011, Delorme2012}, but this allowed to set upper limits to the occurrence of giant planets.

Ongoing surveys target several hundreds stars, with the largest surveys to date due to be completed in the next years. SpHere INfrared survey for Exoplanets (SHINE, \citealt{Chauvin2017}), Gemini GPIES \citep{Macintosh2015} and SCExAO \citep{Jovanovic2016} now combine dedicated extreme adaptive optics systems with coronagraphic and both angular and spectral differential image processing. With enhanced detection performances, the objective is to significantly increase the number of imaged planets to characterize, but also to provide better statistical constraints on the occurrence and the characteristics of exoplanets at wide orbits ($>$10 au). This should give us a more global picture of planetary systems architecture at all orbits to improve our understanding of planetary formation and evolution mechanisms. Early-results already confirm the gain in terms of detection performances compared with the first generation of planet imagers \citep{Nielsen2019, Vigan2020}. 

The near-infrared wavelengths (H and K-bands) have been used intensively. They are a good compromise between low-background noise, high angular resolution and good Strehl correction. However, thermal imaging at L or M-bands has been very competitive in terms of detection performances as the planet-star contrast and the Strehl correction are more favorable in those wavelengths despite an increased thermal background and larger inner working angle. For instance, SPHERE could be in some cases less sensitive than NaCo for the detection of giant planets around young, nearby M dwarfs at typical separations larger than 500-1000 mas.
 
Both SHINE and GPIES surveys have recently discovered three new exoplanets \citep{Macintosh2015,Chauvin2017,Keppler2018}, and a few additional higher mass brown dwarfs \citep{Konopacky2016,Cheetham2018}. Smaller surveys using SPHERE and GPI also discovered several substellar companions \citep{Milli2017,Wagner2020,Bohn2020}. They offer unprecedented detection, astrometric and spectrophotometric capabilities which allow us to characterize fainter and closer giant planets, such as the recent discovery of 51 Eri b (2 MJup  at 14 au, T5-type, of age 20 Myr; \citealt{Macintosh2015, Samland2017}), HIP 65426 b, a young, warm, and dusty L5-L7 massive jovian planet located at about 92 au from its host star \citep{Chauvin2017}, and the young solar analogue PDS\,70 now known to actually host two planets PDS\,70\,b discovered during the SHINE campaign  \citep{Keppler2018,Muller18} and PDS\,70\,c by MUSE \citep{Haffert2018a}. Such surveys also provide key spectral and orbital characterisation data for known exoplanets \citep[e.g.][]{DeRosa2016,Samland2017,Chauvin2018,Wang2018,Cheetham2019,Lagrange2019,Maire2019}. Despite these new discoveries, SHINE and GPIES have yielded significantly fewer exoplanet detections than predicted using extrapolations of radial velocity planet populations to larger semi-major axes \citep[e.g.][]{Cumming2008}. This results in setting strong statistical constraints on the distribution of giant exoplanets at separations $>$10\,au from their stars, as well as sub-stellar companions to young stars \citep{Nielsen2019, Vigan2020}. As these systems are young ($<$100\,Myr), and thus closer to their epoch of formation than for instance radial velocity planets (typically 1--10\,Gyr), statistical analysis of large direct-imaging surveys can provide hints into the potential formation mechanism responsible for producing giant exoplanets orbiting at wide orbits. 

This paper is part of a series of three papers describing early results obtained from analysis of a subset of the SHINE data. This paper describes the observations, data quality, analysis methods, and presents results in terms of detections and upper limits for this subset of SHINE data. Two associated papers describe the general characteristics of the survey sample \citep{Desidera2020} and the statistical analysis and a discussion of the implication for formation scenarios \citep{Vigan2020}. 

Following this introduction, we present in section \ref{sec:observations} the observations that were performed to underpin this work. In Sect. \ref{sec:Data_Analysis}, we describe the reduction and calibration of the data sets. Section \ref{sec:perf} is dedicated to the main results of the survey, including the detection limits of the survey. Finally, we summarize, in Sect \ref{sec:detections}, the characterization of newly discovered and known substellar companions along with a description of the circumstellar disks detected within the survey data. The conclusions and prospects are drawn in section \ref{sec:conclusions}. The relative astrometry and photometry of companion candidates from the sample is listed in Annex \ref{sec:astrophot}.

\section{The SHINE survey}
\label{sec:observations}

The SpHere INfrared survey for Exoplanets (SHINE: \citealt{Chauvin2017b}) has been designed for 200 telescope nights, allocated in visitor mode, using the SPHERE consortium guaranteed time. SHINE has been designed by the SPHERE consortium to: (i) identify and characterized new planetary and brown dwarf companions; (ii) study the architecture of planetary systems (multiplicity and dynamical interactions); (iii) investigate the link between the presence of planets and disks (in synergy with the GTO program aimed at disk characterization); (iv) determine the frequency of giant planets beyond 10\,au; (v) investigate the impact of stellar mass (and even age if possible) on the frequency and  characteristics of planetary companions over the range 0.5 to 3.0 $M_{\odot}$.

The SHINE survey started in February 2015 and will be completed in mid-2021, with observations of 500 targets out of a larger sample of 800 nearby young stars aiming at searching for new sub-stellar companions. The sample is oversized with respect to the available telescope time by a factor of approximately two, on the basis of the adopted observing strategy, which relies on observations surrounding meridian passage in order to achieve the maximum field-of-view (FoV) rotation for optimal angular differential imaging. This requires some flexibility in the target list in order to optimize the scheduling. 

The general design of the survey, the sample selection and the simulations performed for building it, the parameters of the individual targets used in this early statistical analysis, and the general properties of the sample used in this series of papers are described in detail in \cite{Desidera2020}. We describe in this paper early results obtained from the analysis of about a third of the large SHINE survey sample, considering only those targets whose first observations were done before Feb 2017.
Several second epochs observations were also performed between 2016 and 2019 to discriminate companions from background stars on the selected sample described in this paper. This selected sample dataset including 150 targets, is already large enough for reviewing the survey efficiency, to discuss the incidence of massive planets at a separation $> 10$~au, and to have new indications about the formation scenarios for giant planets.

\subsection{Observations setup}

All the observations were performed with the Spectro-Polarimeter High-contrast Exoplanet REsearch \citep[SPHERE;][]{Beuzit2019} combining its SAXO extreme adaptive optics system \citep{Fusco2006,Sauvage2016,Fusco14} and its apodized pupil Lyot coronagraphs \citep{boccaletti2008,carbillet2011,guerri2011}. Observations were acquired in either \texttt{IRDIFS} or \texttt{IRDIFS-EXT} mode, i.e. with both NIR sub-systems, IFS \citep{Claudi2008} and IRDIS \citep{Dohlen2008}, observing in parallel \citep{Zurlo2014,Mesa2015}. The IFS covers a 1.7\as$\times$1.7\as FoV and IRDIS covers a more or less circular, unvignetted FoV of diameter $\sim$9\as. The APLC$\_$YJHs and APLC$\_$Ks coronagraphic configurations were used for the \texttt{IRDIFS} and \texttt{IRDIFS-EXT} observations respectively, and in IRDIS all first epoch observations were performed with the DB-H23 and DB-K12 dual-band filter pairs \citep{Vigan2010}. Due to the presence of known companions and/or the detection of (new) candidate companions, some targets were observed multiple times for astrometric monitoring. In addition to the initial selected filters, follow-up observations were performed in different configurations with IRDIS, for example with the broad-band BB-H and dual-band DB-J23 filters, as listed in the filter column of Table \ref{tab:detections}. This resulted in a varying number of observations for each target.

\subsection{Observations Optimized planning: SPOT}

Given the large number of targets, each associated with a priority and an urgency (how soon the observation has to be made), and the various observing constrains, including those connected to Angular Differential Imaging (ADI: \citealt{Marois2006}) observations, we built a dedicated tool, SPOT, to deliver an optimised scheduling of the observations, both on long and short terms. SPOT is based on  simulated annealing. It is described in in details in \citet{Lagrange2016}. 

In brief, SPOT uses as inputs the calendar of allocated observing nights, the list of targets available for e.g. the whole semester, together with their associated instrumental set up (that is associated with specific overheads) and any specific scheduling constrains (needed for second epoch observations), the targets coordinates and magnitude, the minimal coronagraphic exposition duration, the maximum air mass, the maximum proportion (in time) of the observation that is allowed either before or after the meridian crossing, the minimum exposure time, the amount of FoV rotation during the observation. It must be noted that the amount of FoV rotation depends on the target coordinates and on the actual time and duration of observations. As it also drives the exposure time, we also set a maximum exposition time, to avoid extending too much the duration of the exposure. 

Classical, additional scheduling constraints are applied on all targets: minimal angular distance to the Moon, avoidance of zenith observations. Poor atmospheric conditions may also be taken into account through pointing restrictions that usually depend on the wind direction and speed and through magnitude restrictions in case of non photometric conditions. Finally, SPOT also takes into account the various overheads and the need for astrometric and spectro-photometric calibrations. The astrometric calibrations were to be observed as much as possible in the first night of each run, and the spectro-photometric in the following night. After optimisation SPOT returns a schedule, and  produces the Observing Blocks (OBs) that can be automatically transferred to P2 (the ESO observing preparation tool). 

In practice, we generally request the coronagraphic data to be obtained while the target is crossing the meridian at least 15 min before and after the meridian passage, and with at least 30 degrees of FoV rotation, unless it required more than 7200 s. The minimum exposure time was set to 3600 s. The scheduling of the targets was always optimal; the efficiency of the night could be as high as 100\% when typically 2-3 time more targets were available than actually those scheduled. Yet, when not enough targets were available as inputs, short (typically 30 min) holes could be present in the schedules during the nights. These holes were used to observe fillers requiring short exposure times (binaries, stars with RV trends, astrometric and spectrophotometric standards). By comparing with other surveys executed in service such as BEAST (Janson et al., in preparation), we conclude that use of SPOT allowed an increase of more than 30\% in the field rotation angles for identical observing time with respect to the service scheduling routinely used at ESO.

\subsection{Observing conditions and Data quality}

The SPHERE Instruments are fed by an extreme adaptive optics (AO) system called SAXO. It delivers a very high Strehl ratio, which reaches above 90\% in H-band for very good observing conditions by correcting both perturbations induced by the atmospheric turbulence and from the internal aberrations of the instrument itself. A comprehensive  description of the SAXO design can be found in \cite{Fusco16} and \cite{Beuzit2019}. We derived here the overall statistical AO telemetry data from SAXO and from the ESO MASS-DIMM measurements for the survey observations and relate these parameters to the Strehl ratio, raw contrast and processed contrast in order to evaluate the performance constraints from these observations. The telemetry data points are spread over 130 different nights and cover more than 200 different observations. The AO telemetry data (available for a large number of our survey observations) includes estimates from the real-time computer (called SPARTA) several quantities of interest that could be related to the final performances (hereafter SPARTA data) such as: the Strehl ratio and additional atmospheric parameters including the seeing and the coherence time. The Strehl is defined  at 1.6\,$\mu$m, while the seeing and coherence time are defined at 500 nm.  

\begin{figure*}
\begin{tabular}{cc}
    \includegraphics[width=0.87\columnwidth]{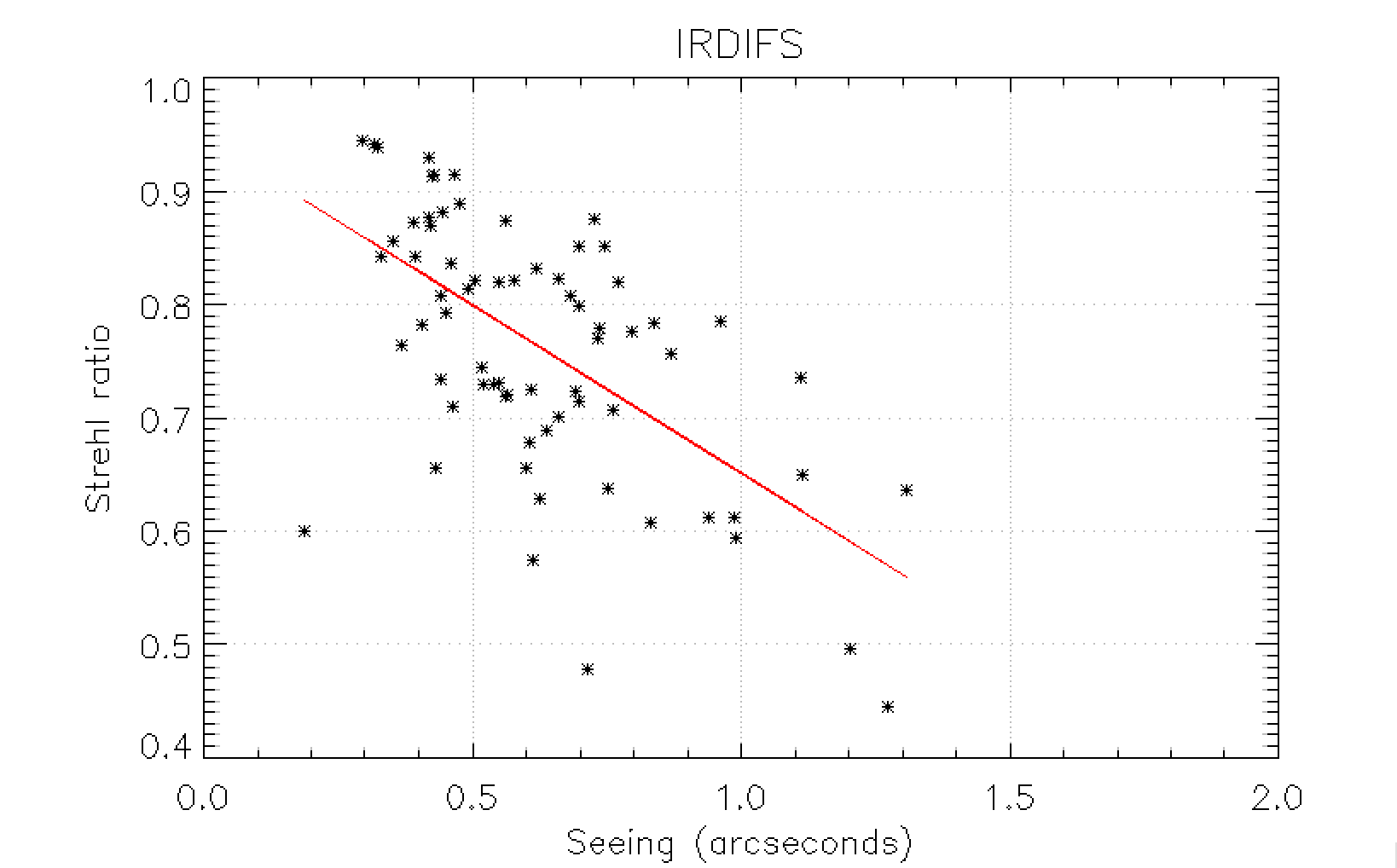} &
    \includegraphics[width=0.87\columnwidth]{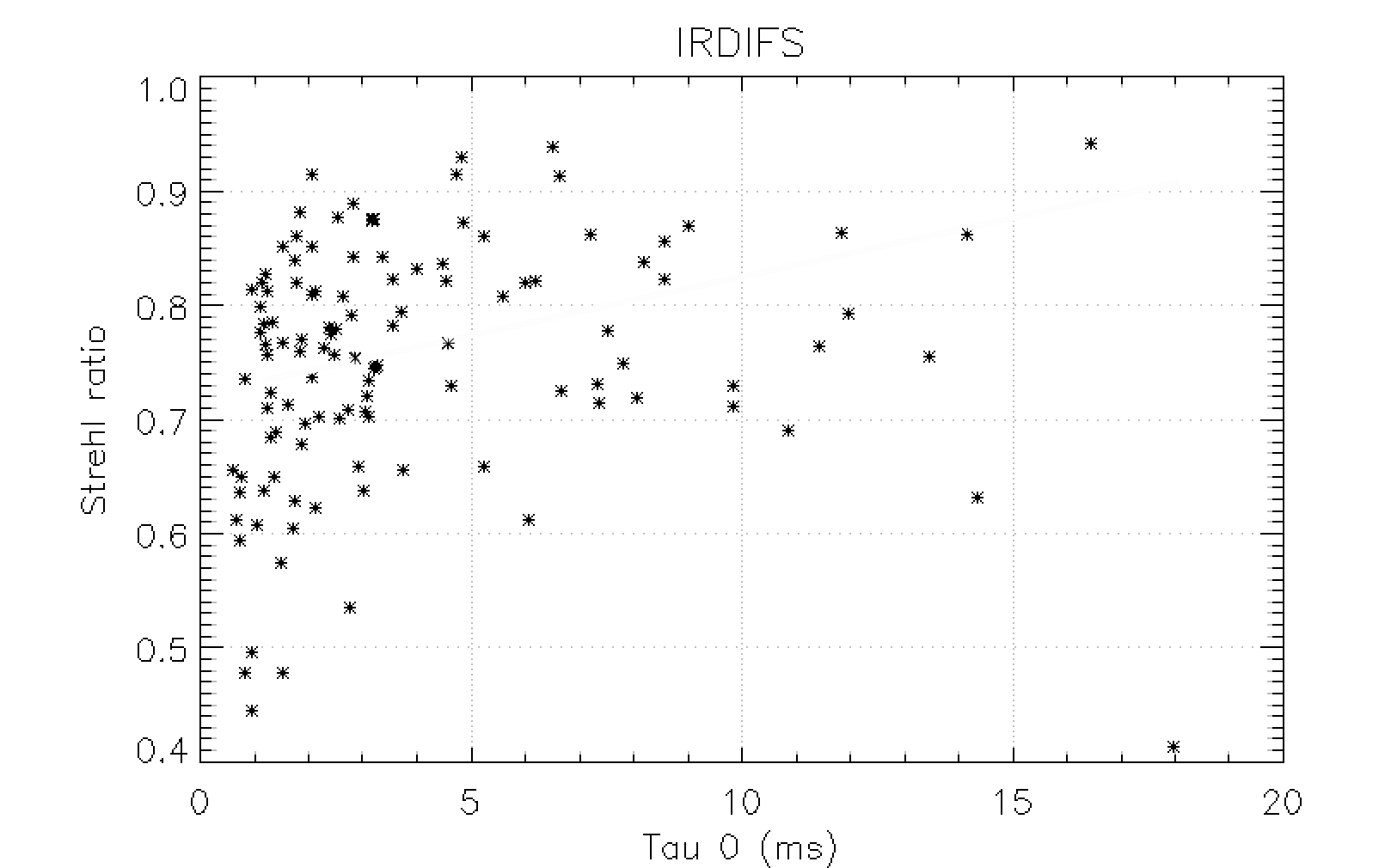} \\
    \includegraphics[width=0.87\columnwidth]{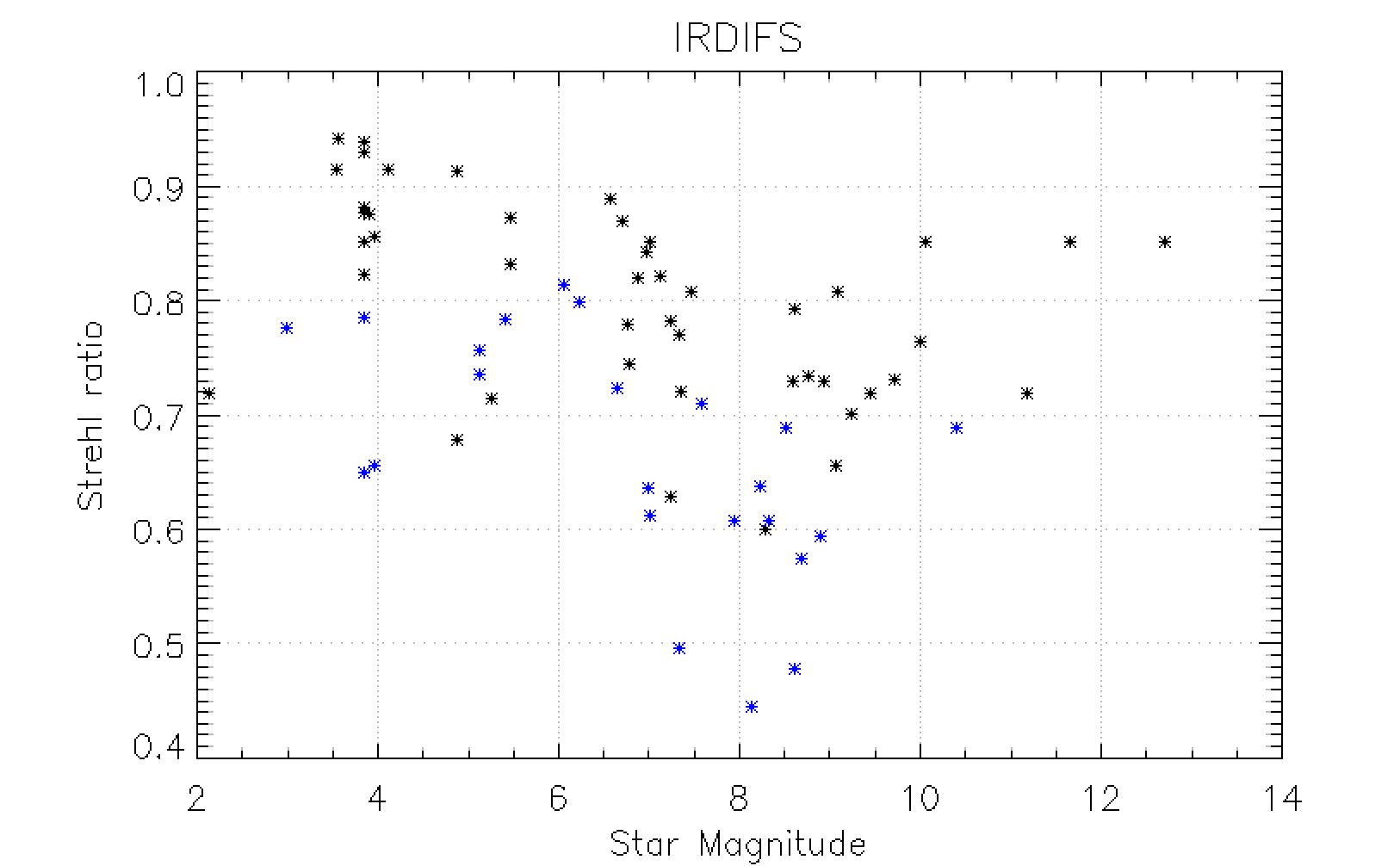} &
    \includegraphics[width=0.87\columnwidth]{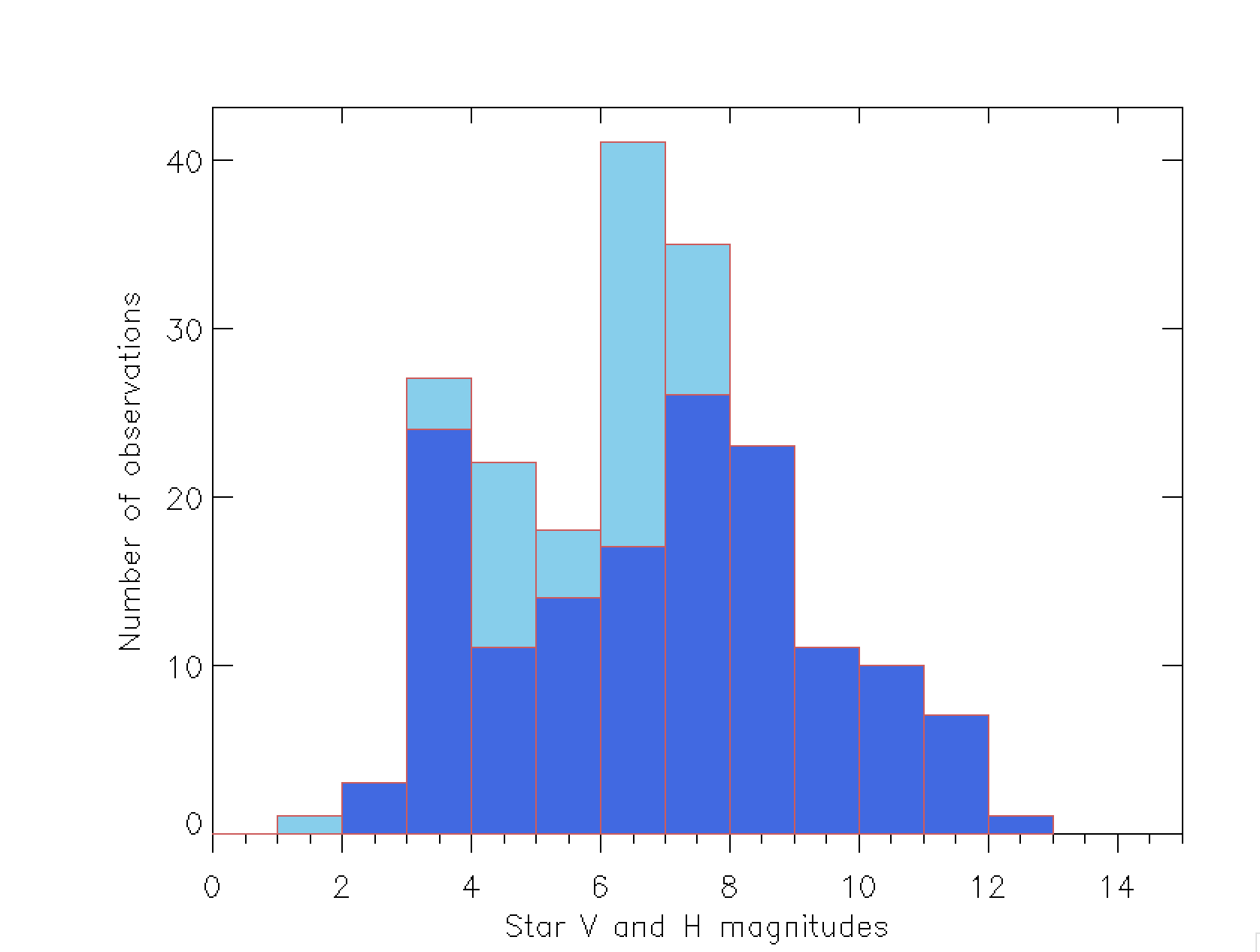} \\
    \end{tabular}
    \caption{Observing conditions and AO performances in IRDIFS observing mode: (Top-Left) Strehl ratio as a function of the seeing measured by SPARTA. (Top-Right) Strehl Ratio as function of the atmospheric coherence time estimated by SPARTA. A linear fit is overlaid in red on the Top figures. (Bottom-Left) Strehl ratio as function of the star V magnitude. The blue data points represent the data taken with coherence time smaller than 3 ms.(Bottom-Right) Histogram illustrating the repartition of the observations (including multiple observations of the same target) as function of the star magnitude in V (dark blue) and H-bands (light blue). The apparent large number of observations of R=3.5 stars is an artificial effect due to the large number of observations targeting Beta Pictoris. }
    \label{fig:sparta}
\end{figure*}

These quantities are also connected here with the brightness of the target (V and H magnitudes), retrieved from the Simbad database, and used as a proxy in V-Band for the number of photons received by the WFS\footnote{For stars with a magnitude $R<10$ (value read from the SPHERE acquisition template), the WFS arm uses a red filter called $LP\_780$ blocking wavelengths below 780nm. As a result the central wavelength of the WFS arm is shifted to redder wavelengths, approximately 850 nm, and the nearest broadband filter is the I filter ($\lambda_c=806$nm, $\Delta\lambda=149$nm). However, not all stars have a measurement of their I magnitude in Simbad, we therefore used the V mag which is most commonly available.}. The distribution in magnitude for the sample considered here (for targets with telemetry measurements) is shown in Fig. \ref{fig:sparta} (Bottom Right). It illustrates the fact that our target selection as described in \cite{Desidera2020} was based on setting a magnitude limit (V$<$12.5) in order to guarantee good AO performances. Our sample selection criteria also favors red targets leading to the faintest H-band observations to be below 8 magnitude. 

The overall good AO performance for the selected range of target brightness is confirmed by the Strehl measurements shown in Fig. \ref{fig:sparta} (bottom left). This Figure shows the distribution of the Strehl ratio as a function of the star magnitude in V-band where the median Strehl is greater than 70$\%$ averaging around 80$\%$ in moderate seeing conditions. This Figure also shows a decrease in the Strehl beyond magnitude 7. The blue data points, representing coherence time smaller than 3 ms, highlight the impact of short coherence time on the performances with Strehl ratio below 80$\%$. It is suggested in \cite{Milli2017b} that the Strehl or its estimation could be affected by the small number of photons reaching the WFS (for 7.5 - 9 magnitude guide stars) while the AO loop is operating at full speed, wich could explain the strehl decrease around magnitude 8. For fainter stars, a slower AO loop frequency of 600 Hz is used in order to still collect a sufficient number of photons per AO cycle for high-order wavefront determination and correction. 

This hypothesis is supported further by the greater decrease in Strehl for the low coherence time data points which explain most of the performance inflection around V=8 magnitude. The large scatter in the Strehl distributions indicates that the seeing is also, as expected, a parameter influencing the Strehl among others such as $\tau_0$ and the star magnitude. Despite this large scatter the (Top Left) plot from the same Figure shows a linear decrease trend of the Strehl with the seeing. This decrease is on average 1.8$\%$ Strehl for an increase in seeing of 0.1$"$ which is similar to the Strehl behavior versus seeing  obtained by \cite{Milli2017b} when correction is applied according to the SPARTA versus DIMM measurements dependency. While the SPARTA seeing provides an estimate which is closer to the image quality in the science frames, it has been shown that the SPARTA seeing estimations are smaller than the ESO DIMM measurements, a fact that may be accounted for the turbulence outer scale. The overall impact of the performance limitation from the seeing on our survey is clearly important as illustrated by the large number of observations occurring for seeing greater than 0.8$"$ (Fig. \ref{fig:sparta} - Top-Left).

The Top-Right part of Fig. \ref{fig:sparta} shows the distribution of the Strehl as a function of the coherence time for our full sample. There is a steep rise in Strehl ratio with the coherence time (for $\tau_0 < 3$ ms)  followed by plateau for larger coherence times (for $\tau_0 < 5$ ms). This shows that for low coherence times, the AO performances are limited by the temporal bandwidth error as mentioned in \cite{Cantalloube2020} which also derives a 3 ms threshold from the system point of view. This is in agreement with the laboratory and first on-sky measurements described in details in \cite{Petit2012}. The overall impact of this performance limitation on our survey is likely important as illustrated by the large number of observations occurring when $\tau_0 < 4$ ms (i.e 70\% of the observations). Further analysis of the impact of this effect should include the seeing contribution in the Strehl error budget to disentangle the seeing and coherence time correlations.

Summarizing, we have shown in this section that the AO performances are clearly related to the observing conditions for our sample. We will further describe the impact of these observational parameters on the high contrast performances in section \ref{sec:perf}.

\section{Data Reduction and Analysis}
\label{sec:Data_Analysis}

Large surveys such as SHINE (especially because of the large IRDIS FoV) unveiled a large number of point sources (mainly background sources) from which true sub-stellar mass companions need to be distinguished. For this reason, both astrometry and photometry of the detected points sources need to be precisely and homogeneously calibrated and extracted to test their companionship and charactezerize their nature. We detail in the following sub-sections the SHINE strategy: i/ to calibrate both IRDIS and IFS on various important aspects (distorsion, plate scale, True North, parallactic angle determination, central start position) in sub-section\,\ref{surveycalib}, ii/ to pre-process the scientific observations and apply basic reduction steps (cosmetics, flat, frame registration and recentering, cross-talk and wavelength calibration) as described in sub-section\,\ref{sec:preprocessing}, iii/ to apply advanced ADI/ASDI algorithms summarized in sub-section\,\ref{sec:postprocessing}, and ultimately extract the companion relative astrometry and photometry as detailed in sub-section. The final error budget for both astrometry and photometry is described in this last sub-section.

\subsection{SPHERE Calibration}
\label{surveycalib}

\begin{figure*}
    \centering
\begin{tabular}{cc}
    \includegraphics[width=\columnwidth,trim={0cm 2cm 0cm 0 cm},clip]{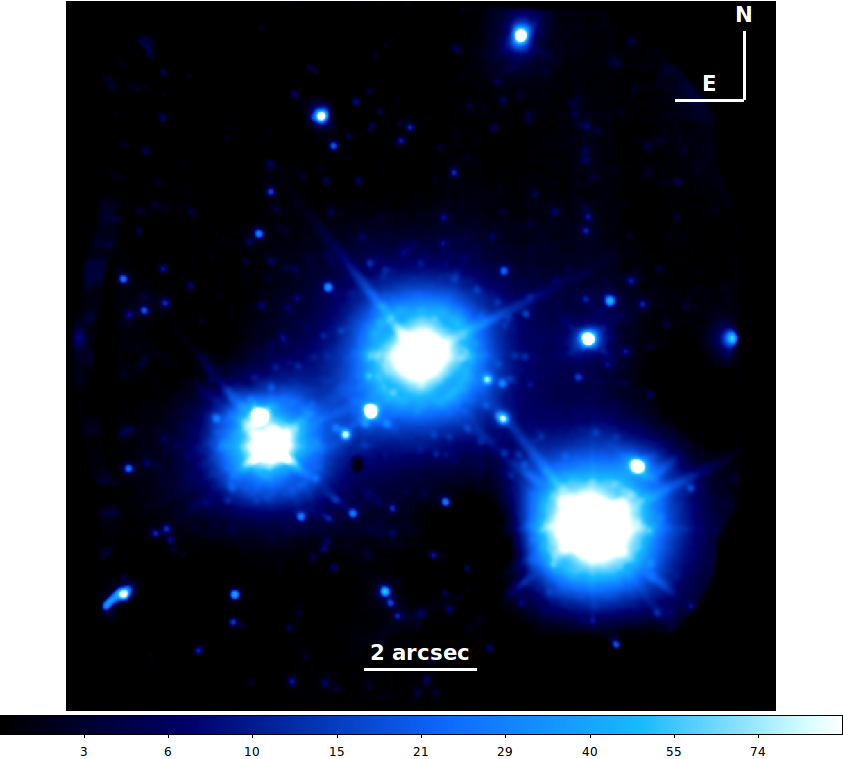} &
    \includegraphics[width=\columnwidth,trim={0cm 2cm 0cm 0 cm},clip]{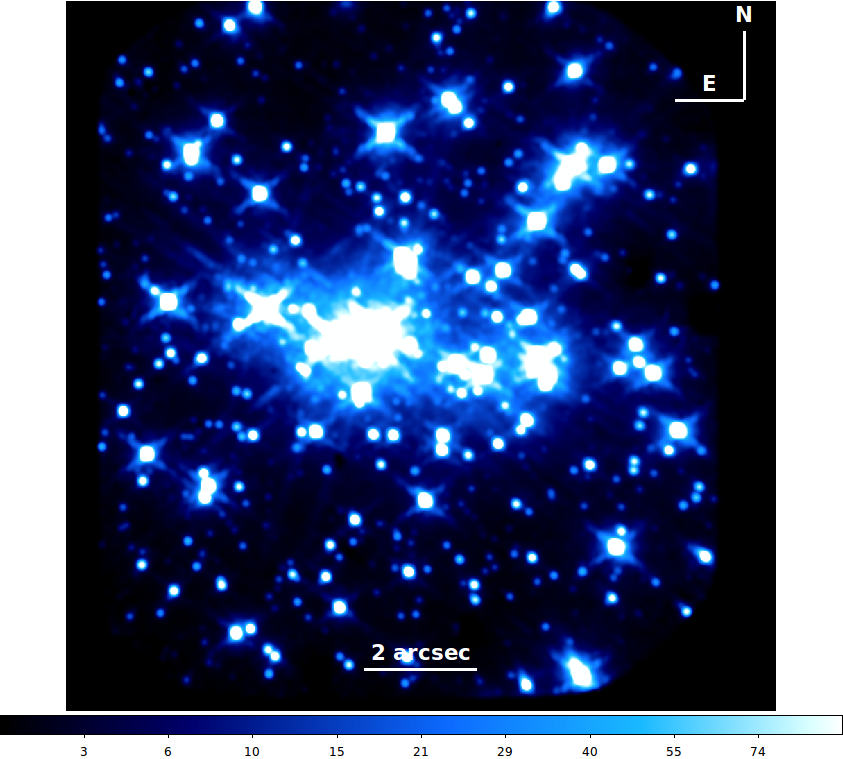} \\
    \includegraphics[width=\columnwidth,trim={0cm 2cm 0cm 0 cm},clip]{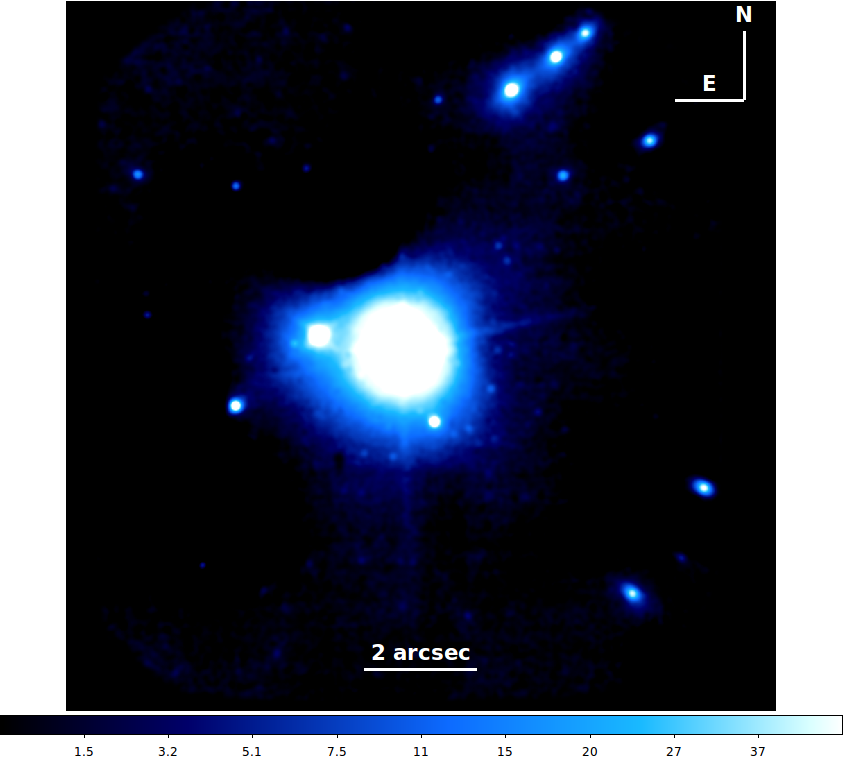} &
    \includegraphics[width=\columnwidth,trim={0cm 2cm 0cm 0 cm},clip]{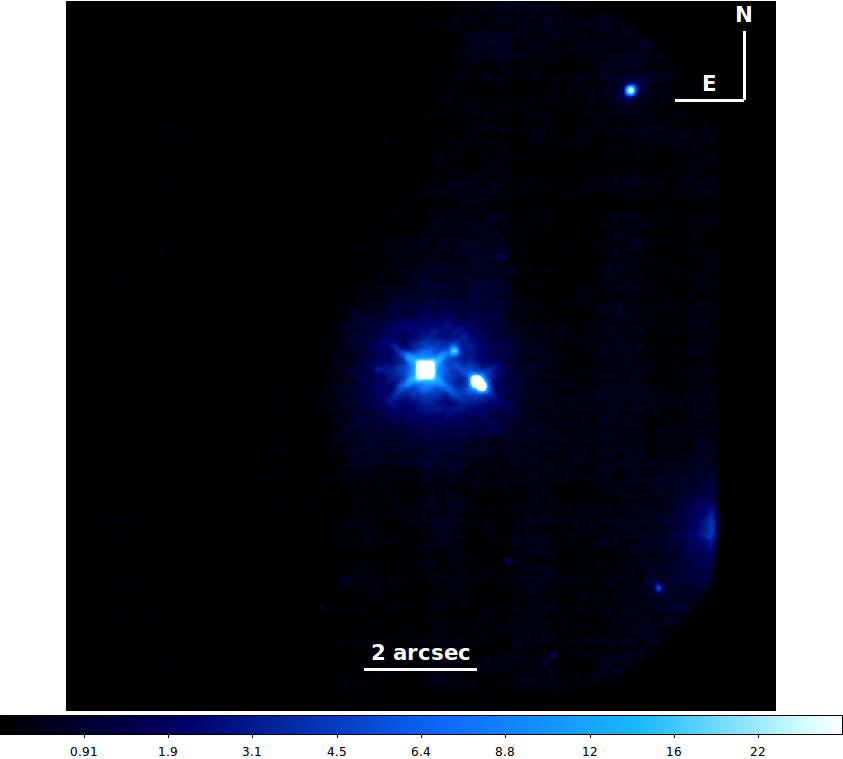} \\
    \end{tabular}
    \caption{IRDIS H2 images of the fields used for the astrometric calibration. Upper left panel: 47 Tuc; upper right panel: NGC~3603; lower left panel: NGC~6380; lower right panel: Orion B1B4. A logarithmic intensity scale is used to show also faint stars.}
    \label{fig:astrometry}
\end{figure*}

\subsubsection{Astrometry}
\label{sec:astrometry}
We present in this section the astrometric calibration of the SHINE survey which depends on several factors described in detail in the following paragraphs: i/ platescale, True North and distorsion correction, ii/ angular offset (between field and pupil-tracking and parallactic angle correction), iii/ star-centering. This section also presents as a conclusion, iv/ a sanity check of the calibration startegy based on Gaia-DR2 results.

\paragraph{Platescale, True North and distorsion}

An extensive description of astrometry with SPHERE can be found in \citet{Maire2016} and \citet{Beuzit2019}. For what concerns in particular the SHINE survey, the astrometric calibration consisted in the correction for the instrument anamorphism ($0.60\pm 0.02$\% between the horizontal and vertical directions of the detector, i.e. 6 mas at 1 arcsec), correction for constant offset angles (between the IFS and IRDIS fields of view, between pupil-tracking mode and field-tracking mode), and determination of the values for the pixel scale and the correction to the true north. We find small but non negligible variations of the last quantities with time requiring dedicated calibration for each observing runs. Appropriate values were estimated for each run using several reference fields of view in clusters with a large number of stars and having accurate (sub-mas) astrometry from Hubble Space Telescope (HST) or ground-based diffraction limited observations: 47~Tuc \citep{Bellini2014, Soto2017, Bellini2017}; NGC~6380 \citep{Bellini2011, Soto2017} and Noyola (private communication); NGC 3603
\citep{Harayama2008, Khorrami2016, Rochau2010}; and Trapezium B1-B4 \citep{Close2012, Close2013} (see Figure~\ref{fig:astrometry}). 

IRDIS was used for all astrometric calibrations, because its field of view allows observation of a large number of stars, in most cases between 50 and 100. Only seven stars were available for the Trapezium B1-B4 field, leading to a less accurate calibration of the field orientation (by 1\%) and on the plate scale (by 0.1\%), and was used only when the other calibration fields were not accessible from February 2015 to March 2016. Due to the less good accuracy of the catalog positions for NGC~6380 and the smaller number of calibrating objects, we used this field only twice in May-June 2015 and once in June 2017. Using several calibrators enable to perform calibrations throughout the year and to do cross-calibration. Because the coronagraph has a small effect on the pixel scale, the astrometric fields were observed with a coronagraphic plate (including an offset for the Trapezium B1-B4 field observation, to shift the B1 star out of the coronagraphic mask).

The pixel scale is slightly different for the H2, H3, K1, and K2 filters, with mean values of 12.255, 12.250, 12.267, and 12.263 mas/pixel, respectively. For IFS, we used a constant value of the pixel scale of $7.46\pm0.02$~mas/pixel. Variations in this case have less impact on the results due to the small field of view. The typical measurement accuracy of the pixel scale is $\pm 0.012$ mas/pixel using our two best fields 47 Tuc and NGC 3603, while the true north correction (weighted value -1.77~degree) typically has an error of $\pm 0.07$~degree. For IFS, an additional offset of 100.48~degree in the clockwise direction is applied to account for the orientation of the instrument FOV. This leads to uncertainties in the position of 3--4~mas at the edge of the IRDIS field of view and less than 1~mas for IFS. For comparison, typical accuracy of GPI using their best calibrations are $\pm 0.021$~mas/pixel in the pixel scale and $\pm 0.12$~degree in the true north correction \citep{DeRosa2019}. The better calibration accuracy obtained for SPHERE is due to the wider field of view of IRDIS that allows the use of stellar cluster fields as calibrators with more accurate catalog positions. 

\paragraph{Field/Pupil tracking and parallactic angle calibration}
While the SHINE science observations are taken in pupil-tracking mode for ADI purposes, the astrometric fields were observed in field-tracking mode. The offset between the pupil-tracking and field-tracking modes was measured at the beginning of the survey to be equal to -135.99$\pm$0.11~degree.
The parallactic angle is computed from the data FITS header: timestamp, RA/DEC of the derotator (INS4.DROT2.RA/DEC) which are more accurate than the "RA" and "DEC" keywords (J2000 coordinates of the target). In case these parameters are used it is possible to correct for the precession of coordinates
between 2000 and the date of observation. In order to derive the precise parallactic angle of each DIT from the timestamp we also included a parametric model of the overheads.  We observe a systematic error in the parallactic angle estimation due to backlash in the derotator mechanism of $\sim$0.05\degre, as demonstrated in \cite{Beuzit2019}. In pupil-stabilized mode, this leads to a $\sim$0.4\,pixel difference in the position of an object located at the edge of the IRDIS FoV on either side of the meridian.

\paragraph{Star centering}
The astrometry we consider in this paper is relative to the star. Since the star point spread function peak is hidden by the coronagraphic mask and to avoid concerns due to the non uniform intensity distribution of the coronagraphic leakage, its position was determined using a special calibration (\texttt{STAR-CENTER}) where four faint replicas of the star image are created by giving a bi-dimensional sinusoidal profile to the deformable mirror (see \citealt{Beuzit2019,Makidon2005}). The \texttt{STAR-CENTER} calibration was repeated before and after each science observation, and results center estimations were averaged. While this calibration greatly reduces uncertainties in the exact position of the star and the center of the field rotation, experience shows that small drifts of a few mas of the star center during long sequences of $\sim$1--2h can occur. For this reason in order to increase the star centering accuracy, we generally used, for the second epochs, when doing (candidate) companions follow up, the \texttt{STAR-CENTER} setup solely for the science exposure. 

We performed specific measurements to estimate the accuracy of the central star position when the \texttt{STAR-CENTER} setup is not used for the science exposure (i.e most cases presented in this paper). To do so we measured the position of the central diffraction peak on the IFS datacubes collapsed in wavelengths, and making the mean over all Detector Integration Time (DIT); as mentioned above, errors are likely independent of the errors in the \texttt{STAR-CENTER} procedure. We found that the mean position of the center is offset with respect to the nominal position along the Left-Right direction in the pupil reference frame by
a small but significant amount $0.52\pm 0.10$~mas for Y-H observations, while there is no offset for the Y-J mode ($0.03\pm 0.05$~mas) or along the Top-Bottom direction in the pupil reference frame in both modes. The root mean square (rms) scatter of the residuals after a 3-$\sigma$ clipping are 1.35 mas (1.23 mas) in RA and 1.24 mas (1.43 mas) in declination, for the Y-H (Y-J) mode. Most outliers are found in data sets that were not validated and a few of them are binaries. There are 5\% of the validated observations that have much larger dispersion of the position of the peak than usual; these anomalous cases make $\sim$10\% of the observations acquired  before February 2016, while the fraction reduces to $\sim$2\% after that epoch. This is likely due to improvements in the AO calibration, that resulted in less distorted diffraction peaks. A few of these residual cases may be unresolved binaries. We conclude that a reasonable estimate for the accuracy of the absolute central star position is $\pm 1.5$~mas for both IFS and IRDIS.

\paragraph{Cross-check with Gaia DR2}

An external check of the accuracy of our astrometry is provided by wide companions in the Gaia Data Release 2 catalogue \citep{Gaia2018}. The number of sources in the IRDIS field of view with a contrast adequate to be detected with Gaia is limited because stars with known bright companions were not included in our sample to avoid problems for the AO and heavy saturation of the detector. By comparison with the SPHERE results, we find that Gaia DR2 limiting contrast (in the visual G-band) corresponds to contrasts in the near infrared of $\Delta$H=2.5~mag for separation sep$<2.5$~arcsec and 6 mag for sep$>4$~arcsec\footnote{Since the companions are typically much redder than the central star, the contrast in the visual band is a few magnitudes deeper.}. For this comparison, we considered 34 IRDIS close companion candidates (some of them not included in the sample described in this paper) with $\Delta$H2$<$6~mag; out of which twelve are in Gaia DR2. We did not considered one of the Gaia DR2 data because it has very large error bars. In addition we removed some SHINE objects because they are saturated, or located at the edge of the IRDIS FOV (sep$>$6.5~arcsec), both leading to inaccurate astrometry.  Gaia measurements are most likely unreliable for very high contrasts. This is the case of HD1160C, for which the Gaia contrast is 8.24 mag. Binary periods for these stars are so long that they should not affect the result. However, a couple of the objects are not binaries but rather background stars (HIP82430 and PDS\,70). We took into account the relative proper motion between the SHINE observation and the Gaia epoch (2015.5). At the end we have seven good comparative measurements.The separation measured by SHINE is slightly larger than that measured by Gaia DR2. The mean offset is $-2.8\pm 1.5$~mas, with a root mean square (rms) of 3.9~mas. The position angle measured by SHINE is similar to that measured by Gaia DR2: the mean offset is $0.06\pm 0.04$~degree, with an rms$=0.11$~degree. The rms agrees well with the expected uncertainties in these quantities in the SPHERE data.

We conclude that the minimum accuracy of the astrometric calibration of SHINE data is $\pm 2$~mas at separations $<1$~arcsec, and $\pm 3$~mas at larger separations. These values are similar to the scatter typically observed for GPI astrometric calibrators \citep{DeRosa2019}. In practice, the astrometric accuracy on the position of faint substellar companions detected in the SHINE data is limited by the measurement uncertainties from the image post-processing (Sect.~\ref{sec:postprocessing}) and by the companion magnitude.

\begin{figure*}
    \centering
    \begin{tabular}{ccc}
    	\includegraphics[width=0.66\columnwidth,trim={0cm 9.5cm 21cm 0cm},clip]{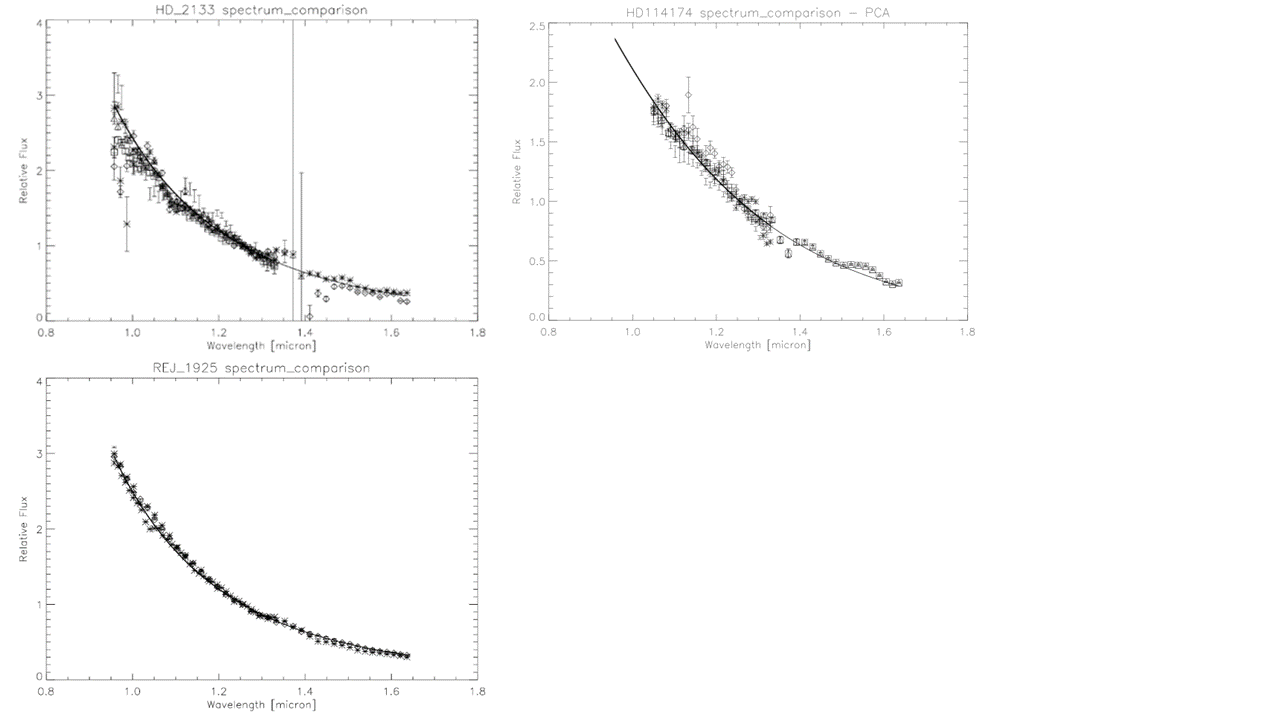} &
    	\includegraphics[width=0.69\columnwidth,trim={0cm 0.2cm 0cm 0cm},clip]{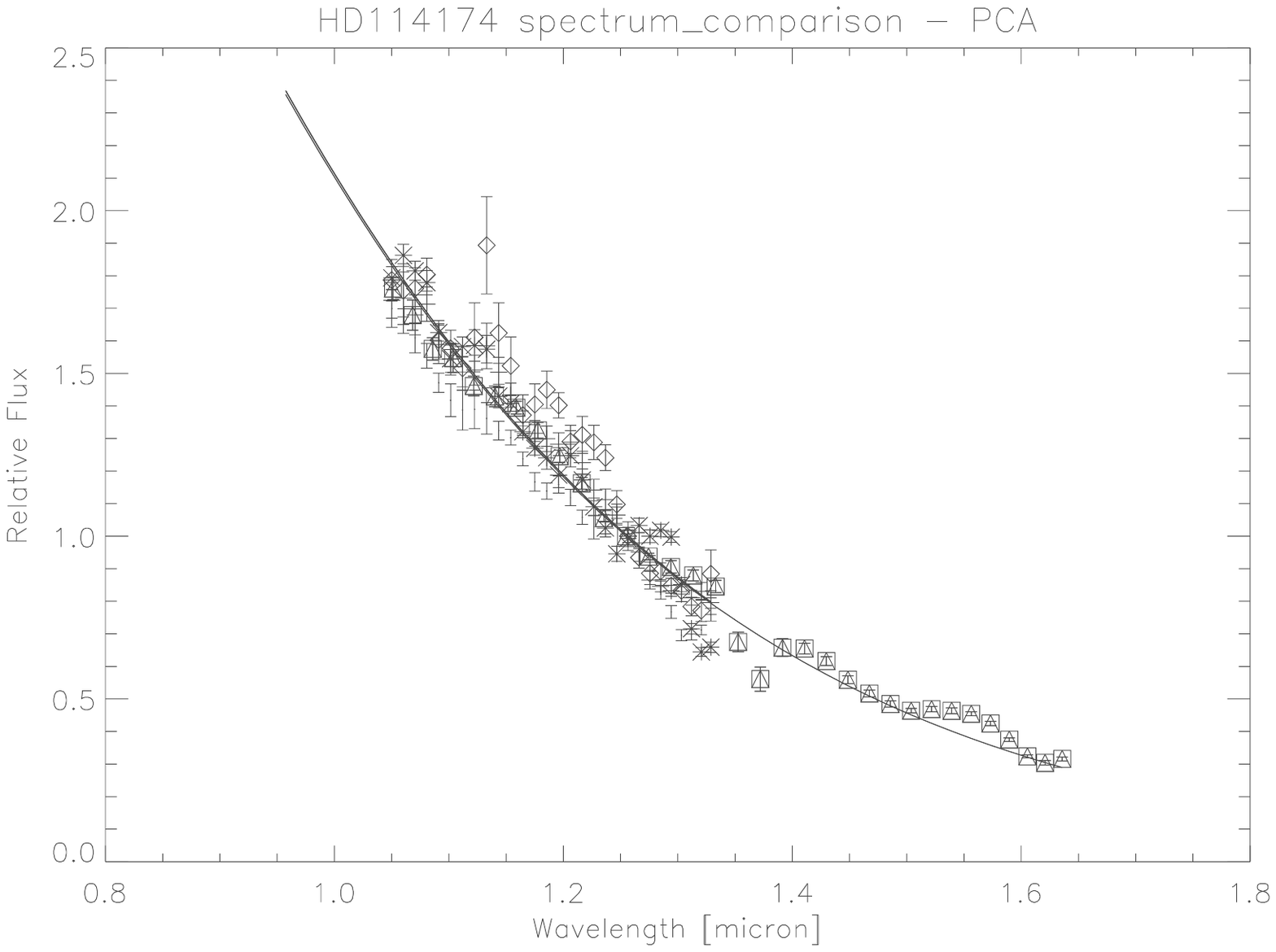} &
    	\includegraphics[width=0.69\columnwidth,trim={0.5cm 0cm 20cm 9.5cm},clip]{spectrophotometricb.png} \\
    \end{tabular}
    \caption{Spectra of the white dwarfs used as spectrophotometric standards compared with predictions from model atmospheres (solid lines). Different symbols are results obtained from different epochs. We normalized the observed spectra at the median flux in a narrow range around 1.25 micron.}
    \label{fig:spectrophotometry}
\end{figure*}

\subsubsection{Photometry}

We describe below the strategy to derive the relative photometry of the SHINE candidates considering the unsaturated and coronographic observations of a scientific target, and two tests done by using IFS to validate our strategy for both IRDIS and IFS which have very similar photometric biases.

\paragraph{Strategy}

In the SHINE survey, photometry of candidate companions and limiting magnitudes for non detections are relative in contrast to the central star. Since the star is behind the coronagraphic mask, simultaneous photometry is not possible. We thus include a flux calibration (\texttt{STAR-FLUX}) for both IRDIS and IFS that is acquired just before and after the science exposure by offsetting the Differential Tip-Tilt Stage (DTTS) by about 0.5 arcsec with respect the coronagraphic mask using the SPHERE tip/tilt mirror \citep{Beuzit2019}. When performing this calibration, suitable neutral density filters are inserted to avoid detector saturation. The transmissions of these neutral density filters were carefully calibrated from 0.9 to 2.3 microns (see \citealt{Beuzit2019} and ESO website \footnote{\url{https://www.eso.org/sci/facilities/paranal/instruments/sphere/inst/filters.html}}) and are taken into account for the contrast estimation. This procedure works very well in stable conditions, but it may be affected by variations of the Strehl ratio from evolving observing conditions. Higher photometric accuracy can be achieved by using the waffle pattern  continuously (for companion candidates follow up) during the observing sequence to monitor the Strehl variations.

\paragraph{Cross-check with catalogues and binary companions}
A first test of the photometric calibration accuracy is provided by comparing the peak counts of the diffraction image of the flux calibration (corrected for the integration time and for the neutral density filter transmission) to the apparent magnitude of the stars from the 2MASS catalogue \citep{Skrutskie2006}, corrected for atmospheric extinction. This procedure neglects stellar variability and the impact of variability on the Strehl ratio. Using J-band IFS data, we obtain an rms of residuals of 0.15 mag for observation in the Y-H mode, and 0.19 mag for those in the Y-J mode. This has been computed after eliminating the expected variations due the Strehl ratio and after clipping outliers deviating greater than 2.5 standard deviation from the mean with an iterative procedure (about 12\% of the data). A similar value is obtained by considering the rms scatter of the estimate of the coronagraph central transmission - measured as the ratio between the maximum counts of the diffraction peak in the coronagraphic image and in the flux calibration (corrected for the length of the integration and neutral density filter transmission). The average coronagraphic transmission measured in this way are $0.00168\pm 0.00005$ for the Y-H mode, and $0.00208\pm 0.00001$\ for the Y-J mode, respectively, for a median SHINE observation. 

A second test of the photometric calibration accuracy is provided by the photometry of three standard systems (HD2133, HD114174, and REJ1925-563) composed of a main sequence star and a white dwarf with separations in the range 0.2-0.7 arcsec, we have observed during the survey. These systems were observed several times, with typically at least one observation per run. For schedule optimization, these observations were acquired in ADI with the star quite far from meridian and in poorer observing conditions, than typical survey targets: we consider here only those observations that were obtained in fair to good atmospheric conditions. Photometry of the faint companions was obtained by inserting a negative scaled point spread function at the position of the companion in the raw datacube, i.e before combining the various DITs. We then measured the rms residuals of the differential image obtained through monochromatic principal component analysis (PCA: \citealt{Soummer2012}) around the companion. The procedure then minimizes these residuals by simultaneously adjusting the contrast and the position. The values we consider are the mean of the results obtained using 2 to 6 PCA modes. The average contrasts in the J-band obtained with this procedure using the main sequence star and the white dwarfs are 7.68, 10.04, and 6.10 mag for HD2133, HD114174, and REJ1925-563, respectively. The rms value for the contrasts in the J-band/H-band are 0.11/0.05, 0.18/0.08, and 0.05/0.07 mag for HD2133 (8 observations), HD114174 (14 observations) and REJ1925-563 (6 observations). The photometric errors as expected increase with magnitude and as a consequence are smaller in H-band than in J-band. This may be also attributed to a higher impact of the variation of the SR and of the speckle noise for fainter targets and shorter wavelengths. A comparison of the spectra determined by our method with atmospheric models is given in Figure~\ref{fig:spectrophotometry}.

We conclude from these measurements that a reasonable photometric errors estimate for both the limiting contrast and for companions characterisation (not including the ADI post processing error contribution), is around $\pm 0.2$~mag. It is dominated by the STAR-FLUX variation which is also the case for the survey data photometric error which is on average equal to $\pm 0.25$~mag. 


\subsection{SPHERE Data Center preprocessing}
\label{sec:preprocessing}

\subsubsection{IRDIS-only steps}
The SHINE survey was reduced by the SPHERE Data Centre (hereafter SPHERE-DC) \footnote{\url{ http://sphere.osug.fr/spip.php?rubrique16\&lang=en}}. For IRDIS data, the first reductions steps (dark/background, flat, and bad pixel correction) rely on the SPHERE Data Reduction and Handling (hereafter DRH) pipeline \cite{Pavlov2008} provided by ESO. We used the on sky background because there is a significant difference between day-time background calibrations and the on sky background recorded in the science frames. This is the case mostly in K-band with a systematic, spatially variable, offset (typically of the order of 100~ADU for 64s exposure time) which is due to the sky background contribution itself.  A similar effect is also visible in H-band, on a smaller level (typically of only few ADU).
 
Since most of the SHINE observations use the pupil-tracking observing mode, a very accurate determination of the star center is needed in order to successfully use both the angular differential imaging and spectral differential imaging methods. Also since IRDIS is used in dual-band imaging, the star center in both IRDIS channels is used to combine them. By default, we use the DRH \texttt{sph-ird-star-center} routine to determine the star center position, using the waffle images acquired for this purpose just before and after the science observations. This very fast routine provides an accurate centering in many cases, but could fail for weak waffle spots, especially in the K-band where these spots can be hidden beneath a much stronger thermal background noise but also when the deformable mirror offsets to create this pattern are set too low. We therefore designed an automated way to check the quality of the DRH centering by comparing the two center positions derived out of the four waffles. When the distance between these two possible center positions was found to be greater than 0.9 pixel, we used a SPHERE-DC made IDL routine that is more robust to identify weak waffle spots. This routine is able to detect weaker waffles by locating them within small circular apertures located at the expected (wavelength-dependent) position and by using in combination a high-pass spatial filtering, sky-background subtraction and median stacks of all waffle images available in order to increase their SNR.

A small fraction of the SHINE datasets use continuous waffle mode observations, meaning that the waffle spots are activated during the entire observation, for better astrometric monitoring (with some localized loss in the limiting contrast because of the secondary spots). In this specific cases, we perform individual re-centering of each frame in the sequence using the SPHERE-DC dedicated star-centering routine described above. This improves the quality by correcting any drift or jitter of the targeted star behind the coronagraph, and as a result improves the quality of the astrometric measurements by removing these sources of error.

 \subsubsection{IFS-steps}

For IFS data we use the Data Reduction and Handling (DRH \citealt{Pavlov2008}) pipeline but complement it with additional steps implemented at the SPHERE Data Center \citep{Mesa2015,Delorme2017} to improve the wavelength calibration, apply a correction for cross-talk, and improve the handling of bad pixels. The improvements of the wavelength calibration are obtained by using a cubic fit whose coefficients are estimated from the wavelength shifts of the spots generated by the \texttt{STAR-CENTER} calibration - that are known to scale linearly with wavelength. The cross-talk correction is performed using an iterative procedure that corrects for the spectrograph PSF, using coefficients derived using appropriate tests performed in the laboratory during the instrument assembly. Bad-pixels are corrected using a dedicated sky observation acquired at the end of each science exposure. This procedure yields more accurate results than the one based on the flat field calibration within the DRH.
 
\subsubsection{Final steps common to IRDIS and IFS}

After these pre-reduction steps, both IRDIS and IFS datasets are corrected for the instrument anamorphism and the astrometric solution (pixel scale and True north) estimated from the calibration described in Sec. \ref{sec:astrometry} is applied to the dataset. The output data is composed of a pre-reduced master cube combining all frames obtained during a given observation sequence, that is used as input for all Angular Differential Imaging \citep[ADI][]{Marois2006} algorithms. We also associate to this master cube a vector of de-rotation angles for each frames, using the accurate timing and overheads for each frame, to produce a frame-to-frame  determination of the parallactic angle. 

\subsection{ADI and ASDI postprocessing}
\label{sec:postprocessing}

For both IRDIS and IFS, we obtain a good-quality re-centered images gathered in a single master cube associated with their parallactic angle values. Subsequent steps follow to estimate and subtract the stellar halo from each images, followed by derotation and stacking of the residuals. The most critical step is the estimation of the stellar halo that drives the level of the noise residuals. We applied different Angular Differential Imaging (ADI) algorithms to optimize the detection performances and to identify associated biases. We rely mainly on the SpeCal (\cite{Galicher2018} software which offers various ADI options. We selected for the homogeneous reduction TLOCI \citep{Marois2014, Galicher2018} and Principal Component Analysis (PCA: \citealt{Soummer2012}) for IRDIS and PCA Angular and Spectral Differential Imaging (ASDI: \citealt{Mesa2015}) for IFS. 

In the TLOCI approach \citep{Lafreniere2007}, the PSF-reference is estimated for each frame and each location. Linear combinations of all data are computed to minimize the residuals into an optimization zone, which is much bigger than the subtraction zone to avoid the self-removal of point-like sources.The SpeCal version of the TLOCI algorithm is derived from the one described in \citet{Galicher2018} assuming a flat planet spectrum in contrast. Adjustable parameters are used to select the frames and to describe the regions of interest. In SpeCal, the gap between this region and the region of interest is set to 0.5 Full Width Half-Maximum (FWHM). Hence, the optimizing region is far enough from the region of interest so that the flux of a source in the latter does not significantly bias the linear combination. Finally, an additional parameter sets the radial width of the optimizing region. We considered here a radial width of the subtraction zone of 1 FWHM in radius; a radial-to-azimuthal width ratio of 1.5; a standard surface of the optimization zone was N = 20 PSF FWHM and 10$\%$ is used for the minimum residual flux ratio due to self subtraction compared to the flux of a putative candidate.

For historical reasons, two PCA algorithms are implemented in SpeCal. The first version can be applied on IRDIS or IFS data using ADI or ASDI. This algorithm follows the equation of  \citet{Soummer2012}. In the ADI case, which is the option selected to reduce IRDIS data in this paper, the principal components are calculated for each spectral channel independently. Each frame is then projected onto a limited number of modes. In the ASDI case, the algorithm is the same but it works simultaneously on the spatial and spectral frames. The second version of PCA used to reduce IFS data in this paper, is very similar to the first version we have described but it was applied only on IFS data using the ASDI option \citep{Mesa2015}. In addition, it is worth mentioning that both PCA algorithms we use have no frame selection to minimize the self-subtraction of point-like sources when deriving the principal components.  

For PCA and TLOCI algorithms that bias the photometry of off -axis point sources, SpeCal estimates the throughput at each position in the field by generating a datacube of fake planets for which the ratio of the flux in the resulting image to the flux of the fake planet is calculated to obtain the centro-symmetrical 1D-throughput as a function of the angular separation which is then applied to the images. 
For PCA, Specal estimates the throughput by inserting fake planets injections which are only 10 times brighter than the local stellar residuals level after PCA. This makes it possible to be close to the level of interest while being sufficiently above the residual to minimize the bias in the throughput estimation. For TLOCI, the throughput is calculated from an analytical formula (see section 2.10.2 of \cite{Galicher2018}.
The contrast curves, we derived in the following, for each spectral channel are based on the azimuthal standard deviation calculated in annuli of 0.5 FWHM width. Finally, the 5$\sigma$ detection limits are derived by taking into account the flux loss from ADI self-subtractions, the transmission of the coronagraph at short separations (close to the inner working angle (IWA: $~0.1"$) and the transmission of the neutral-density filter if used when registering the PSF. These detection limits are thus  normalized by the unsaturated PSF flux. Both 1D and 2D contrast maps are estimated following these steps for each star for each reduction technique.

All target stars were processed for each instrument in a homogeneous way using at least TLOCI or PCA with similar sets of parameters. We considered 50, 100 and 150 PCA modes for IFS and 5 for IRDIS. We inspected  by eye at least three residual maps for each star and for both IFS and IRDIS to look for candidate companions (CC). We also included small number statistic correction. In addition to the standard SpeCal reductions, we also used, on several cases, ANDROMEDA \citep{Cantalloube2015} and PACO \citep{Flasseur2018,Flasseur2020b} algorithms to search for points sources. Given their statistical robustness to derive detection limits and to better identify false detections these algorithms will become the main algorithms for the final analysis of the SHINE survey.

\subsection{Astrometric and Photometric extraction for point sources }

PCA and TLOCI algorithms are known to distort the images of any off-axis point-like source. To retrieve accurately the relative photometry and astrometry of the detected candidate companions (CC) with their uncertainties, SpeCal fits a model of an off-axis point source image to the detected point source and adjusts its position and flux to locally minimize the flux. After building a model of the point source using the technique described in \citep{Galicher2011}, the flux and the position of this synthetic image are adjusted within a disk of diameter 3 FWHM so that it includes the positive and the negative parts of the point source image. To optimize the computation time, instead of calculating the synthetic image each time we test a new planet position, we shift the synthetic planet image to its rough position. Once the optimization is completed, we measure the excursion of each parameters that increases the minimum residual level by a factor of 1.15. These excursions correspond to the 1 $\sigma$ accuracy due to the fitting errors in the SpeCal outputs. The spectrum extraction from IFS data is performed similarly by processing the wavelength channels separately. For the channels with no detection above 1 $\sigma$ we provide an estimation of the upper limit only.

In conclusion, we summarize the astrometric and photometric error budget given by Specal considering both calibrations and ADI/ASDI extraction errors.

For astometry, this budget includes:
\begin{enumerate}
    \item calibration with uncertainties of the detector distorsion, plate scale and True North,
    \item determination of the correction for constant offset angles (between the IFS and IRDIS fields of view, between pupil-tracking and field-tracking modes); 
    \item calculation of the parallactic angle variation (correction of precession and timestamp),
    \item central star position,
    \item determination of the companion/candidate relative position with Specal \citep{Galicher2018}.   
\end{enumerate}

For photometry, it includes:
\begin{enumerate}
    \item Contrast estimation with proper calibration of neutral density, exposure time, and associated error related to the variation between Start/End STAR-FLUX calibrations but not including the strehl variation during the observing sequence (psf),
    \item Temporal variations of the stellar coronographic flux between 30-50 pixels during the observing sequence (Seq),
    \item Companion flux determination considering ASDI signature, taking into account flux cancellation from the algorithm used, and coronograph attenuation correction as described in \citep{Galicher2018}.  
    .
\end{enumerate}

\section{High contrast Performances}
\label{sec:perf}

\subsection{Contrast curves}

\begin{figure*}
	\centering \includegraphics[width=\textwidth,trim={1cm 6cm 0 -2cm}]{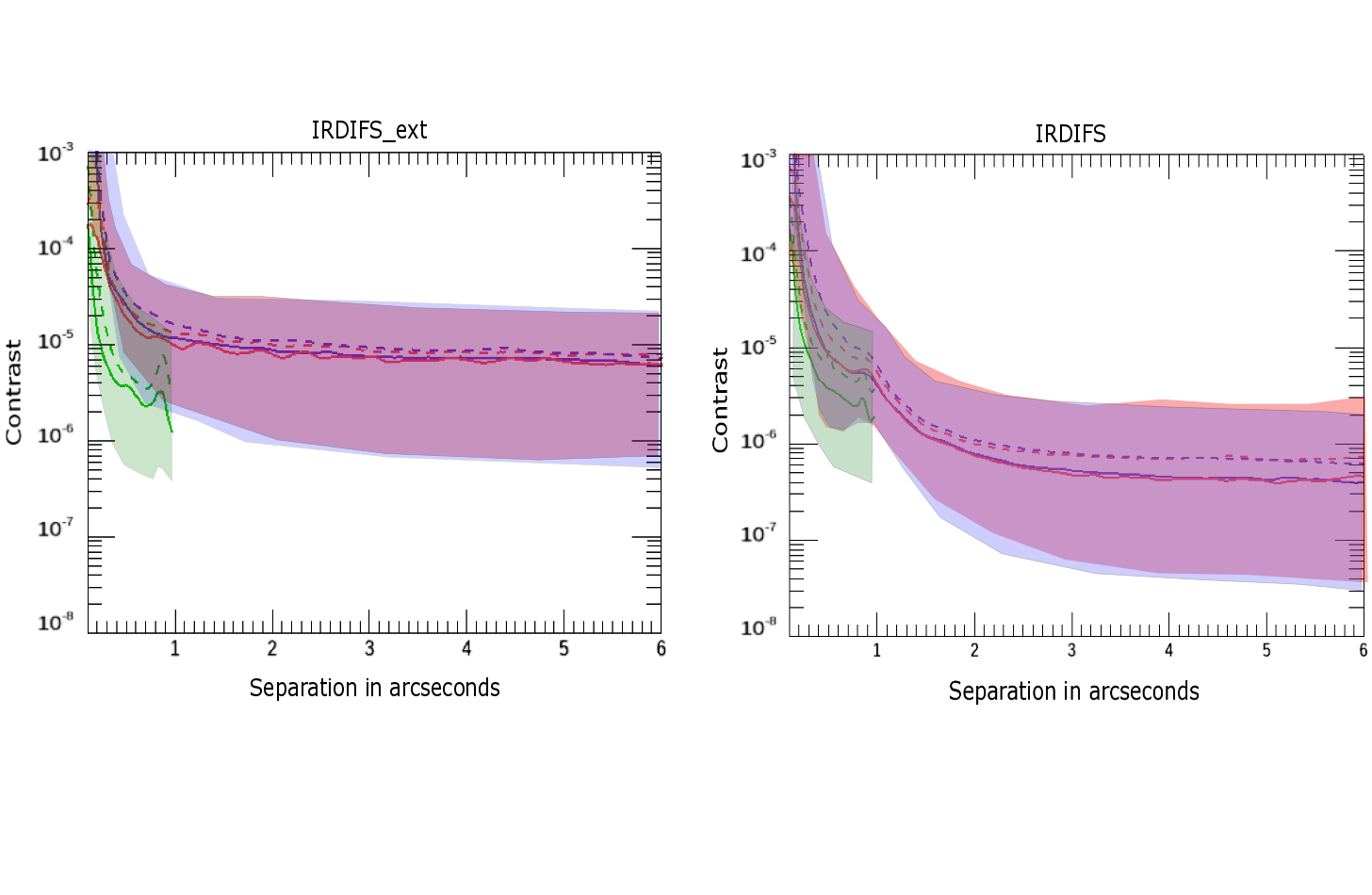}
	\caption{ Contrast curves at 5\,$\sigma$ obtained for the full sample for irdifs-ext (Left) and irdifs (right) modes observations. The solid line gives the median value of the contrast. The dash line gives the mean value of the contrast. Red color is for IRDIS data reduced in PCA ADI, Blue color is for IRDIS data reduced in TLOCI ADI, and green color is for IFS with PCA ASDI reduction }
	\label{fig:contrast IRDIFS}
\end{figure*} 
\begin{figure*}
	\centering 
	\includegraphics[width=1.05\textwidth,trim={1cm 6cm 0 6cm},clip]{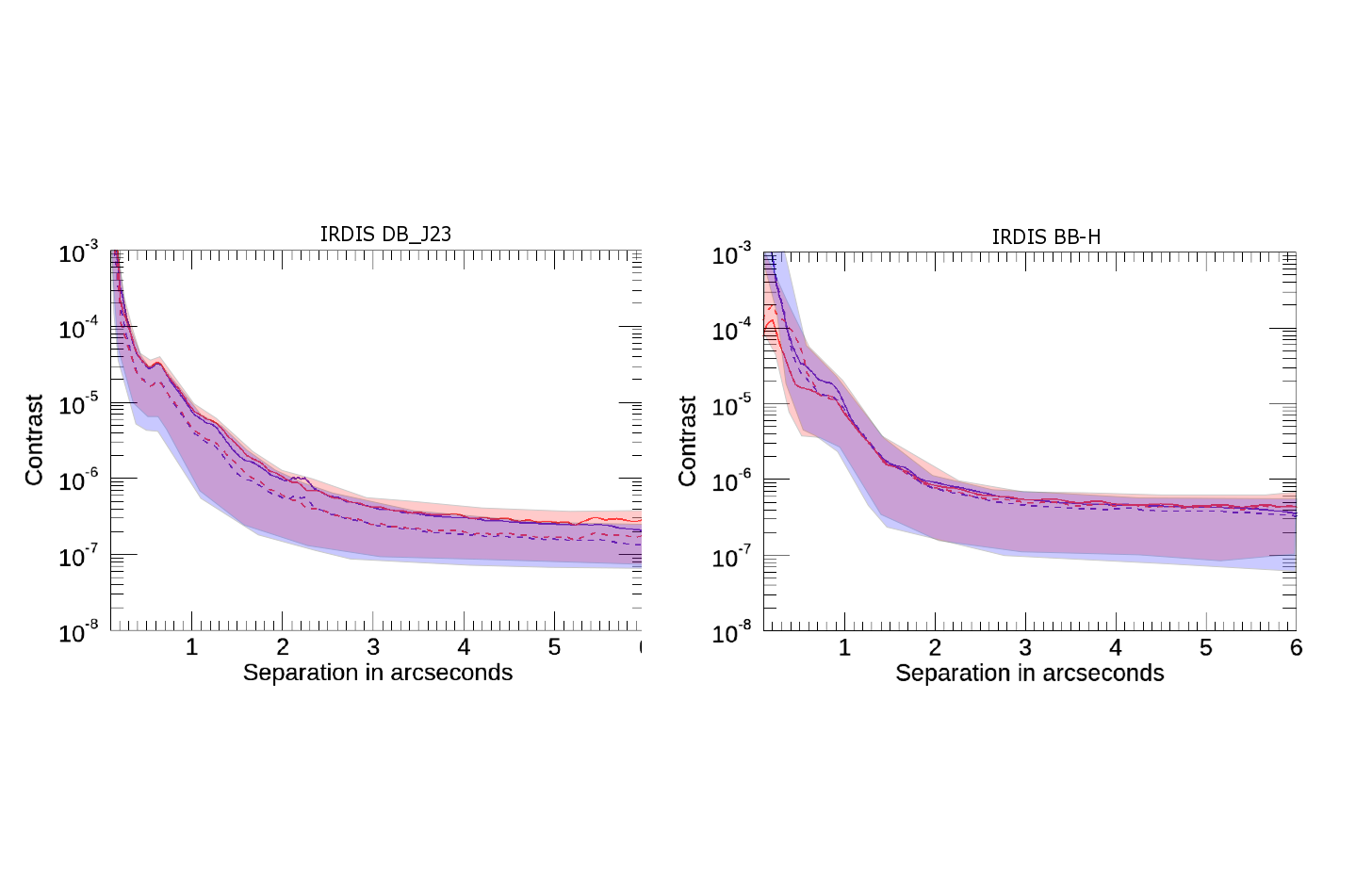}
	\caption{Contrast curves at 5\,$\sigma$  obtained for a small subsample of targets dedicated to follow up for IRDIS mode observations. The solid line gives the median value of the contrast. The dash line gives the mean value of the contrast. Red color is for IRDIS data reduced in PCA ADI, Blue color is for IRDIS data reduced in TLOCI ADI}
	\label{fig:contrast IRDIS}
\end{figure*}

\begin{figure*}
	\begin{tabular}{cc}
		\centering
		\includegraphics[width=0.95\columnwidth]{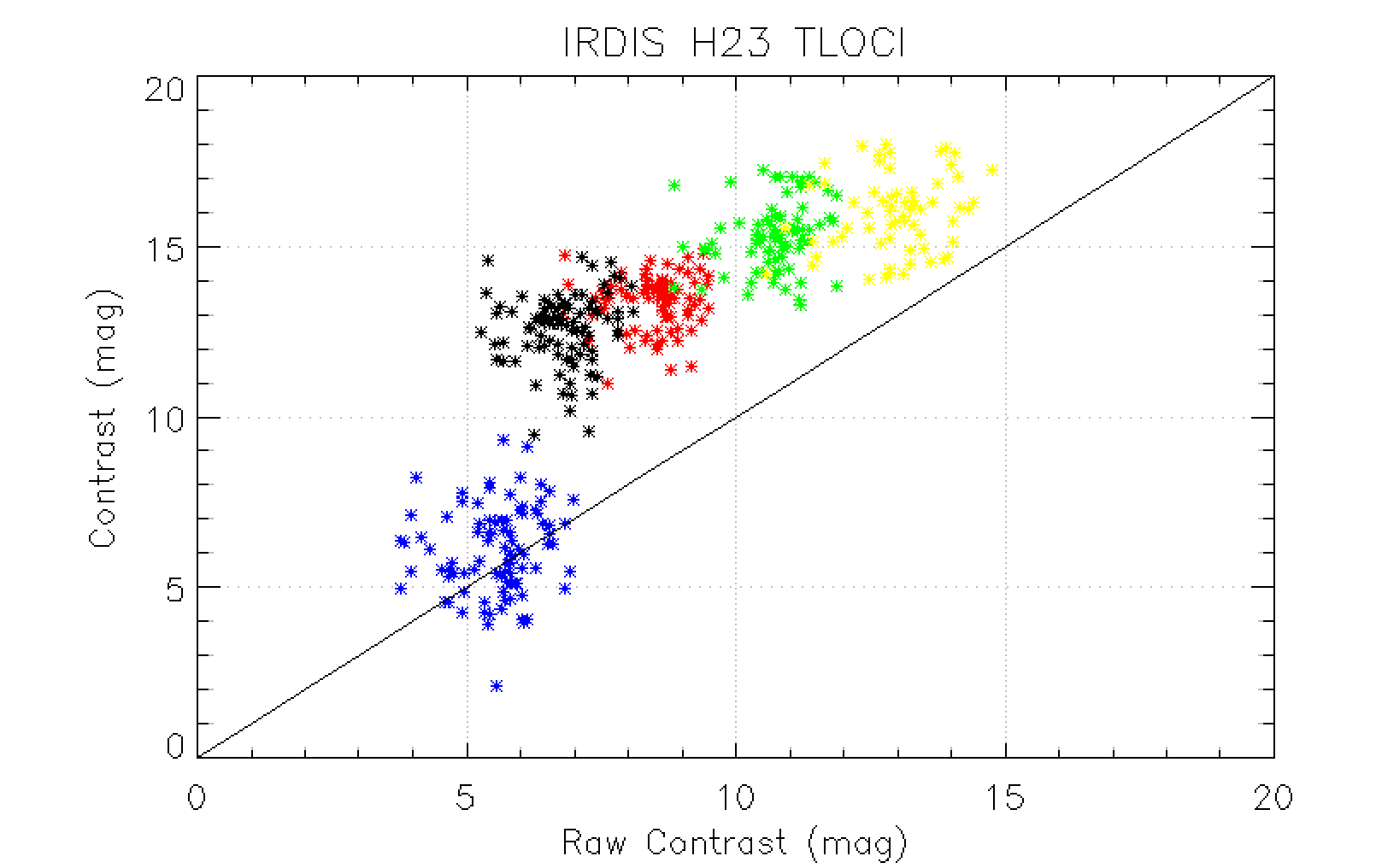} &
		\includegraphics[width=0.9\columnwidth]{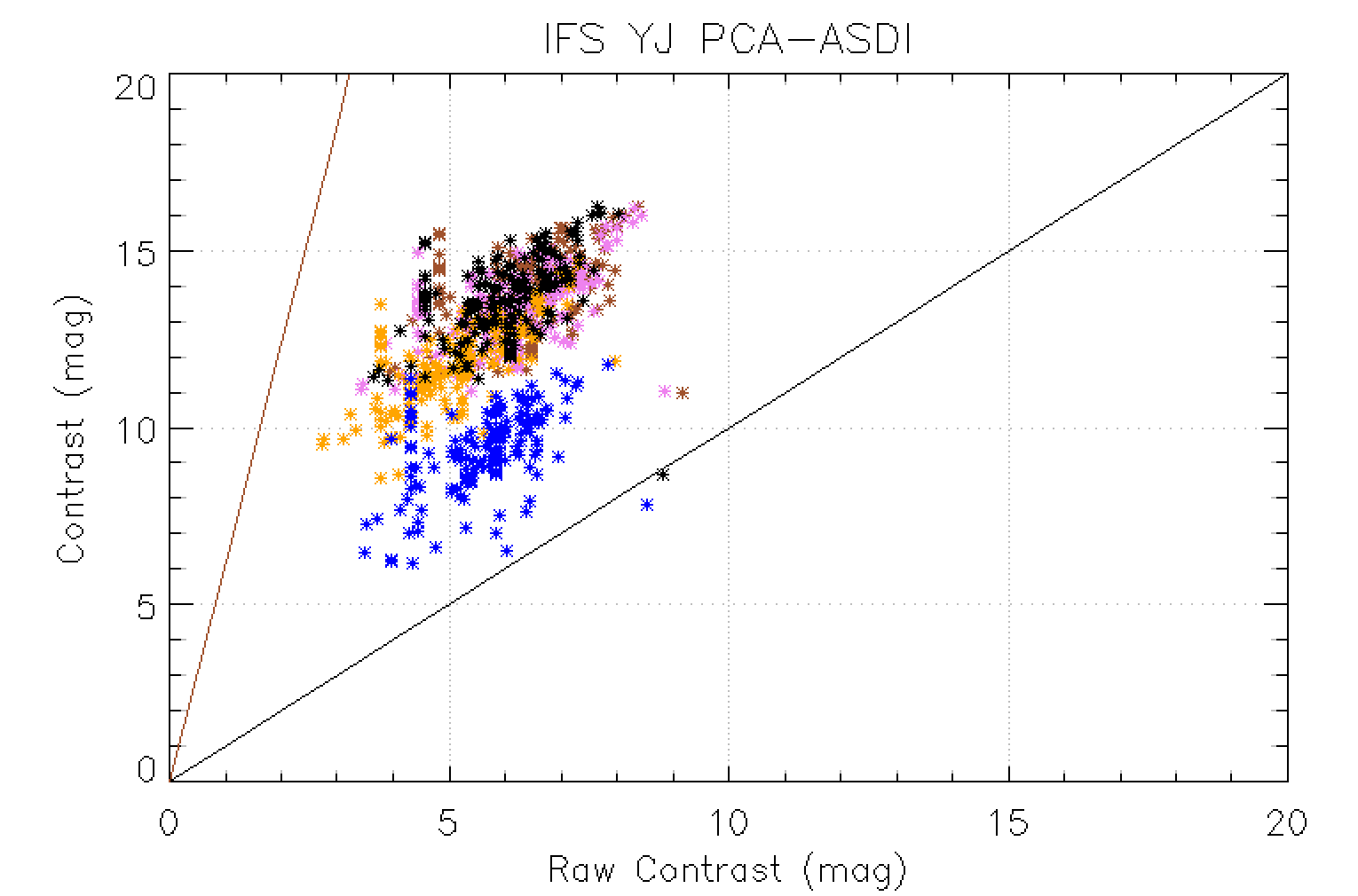}\\
	\end{tabular}
	\caption{IRDIS and IFS processed contrast computed using respectively TLOCI and PCA-ASDI as function of raw contrast at various separations in irdifs mode. Blue is for 100 mas separation, orange for 200 mas separation and pink for 400 mas separations, black for 500 mas separation, , brown for 700 mas separation, red for 1000 mas separation , green for 2000 mas separation and yellow for 4000 mas separations. The black and brown line represent respectively a gain of 1 and 6.25 assuming 39 spectral channels. In fact So there are 15 independent spectral IFS channels leading to a maximum gain of 3.9 when neglecting the possible combination of the two IRDIS channels.}
	\label{fig:cont_gain}
\end{figure*}

We derived the 5-$\sigma$ IRDIS and IFS contrast curves of each observation for all the targets in the sample as presented on Figures \ref{fig:contrast IRDIFS} and \ref{fig:contrast IRDIS}. These detection limits are derived based on the noise in the speckle-subtracted image, compensated for the throughput of the algorithm (calibrated with fake planet injections), the transmission of the coronagraph (calibrated from measurements in SPHERE), and the small sample statistics \citep{Mawet2014}. More details are provided in \citet{Galicher2018}. Two types of contrast are discussed in detail in this section: the raw contrast, computed on the median coronagraphic image of each observation and the contrast after post-processing described above. As illustrated on Fig \ref{fig:cont_sparta_IRDIS} and \ref{fig:cont_sparta_IFS}, the raw coronagraphic contrast at various separations shows a strong dependency with the Strehl and seeing both estimated by the average of the SPARTA values during the coronagraphic sequence. The two smaller separations (from 100 to 700 mas) are within the AO control radius located at 840 mas separation radius in the H-band. 

This dependency remains clearly visible on post-processed data especially at 500 mas, despite other factors also coming into considerations such as the field rotation and the stability of the conditions (see Fig \ref{fig:cont_sparta_IRDISb}). At small separation, these scatter plots shows that one can easily gain one magnitude in raw contrast by increasing the Strehl by 10\% or the decreasing the seeing by 30\%. As a consequence it is clear that conducting the survey in visitor mode which does not offer the best seeing condition had some impact on the survey ultimate contrast performances. Outside the AO correction radius, at separation greater than 900 mas, there is still a smaller dependency of the contrast with the seeing due to the residual light scattering outside the AO control radius and the lower height of the diffraction peak. As part of this study, the dependency of the post-ADI contrast on parameters tracing the quality and stability of the conditions were also investigated as illustrated on Fig \ref{fig:cont_sparta_IRDIS}, \ref{fig:cont_sparta_IFS} and \ref{fig:cont_sparta_IRDISb}. We considered the dispersion in the seeing, coherence time, Strehl, during the duration of the pupil-stabilised sequence. No significant correlation could be drawn from this sample. It is worth noticing that at small separation (100 mas) the processed contrast is lower than the raw contrast because of the very small angular rotation, the small throughput of the algorithm (from self subtraction) and the stronger coronagraphic residuals. 

The gain from raw contrast to post-processed contrast is clearly visible on Fig \ref{fig:cont_gain} which shows a typical improvement greater than 5 magnitudes at short separations for both IRDIS and IFS. It is worth noticing that the contrast gain in the IFS field of view is consistent for all separations and reach at least 7 magnitude. For the shorter separation located at the edge of the coronagraph (100 mas) there is marginal improvement for IFS from ASDI and no improvement for IRDIS in contrast due to the very small angular rotation, the small throughput of the algorithm and the stronger coronagraphic residuals for both instruments. At larger separation than the AO cutoff, the improvement from the post-processing range from 2 to 4 magnitudes. 

We also plot on Fig \ref{fig:contvsmag} the processed contrast as function of the star magnitude for both IFS and IRDIS at various separations. From these figures it is clear that there seem to be small correlation between these parameters when the post-processing if performed using TLOCI. On the contrary the PCA method for both IRDIS and IFS seems to be affected by the magnitude of the star at least for the largest separations which is likely related to the noise from the instruments background.

We also highlight that one clear cause of contrast degradation in the post-ADI contrast is the presence of a smooth halo within the AO-corrected region which results from either bad seeing or from high-altitude wind related to the jet stream as discussed in \citep{Cantalloube2018} leading to a non symmetrical halo in the direction of the wind (called the wind driven halo, WDH). This halo is rotating in the pupil-stabilised data set because it is fixed on the sky as described in \citep{Cantalloube2020}. In post-ADI frames, it therefore appears as a brighter elongation along the wind direction, with negative counterparts at 90 degrees. The impact of this effect does not appear to be very strong on the azimuthal averaged contrasts we have plotted but is noticeable on the 2D contrast maps. Increasing the temporal bandwidth of SAXO is the foreseen solution to help mitigate this effect. This is considered as one option in a forthcoming upgrade of the instrument, SPHERE$+$ \citep{Boccaletti2020}.

\begin{figure*}
	\centering
	\begin{tabular}{ccc}
		\includegraphics[width=0.66\columnwidth]{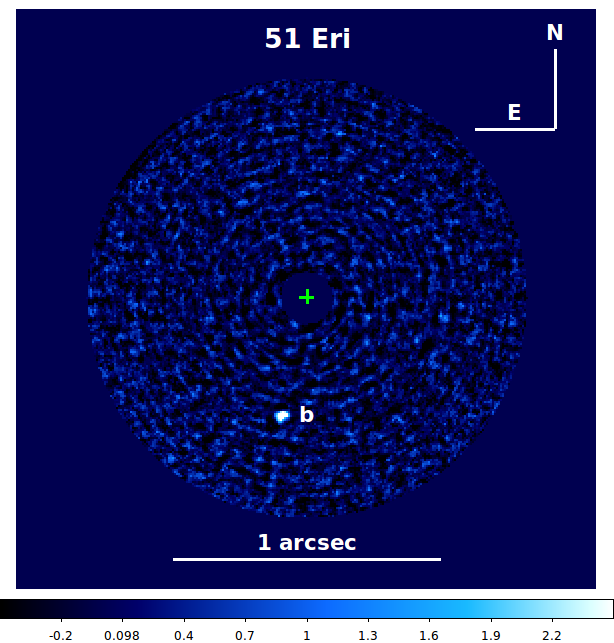} &
		\includegraphics[width=0.66\columnwidth]{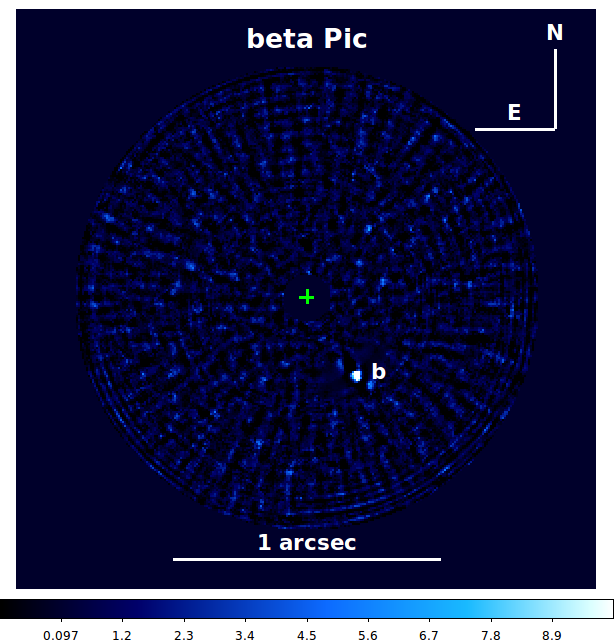} &
		\includegraphics[width=0.66\columnwidth]{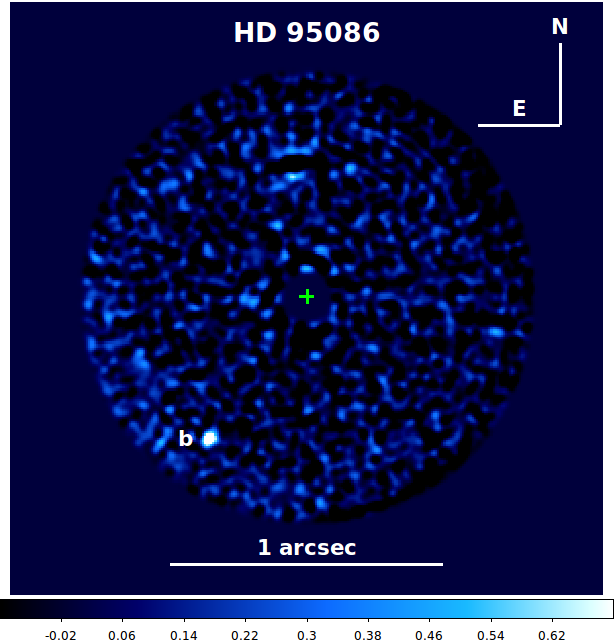} \\
		\includegraphics[width=0.66\columnwidth]{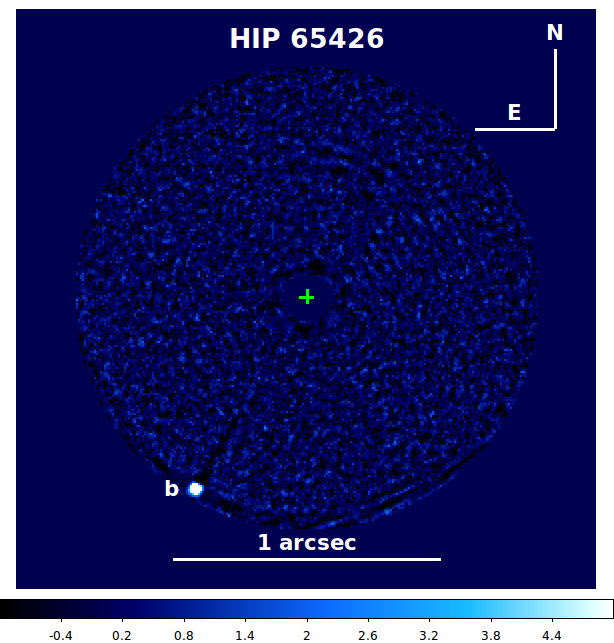} &
		\includegraphics[width=0.66\columnwidth]{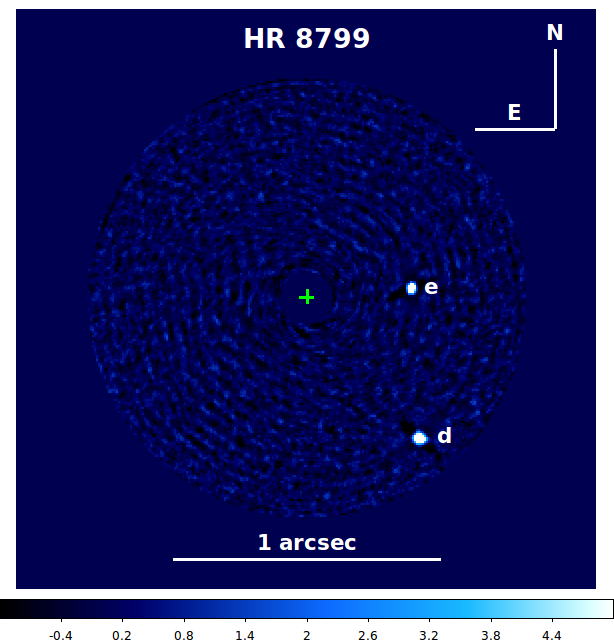} &
		\includegraphics[width=0.66\columnwidth]{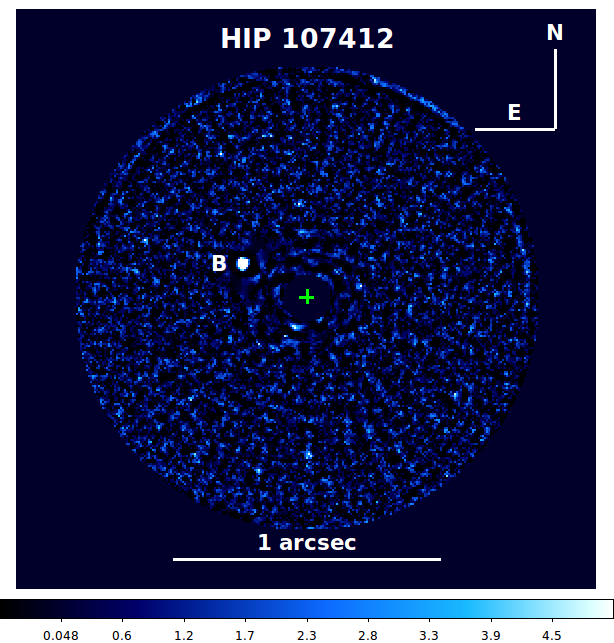} \\
	\end{tabular}
	\caption{SNR maps for the sample data where at least a substellar close companion was detected with IFS. In each panel, the green cross mark the star position}
	\label{fig:detections_ifs}
\end{figure*}

\begin{figure*}
	\centering
	\includegraphics[width=2.2\columnwidth]{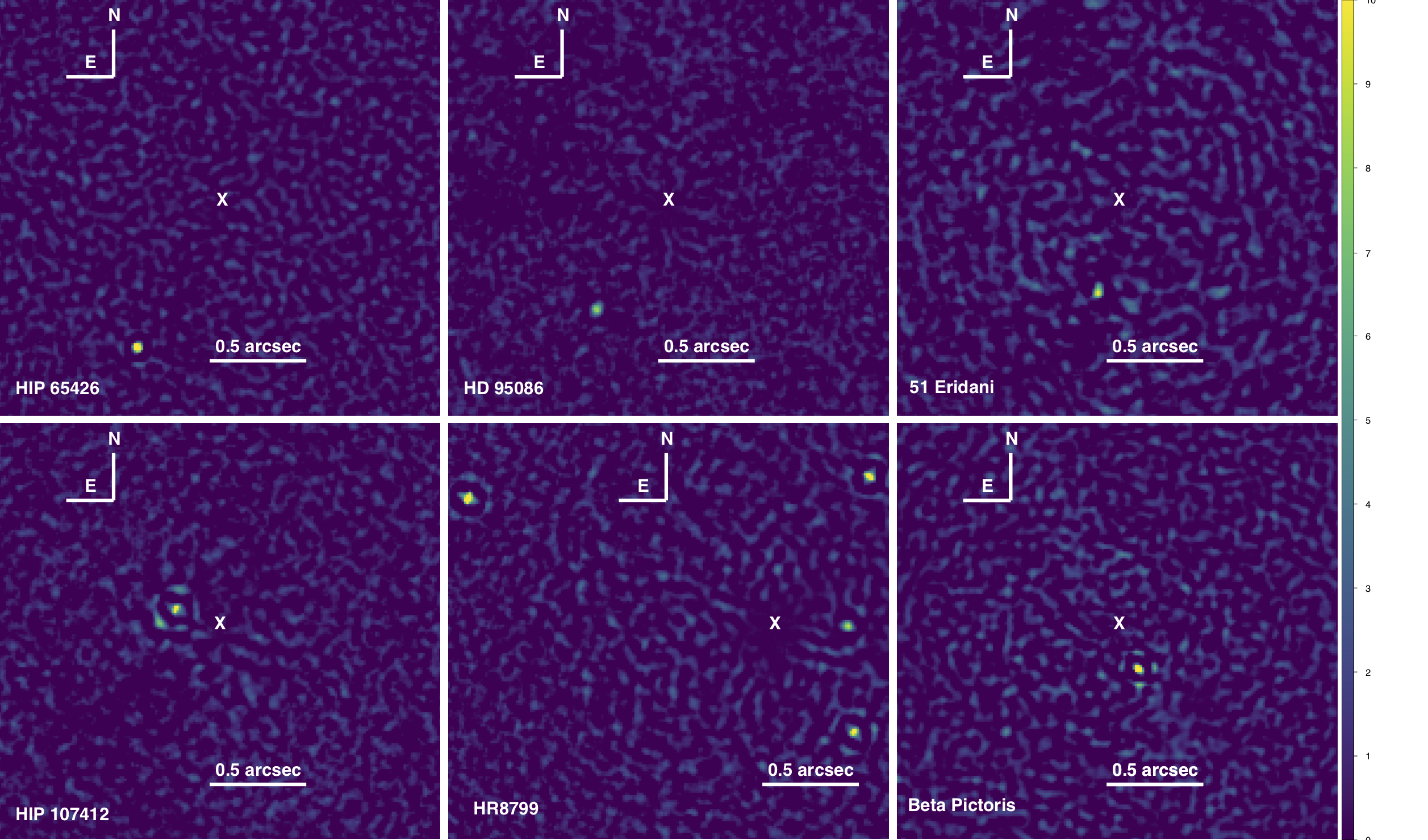}
	\caption{SNR maps for the sample data where at least a substellar close companion were detected with IRDIS. In each panel, the white cross marks the star position}
	\label{fig:detections_irdis}
\end{figure*}

\begin{figure*}
	\centering
	\begin{tabular}{ccc}
		\includegraphics[width=0.66\columnwidth,trim={0.5cm 0cm 0.5cm 0cm},clip]{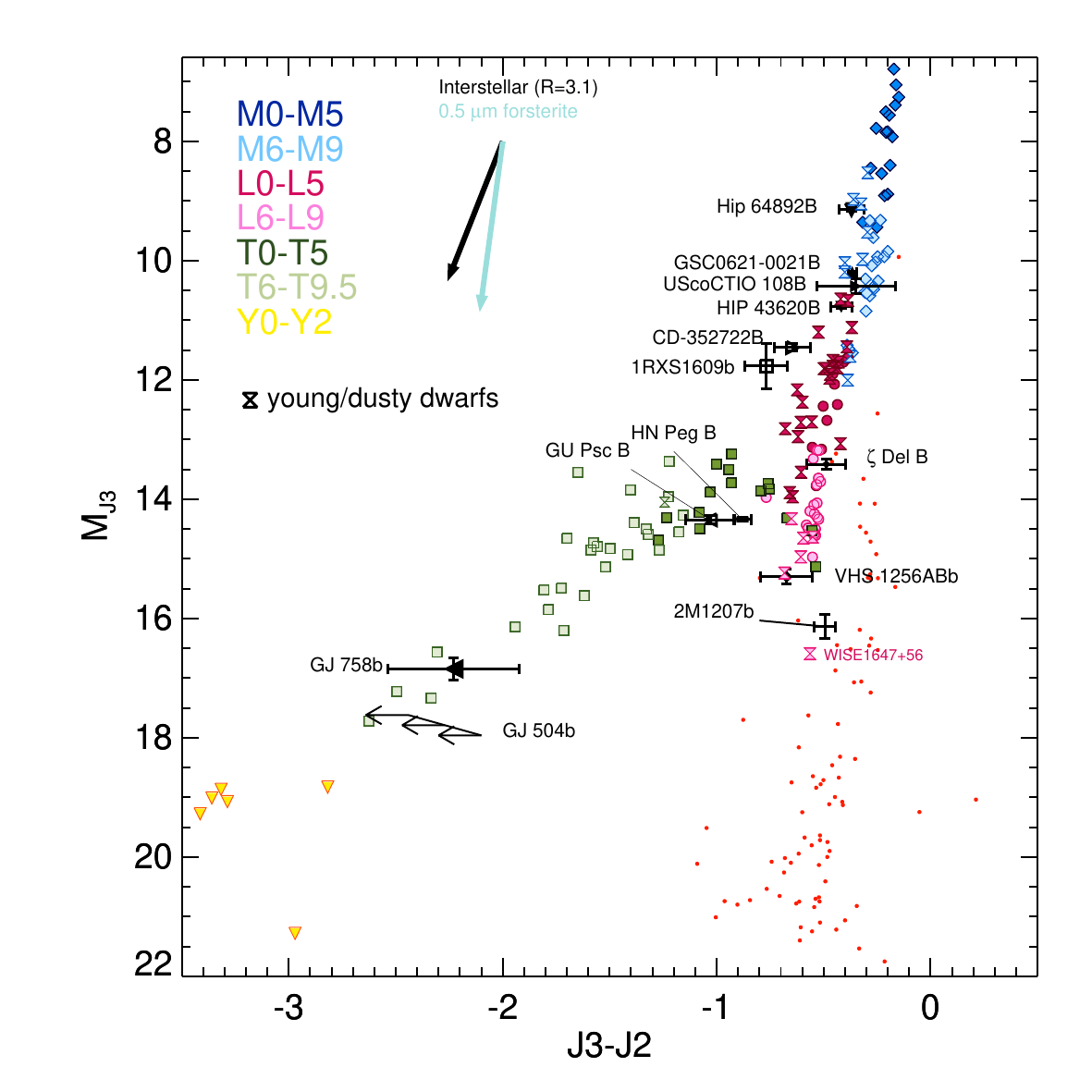} &
		\includegraphics[width=0.66\columnwidth,trim={0.5cm 0cm 0.5cm 0cm},clip]{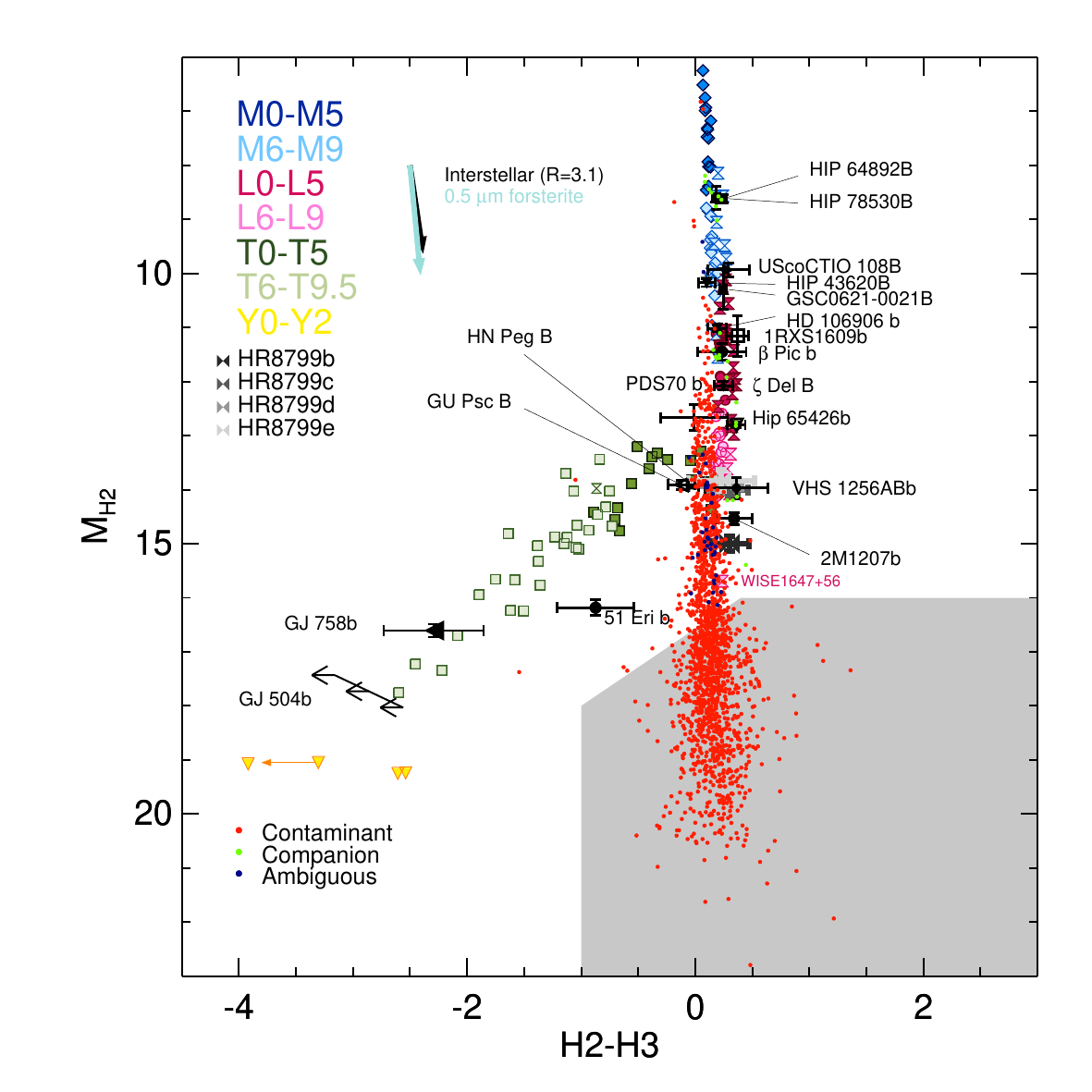} &
		\includegraphics[width=0.66\columnwidth,trim={0.5cm 0cm 0.5cm 0cm},clip]{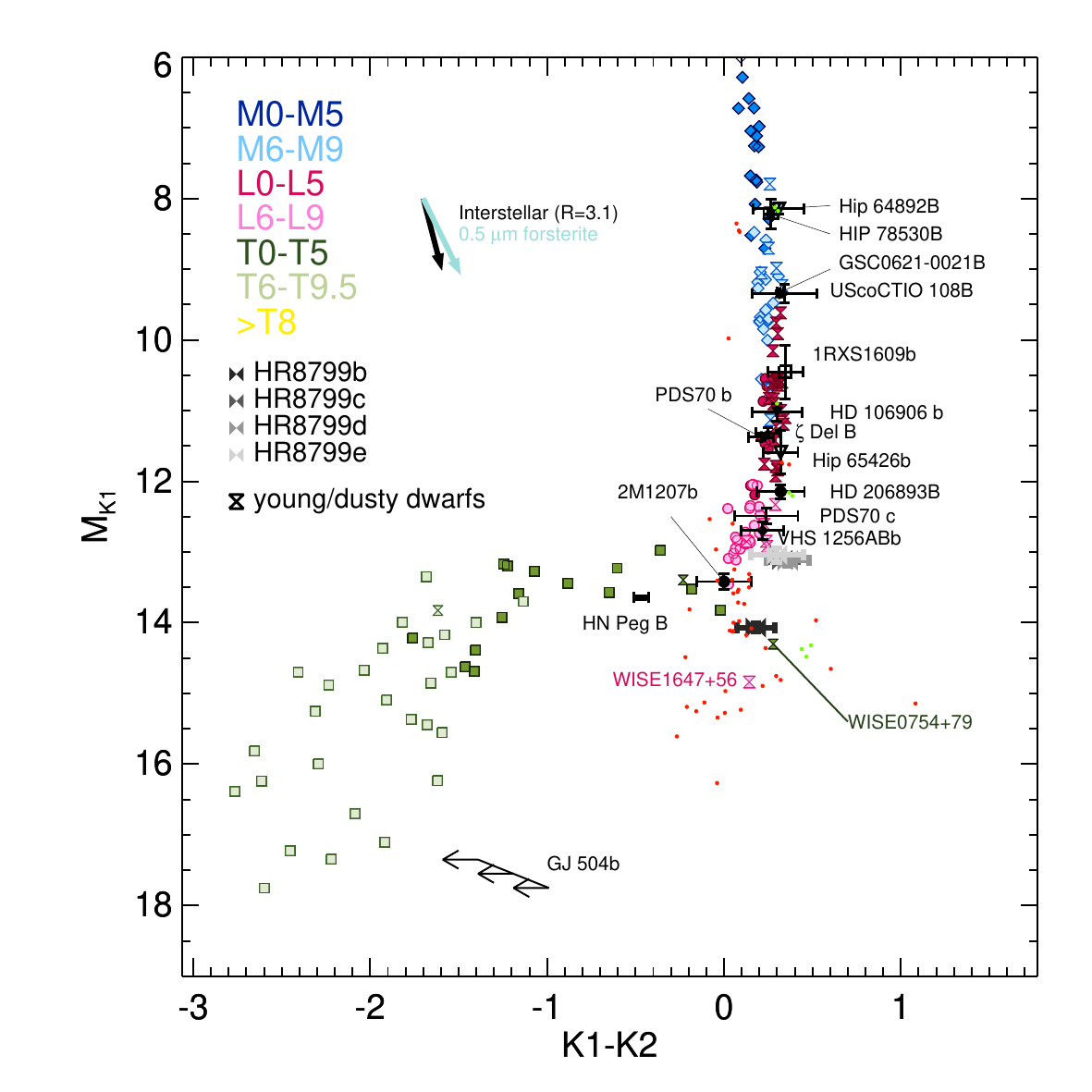} \\
	\end{tabular}
	\caption{Photometry of the candidates (spots) observed in the sample of 159 stars and reported in color-magnitude diagrams assuming a common distance with the host-star. The grey zone represents the excluded background-like point sources.}
	\label{Fig:CMDs}
\end{figure*}

\subsection{Mass detection limits}

The contrast curves are converted into mass limits using mass-luminosity relationships. Whereas for old ($\gtrsim1$~Gyr) systems this relationship is essentially unique for  gas giants at large separation \citep{Burrows1997,Baraffe2003}, at young ages the value of the post-formation luminosity still remains uncertain \citep{Marley2007,Spiegel2012,Marleau2014}. We present here a mass conversion of the contrast curves for a few specific targets using the canonical predictions of the COND-2003 evolutionary models \citep{Baraffe2003}. The impact of using other mass-luminosity relationships on the sensitivity of the SHINE survey are explored in more details in \citet{Vigan2020}. For IRDIS contrast curves, we convert the contrast curves using the evolutionary models computed in the appropriate dual-band filter, while for the IFS contrast curves we use the predictions in the J-band filter for YJ data and in the H-band filter for YJH data. As demonstrated in \citet{Vigan2015}, this approach for the IFS detection limits provides an accurate estimation of the detection limit.

\begin{figure*}[h]
    \centering
\begin{tabular}{cc}
    \includegraphics[width=0.8\columnwidth]{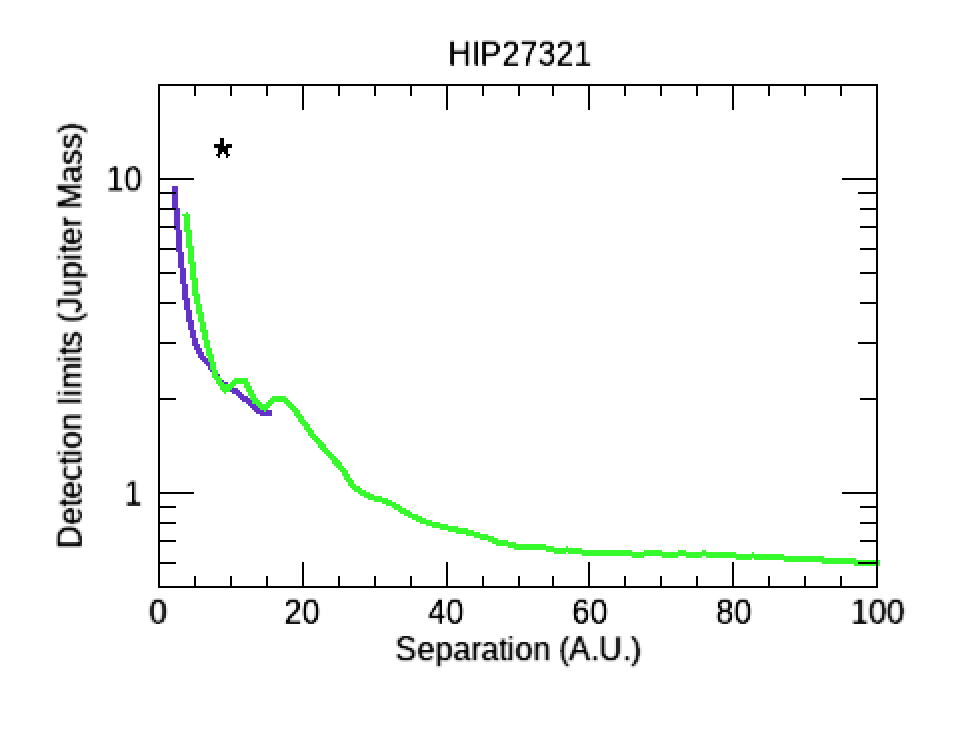} &
    \includegraphics[width=0.8\columnwidth]{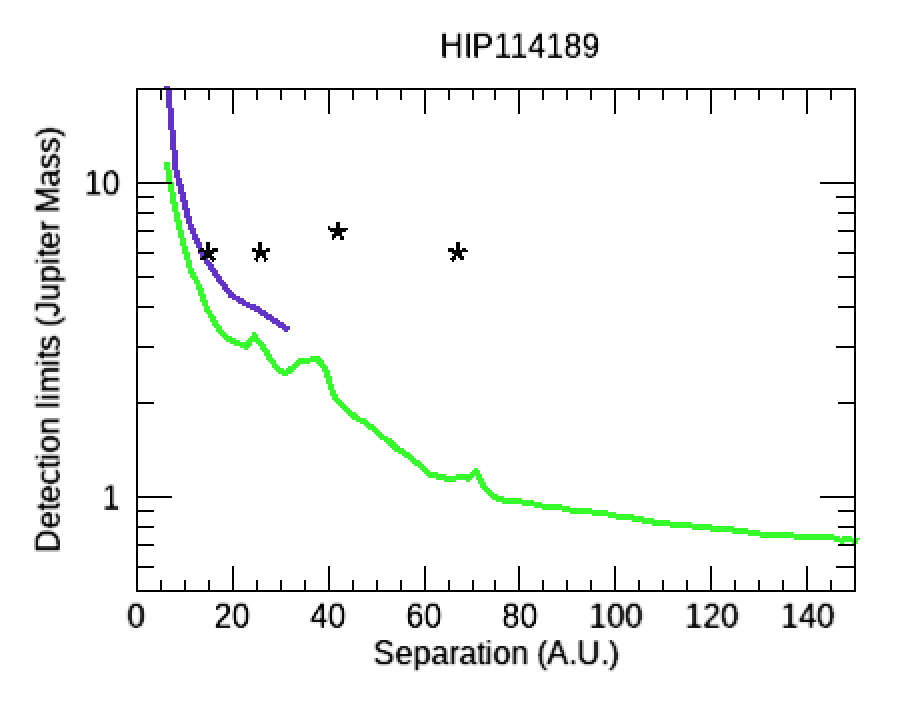} \\
    \includegraphics[width=0.8\columnwidth]{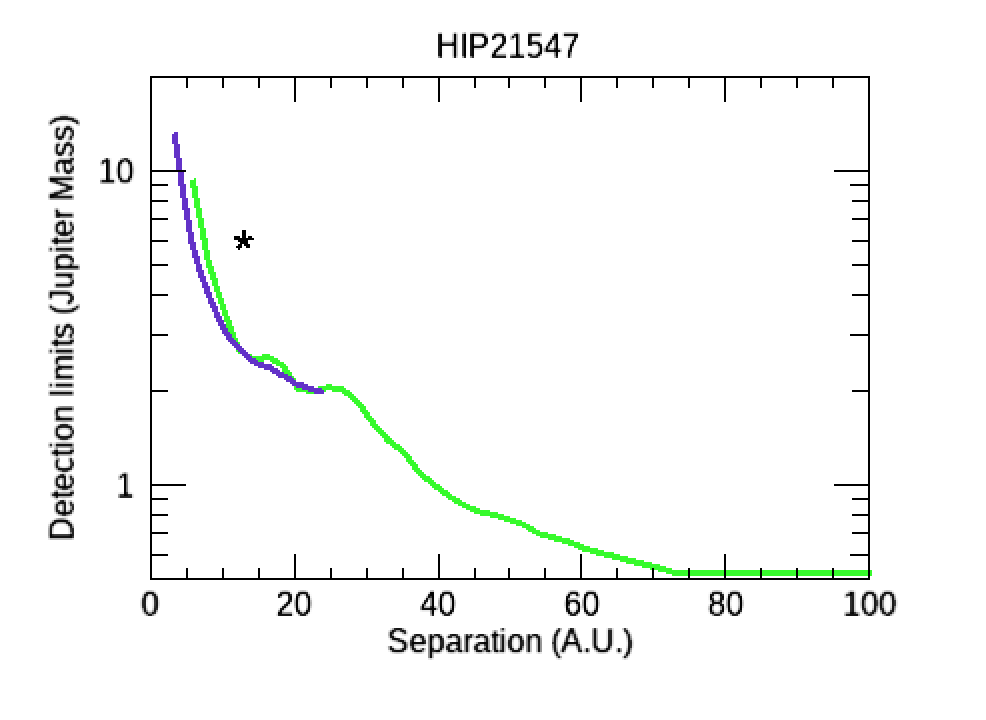} &
  \includegraphics[width=0.8\columnwidth]{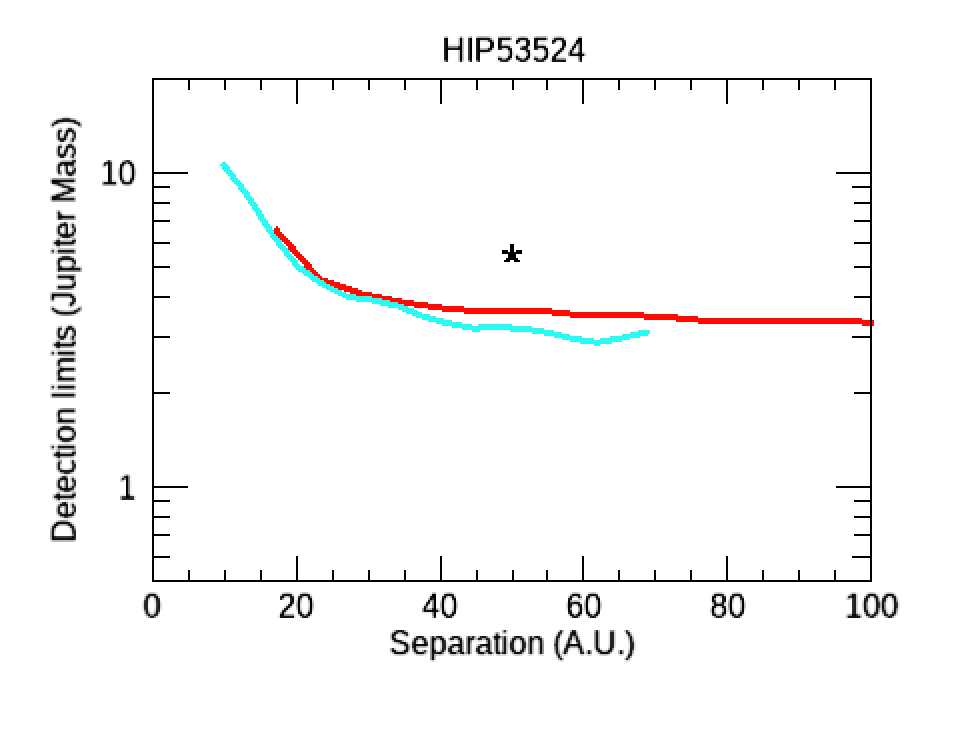}  \\
  \includegraphics[width=0.8\columnwidth]{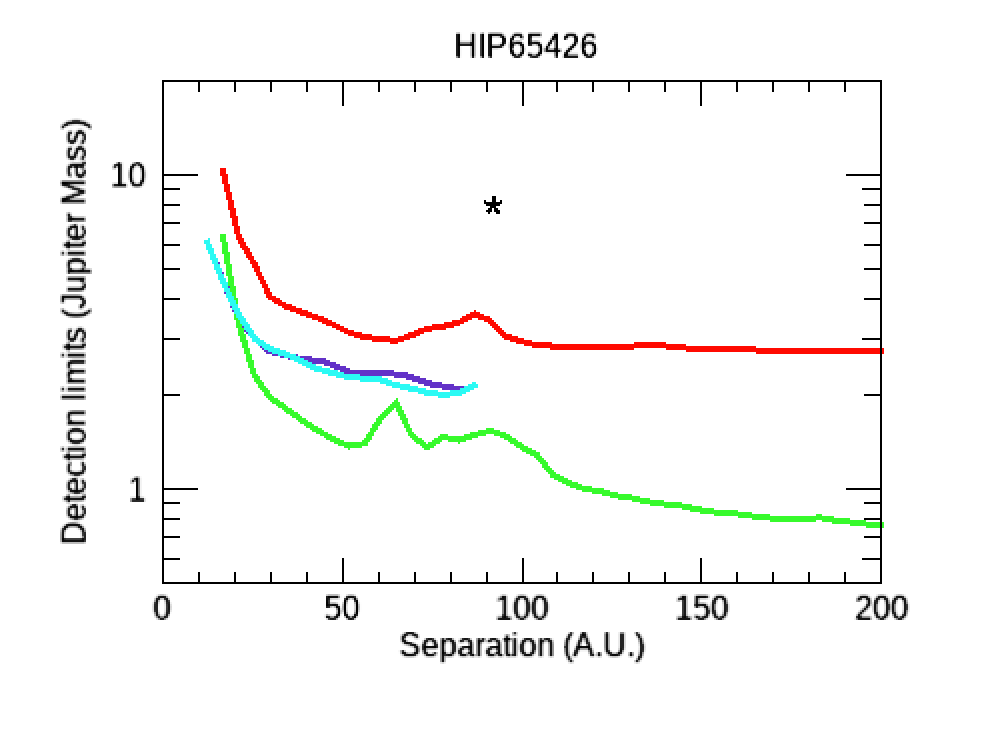} &
   \\
   \includegraphics[width=0.8\columnwidth]{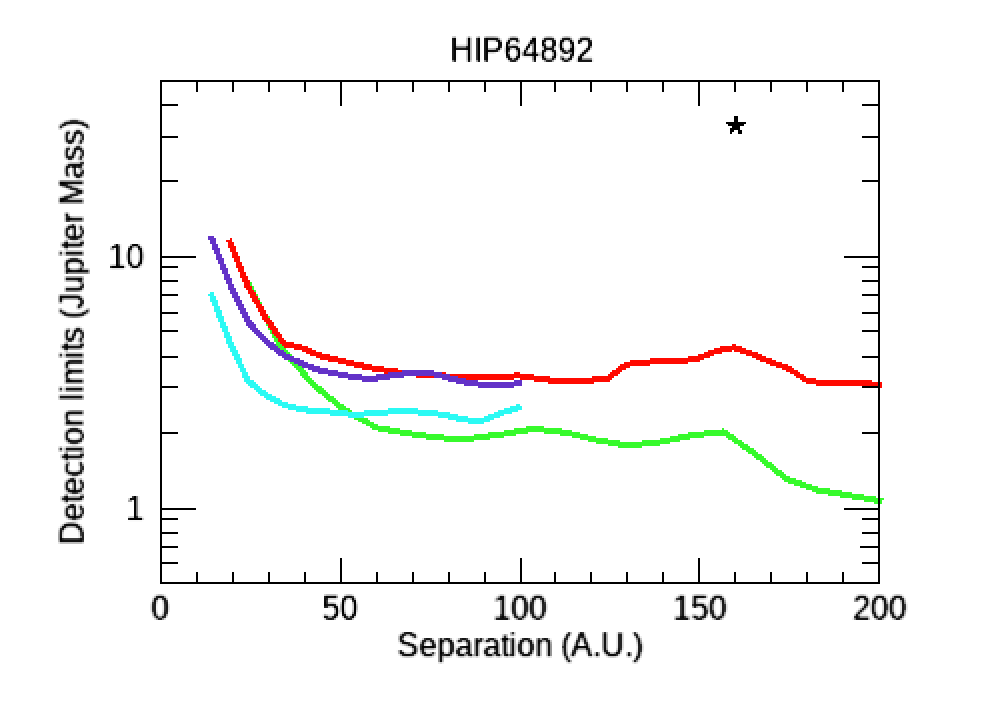}  & \includegraphics[width=0.8\columnwidth]{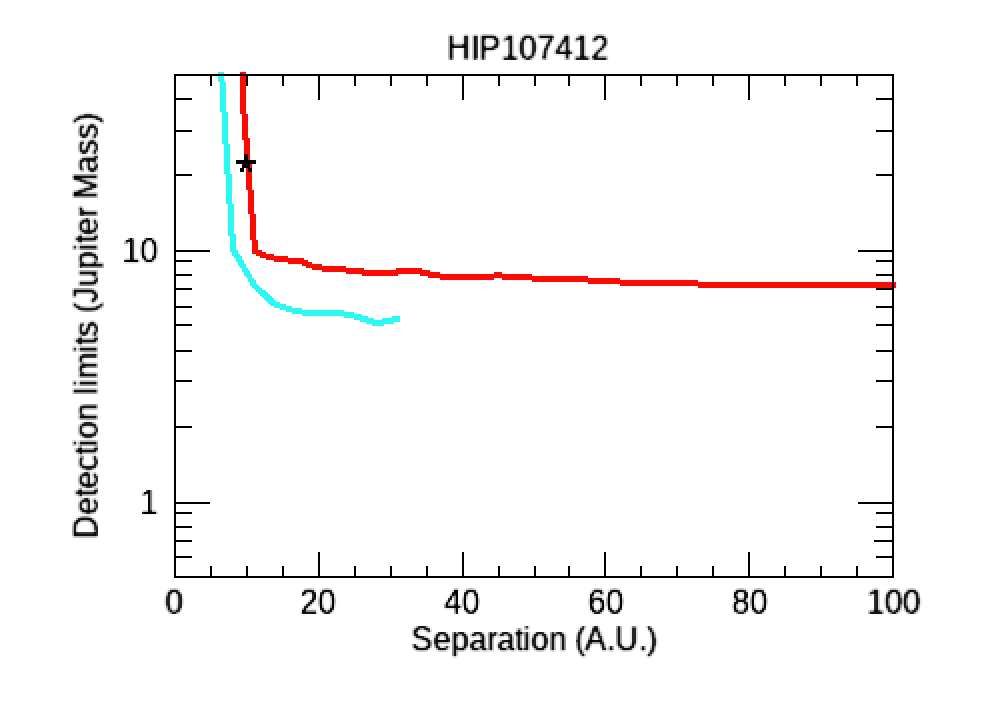}\\
    \end{tabular}
    \caption{SPHERE IRDIS and IFS mass detection limits. The IFS H data are represented in blue, the IFS YJ data are represented in violet, while the IRDIS H23 and IRDIS K12 data are represented respectively in green and red. The limits have been converted into mass using the COND-2003 evolutionary models \citep{Baraffe2003}.}
    \label{fig:detlim}
\end{figure*}

For a selection of companions presented in \ref{sec:detections_bdp}, we show the detection of planet-like limits achieved by the survey in Fig. \ref{fig:detlim}. Several targets appear to be relatively easy (HR~8799~bcd, HIP~65426 and HIP~64892) for this survey, allowing for exquisite characterisation. Other targets appear to be much more challenging mainly because they are at separations closer than 20 AU. The detection limits for these typical objects reach few Jupiter masses at such small separation while it can reach around 1 Jupiter mass at separations greater than 50 AU.

\begin{figure*}[h]
    \centering
\begin{tabular}{ccc}
    \includegraphics[width=0.66\columnwidth]{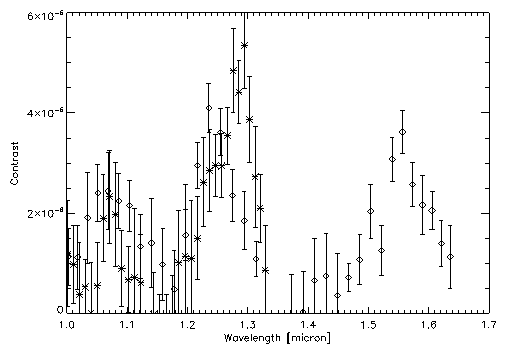} &
    \includegraphics[width=0.66\columnwidth]{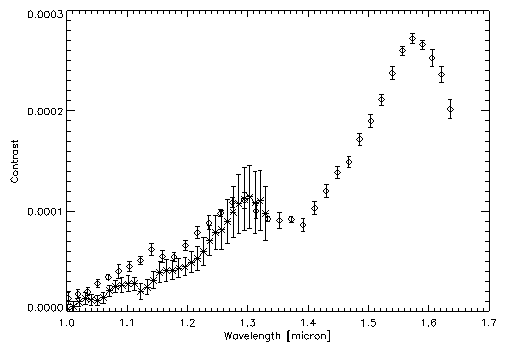} &
    \includegraphics[width=0.66\columnwidth]{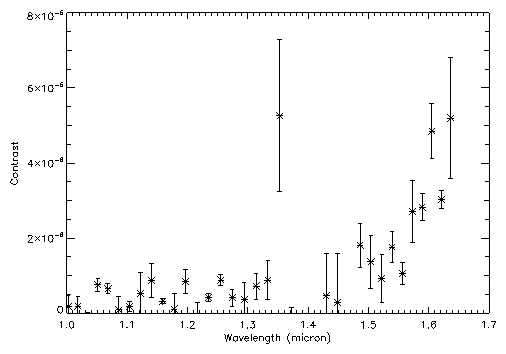} \\
    \includegraphics[width=0.66\columnwidth]{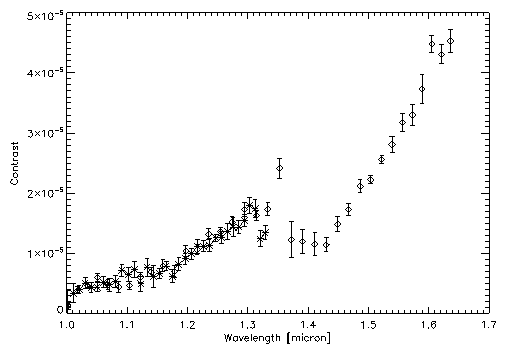} &
    \includegraphics[width=0.66\columnwidth]{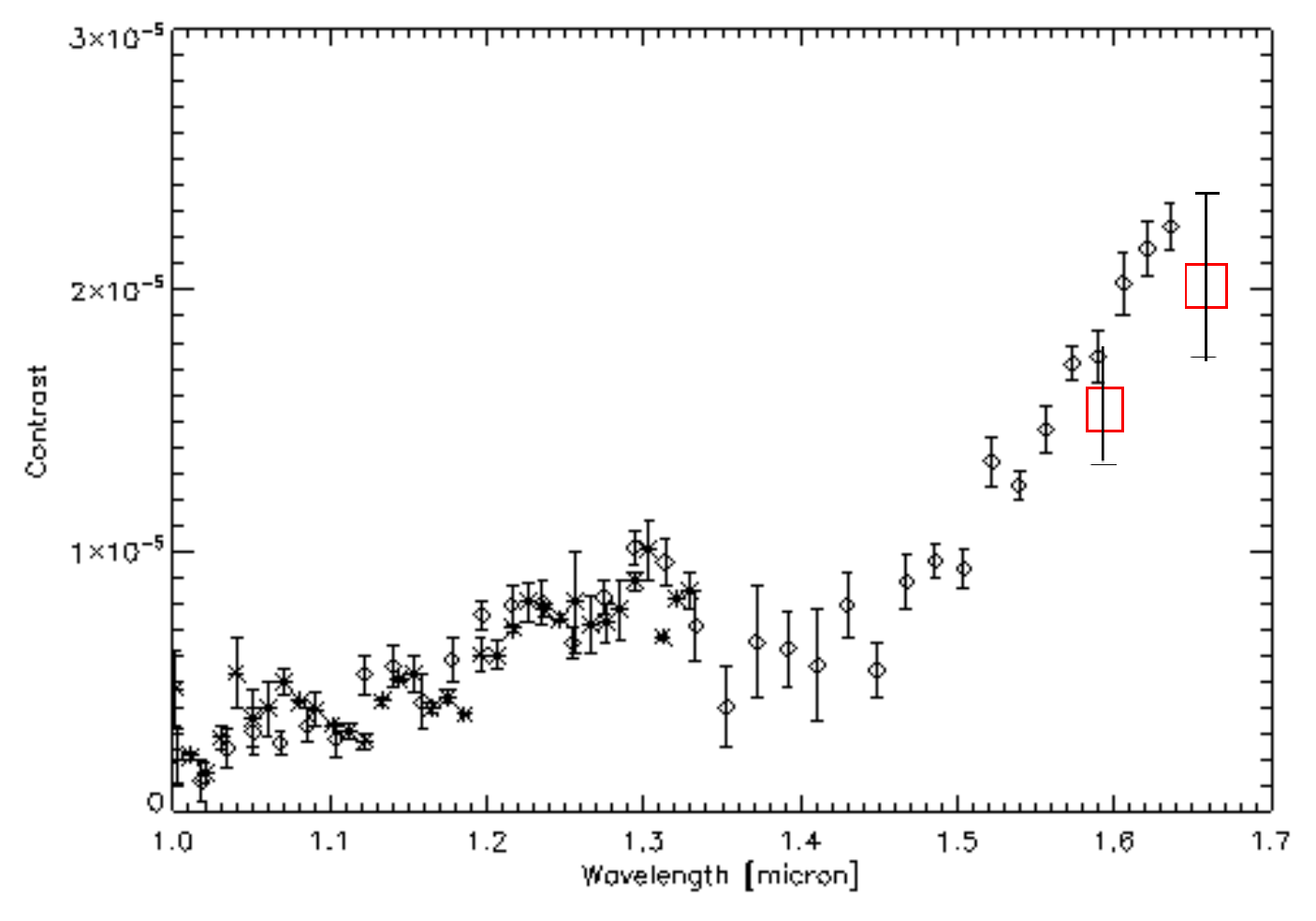}&
    \includegraphics[width=0.66\columnwidth]{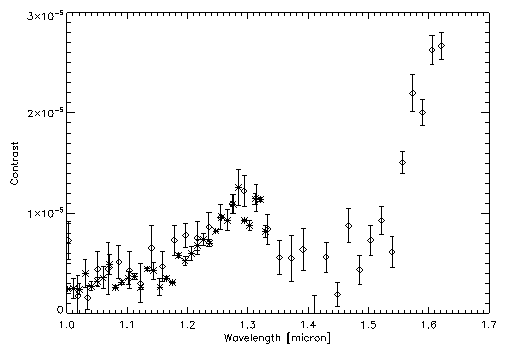} \\
    \includegraphics[width=0.66\columnwidth]{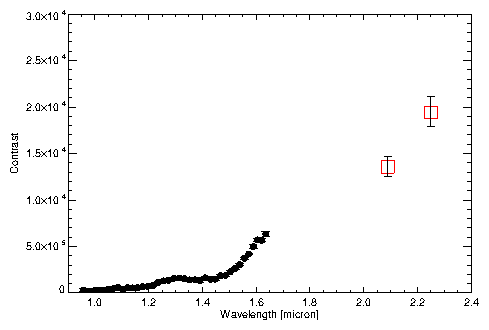} &
    \includegraphics[width=0.66\columnwidth]{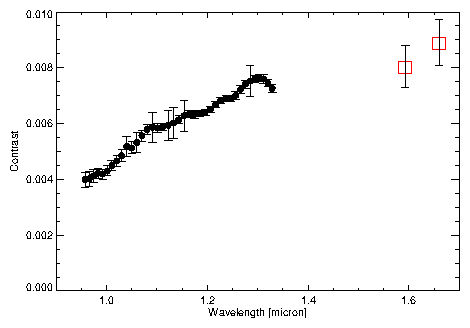} \\
    \end{tabular}
    \caption{Black symbols: SPHERE contrast spectra of substellar companions detected within the sample of stars considered in this paper. Upper row: left: 51~Eri~b; center: $\beta$~Pic~b; right: HD~95086~b; Middle row: left: HIP~65426~b; center: HR~8799~d; right: HR~8799~e. Lower row: left: HIP~107412~B; center: PZ~Tel~B. Results shown are obtained by making a median of results at different epochs with good quality results. Spectra obtained with the YJ and YH mode were kept separate. The error bar is either the result of the scatter of results at different epochs, or if only one was available in that epoch, the error bar obtained by considering other positions at the same separation from the star. The spectra shown are obtained using ASDI PCA, and corrected for the attenuation factor using fake planets at similar separation. This method is reasonable when the contrast is very large, so that the planet cannot be seen well at individual wavelengths. Red symbols: contrast values obtained from IRDIS data.Note that there is some offset between IFS and IRDIS contrasts. This is at least in part due to the non optimal extraction of IFS spectra shown in this Figure. More accurate spectral extractions may be found in the papers about the individual targets.}
    \label{fig:spectra}
\end{figure*}

\section{Exoplanet, Candidate Companion Detections}
\label{sec:detections}

\subsection{Brown dwarfs and exoplanets}
\label{sec:detections_bdp}
A total of sixteen substellar companions were imaged in the course of this part of the SHINE survey, including seven brown dwarf companions (PZ Tel\,B, $\eta$ Tel\,B, CD\,-35\,2722\,B, HIP 78530\,B, HIP\,107412\,B, GSC\,8047-0232\,B and HIP\,64892\,B), and ten planetary-mass companions (51 Eri\,b, $\beta$ Pictoris\,b, HD\,95086\,b, HR8799\,bcde, GJ\,504\,b, AB\,Pic\,b and HIP\,65426\,b). Two new companions have been discovered in this sample: the exoplanet HIP\,65426\,b \citep{Chauvin2017} and the brown dwarf companion to HIP 64892 B \citep{Cheetham2018}. PDS\,70 was not originally part of the SHINE sample and was specifically targeted to explore the properties of the transition disk that led to the discovery of the first planet PDS\,70\,b in this system \citep{Keppler2018,Muller18}, and the confirmation of the second planet PDS\,70\,c \citep{Mesa19}. The other substellar companions, known from previous direct imaging campaigns, were originally blindly selected based on their host star properties, but then carefully characterized to study their orbital, spectral and physical properties in connection sometimes with their environment (the presence of additional planets or disk structures in the system). The SNR maps for a selection of candidates are shown in Fig.\,\ref{fig:detections_ifs} for IFS data and Fig.\,\ref{fig:detections_irdis} for IRDIS data. The extracted spectro-photometric data are reported in Fig.\,\ref{fig:spectra}.  The measured position, contrast, semi-major axis, predicted mass for each companion are listed in Table\,\ref{tab:detections}. The detection limits for a few of these systems are shown on Fig.\,\ref{fig:detlim} as illustration of the SPHERE performances. The Sco-Cen members HIP\,71724 and HIP\,73990 with stellar to substellar-mass companion previously discovered in sparse-aperture masking by \cite{Hinkley2015} were observed, but the close companions were not resolved given their probable small separations at the epoch of our observations. 

In addition to the discovery and characterization papers on HIP\,65426\,b and HIP\,64892\,B summarized below, in-depths studies of various of these companions based on from the SHINE sample have been published in these early years of the survey: PZ\,Tel\,B \citep{2016A&A...587A..56M},
HIP\,107412\,B \citep{Milli2017,2017A&A...608A..79D} 
for the brown dwarf companion, and HR\,8799\,b,c,d,e \citep{Zurlo2016,2016A&A...587A..58B}, 
51\,Eri\,b \citep{Samland2017,Maire2019}, HD\,95086\,b \citep{Chauvin2018}, $\beta$ Pictoris\,b \citep{Lagrange2019}, GJ\,504\,b\citep{Bonnefoy18} for the imaged exoplanets.
 
 \subsubsection{HIP\,65426\,b}

HIP~65426~b was the first planet discovered with SPHERE as part of the SHINE Survey (\cite{Chauvin2017}). This warm, dusty giant planet is orbiting at a relatively large projected angular distance of of 830~mas (92 au projected) from its intermediate-mass primary (Fig.\,\ref{fig:detections_ifs} and \ref{fig:detections_irdis}). Multi-epoch observations confirm that it shares common proper motion with HIP\,65426, a young A2 member of the 17 Myr old Lower Centaurus-Crux association. Spectro-photometric measurements extracted with IFS and IRDIS, as shown in Fig. \ref{fig:spectra}, between 0.95 and 2.2 $\mu$m indicate a warm, dusty atmosphere characteristics of young low-surface gravity L5-L7 dwarfs. Hot-start evolutionary models predict a luminosity consistent with a $6-12$~M$_{\rm Jup}$, $T_{\rm eff}=1300-1600$\,K and $R=$1.5~R$_{\rm Jup}$ for the planet.
These results were later confirmed by \cite{Cheetham2019} combining SPHERE and NaCo observations.

Given its physical and spectral properties, HIP~65426~b represents a particularly interesting case to study the presence of clouds as a function of particle size, composition, and location in the atmosphere, to search for signatures of non-equilibrium chemistry, and finally to test models of planet formation and evolution \citep{Cheetham2018}. The planet location would not favor a formation by core accretion unless HIP~65426~b formed significantly closer to the star followed by a planet-planet scattering event. Dedicated simulations by \cite{2019A&A...624A..20M} showed that core formation at small separations from the star followed by outward scattering and runaway accretion at a few hundred astronomical units succeeds in reproducing the mass and separation of HIP 65426~b. In such a scenario, the planet is predicted to have a high eccentricity ($\geq 0.5$) and to be accompanied by one or several roughly Jovian-mass planets at smaller semi-major axes, which also could have a high eccentricity. SHINE detection limits setting relatively good constraints at close separation (2~M$_{\rm Jup}$ beyond 20~au) as shown in Fig. \ref{fig:detlim} do not exclude the presence of unseen inner massive planets in that system that could have scattered out HIP~65426~b. 

 \subsubsection{HIP~64892~B}
 
 A bright, brown dwarf companion was also discovered around the star HIP~64892, a B9.5V member of the Lower Centaurus-Crux association. The measured angular separation of the companion (about 1.27") corresponds to a projected distance of 159~au. We observed this target with IRDIS dual-band Imaging and in long-slit spectroscopy to estimate its spectral energy distribution and astrometry. More details can be found in the discovery paper by \cite{Cheetham2018}. Luminosity and spectrum are consistent with a young ($16$~Myr), low-surface gravity brown dwarf with a spectral type of M9. From comparison with the BT-Settl atmospheric models its effective temperature was estimated to $T_{\rm eff}=2600$~K, and comparison of the companion photometry to the COND evolutionary models yields a mass between 29 and 37~M$_{\rm Jup}$. Despite this, HIP 64892 is one of the highest mass stars around which a substellar companion has been detected, due to the challenges associated with observing such stars and the tendency for large surveys to focus on solar-like or low-mass stars. HIP~64892 is a rare example of high or intermediate mass stars with extreme mass ratio ($q = 0.01$) companions at large separations and will be useful for testing models relating to the formation and evolution of such low-mass objects.
 
\begin{table*}
    \caption[]{Exoplanets and brown dwarf companions observed during the SHINE campaign}
    \label{tab:detections}
    \centering
    \begin{tabular}{lccccccccc}
    \hline \hline
    Name & SpT$_{\rm host}$ & M$_{\rm \star}$ & $a$ & Mass & Observation  &  Contrast    &     separation & P.A. & Ref.  \\
     &     &                  ($M_{\rm \odot}$)   &      (au)            &     (\MJup)   & dates \tablefootmark{a}   & (H2)  & (mas) & (deg)& \\
    \hline
    \multicolumn{10}{c}{SHINE discoveries} \\
    \hline
    \object{HIP 64892 B}     & B9  & 2.09                & 147--171         & 29--37   & 2016-04-01& $7.24\pm0.08$   &    $1270.5\pm2.3$    &   $311.68\pm0.15$      & (1) \\
    \object{HIP 65426 b}     & A2  & 1.96                & 80--210          & 7--9     & 2016-05-30& $10.73\pm0.09$   &  $830.4\pm4.9$      &    $150.28\pm0.22$      & (2) \\
    \hline
    \multicolumn{10}{c}{Previously known companions} \\
    \hline
    \object{$\eta$ Tel B}    & A0  & 2.00                & 125--432         & 20--50   & 2015-05-05 & $6.99\pm0.22$        & $4214.5\pm22$         & $167.25\pm0.08$     & (8) \\
    \object{CD -35\,2722 B}  & M1  & 0.56                & 74--216          & 23--39    & 2015-29-11& $5.89\pm0.18$       & $2987.5\pm3.51$         &   $241.9\pm0.06$  &  (9) \\
    \object{HIP 78530 B}     & B9  & 1.99                & $\sim$620        & 19--26   & 2015-05-04& $8.05\pm XX $& $4537.5\pm3.0$ & $139.98\pm0.04$ & (18)\\
    $\beta$ Pic b   & A3  & 1.61                & 8.5--9.2         & 9--16    &2018-12-15 & $9.86\pm0.1$  & $177.4\pm4.51$ & $29.07\pm0.6$ & (3,10)\\
    \object{HR 8799 b}       & A5  & 1.42                & 62--72           & 5.3--6.3 & 2014-07-13 &$13.19\pm0.16$  & $1721.8\pm6.0$ & $65.76\pm0.19$ &(4,11)\\
    \object{HR 8799 c}       & A5  & 1.42                & 39--45           & 6.5--7.8 & 2014-07-13 &$11.99\pm0.16$  & $ 947.7\pm4.0$ & $326.58\pm0.24$ &(4,11) \\
    \object{HR 8799 d}       & A5  & 1.42                & 24--27           & 6.5--7.8  & 2014-07-13&$12.00\pm0.17$ & $658.0\pm5.0$ & $216.35\pm0.52$ &(4,11)\\
    \object{HR 8799 e}       & A5  & 1.42                & 14--17           & 6.5--7.8 & 2014-07-13&$11.93\pm0.17$  & $386.1\pm9.0$ &$268.81\pm$1.3& (4,19)\\
    \object{HD 95086 b}      & A8  & 1.55                & 28--64           & 2--9     & 2015-03-02& $12.37\pm0.16$ & $622.0\pm4.0$ & $148.8\pm0.4$ & (5,12)\\
    \object{51 Eri b}        & F0  & 1.45                & 10--16           & 6--14    & 2016-12-12& $13.96\pm0.51$ & $444.8\pm10.6$&$159.4\pm2.6$ & (6,20,13) \\
    \object{HIP 107412 B}    & F5  & 1.32                & 6.2--7.1         & 15--30   & 2016-09-16& $9.67\pm0.09$  & $265.0\pm2.0$ & $62.25\pm0.11$ & (22,14)\\
    \object{PZ Tel B}        & G9  & 1.07                & 19--30           & 38--54   &2014-07-15& $5.07\pm0.61$ & $478.48\pm2.10$ & $59.58\pm0.48$ & (7,15,21) \\
    \object{AB Pic b}        & K1  & 0.97                & $\sim$250        & 13--30   & 2015-02-06 & $8.04\pm0.98 $& $5392.1\pm13.1$ & $175.26\pm0.15$  & (16)\\
    \object{GSC 8047-0232 B} & K2  & 0.89                & 190--880         & 15--35   & 2015-09-25 & $7.67\pm0.15$ & $3207.7\pm2.7$  & $358.82\pm1.03$ & (17) \\
    \object{GJ 504 b} &  G0  & 1.15                 &     40-50      &   4-6  & 2015-05-06 & $14.56\pm0.15$ & $1463.7\pm3.8$ & $323.1\pm0.1$ & (23, 24) \\

    \hline 
    \end{tabular} 
    \tablefoot{\tablefoottext{a}{We only list one selected epoch for each target}}
    \tablebib{SHINE papers: (1) \citet{Cheetham2018}; (2) \citet{Chauvin2017}; (3) \citet{Lagrange2019}; (4) \citet{Zurlo2016}; (5) \citet{Chauvin2018}; (6) \citet{Samland2017}; (7) \citet{2016A&A...587A..56M}; (20)
    	\citet{Maire2019}; (24) \citet{Bonnefoy18}, Discovery papers: (8) \citet{2000ApJ...541..390L}; (9) \citet{2011ApJ...729..139W}; (10) \citet{Lagrange2009}; (11) \citet{Marois2008};  (12) \citet{2013ApJ...772L..15R}; (13) \citet{Macintosh2015}; (14) \citet{Milli2017}; (15) \citet{2010ApJ...720L..82B}; (16) \citet{2005A&A...438L..29C}; (17) \citet{2005A&A...430.1027C}; (18) \citet{lafreniere2011}; (19) \citet{2010Natur.468.1080M}; (21)
    	\citet{Mugrauer2010}; (22) \citet{2017A&A...608A..79D}; (23) \citet{Kuzuhara2013}
   
    }
\end{table*}

\subsection{Colours and spectra of detected substellar companions}

Fig.\,\ref{fig:spectra} shows the spectra of the brown dwarf and exoplanet companions detected within the IFS field of view: 51 Eri b, $\beta$ Pic b, HD\,95086 b, HIP 65426 b, HR 8799 d and e, HIP 107412 B and PZ Tel B. Most of these spectra are very red and can be classified as of L- or early T-type. The only middle T-type object detected in the IFS field of view is 51~Eri~b \citep{Samland2017}, while PZ~Tel~B is an M-type object \citep{2016A&A...587A..56M}. These very red colors were quite unexpected for such faint objects, and indicate that young substellar objects have a spectrum quite different from those of older ones of the same luminosity, which can be interpreted as an indication for the presence of dust in their atmosphere  \citep{Chauvin2004,2014A&A...562A.127B}. From the point of view of planet detection - the focus of this paper - the implication is that young substellar companions may be below detection limit in the Y and J-band and even faint in the H-band, while they may be detected quite easily in the K-band (extension of the datacubes to short wavelengths is however useful to model and better subtract speckles). An extreme case of this behaviour is represented by HD~95086 b \citep{Chauvin2018}. This has a significant impact on the detection limits of our survey and points toward the need of model atmospheres that include dust to properly derive masses from the observed magnitudes. 

\subsection{Ranking and identification of SHINE candidates}
\label{sec:ranking}

We identified 1483 unique candidates in the IRDIS images (all epochs), 1176 of which were re-detected at several epochs or using archival data and 307 of which were only detected at a single epoch. Most of the candidates have been observed with the H23 filters ($>95\%$). Sixty-nine candidates have K1-K2 photometry extracted from the \texttt{IRDIFS\_EXT} observations. They were identified in follow-up observations of known systems (HD 95086, HIP107412, HIP 65426, HIP 64892, CD-35 2722) or for the most part in first epoch observations of stars belonging to the Sco-Cen association for which the detection of  late-L dusty companions with red near-infrared colors is favored at K-band \citep[e.g., HD95086b;][]{Chauvin2018}. 

We used color-magnitude diagrams (CMDs) to identify the most promising candidates and perform an efficient follow-up campaign. The candidate absolute magnitudes were computed assuming they are bound and at the same distance as the observed star. Most contaminants are expected to be M- and K-type stars \citep[e.g., ][]{2011ApJ...726...27P}. Their characteristic colors as well as the range of contrasts probed by the observations  creates a locus which can fall in a distinctive color-luminosity space from the sequence of substellar objects in the CMDs.  The diagrams are described in details in \cite{Bonnefoy18} and are enriched with the recent photometry and distances of reference objects measured by \cite{Muller18}, \cite{Mesa19}, and \cite{Dupuy20}. 

The large number of candidates identified in the H23 observations allows for building a well defined locus of contaminants in the corresponding CMD (see Figure \ref{Fig:CMDs}). The locus shows an elongation along a vector which may be related to the distribution of spectral type of the contaminants combined to interstellar extinction (the fainter the object with respect to the host star, the higher the optical extinction and color reddening). The locus  intersects the  sequence of substellar objects at the L/T transition and therefore can not discriminate  contaminants from bound companions falling in that luminosity range. We could however define an exclusion zone (grey shaded region) where substellar companions are not expected to produce such colors and absolute magnitudes. The zone is defined to avoid missing i/ young mid- and late- T-type planets such as 51 Eri b \citep{Macintosh2015} which can have reddened colors with respect to mature field T-dwarfs and ii/ young planets at the L/T transition such as HR8799 which can appear under-luminous with respect to older objects in the same temperature range. 
This criterion was used to identify 44\% of the candidates with H23 photometry as likely contaminants for which a re-observation was not warranted. Consequently, we decided to prioritize the observations of candidates falling right on the sequence of late-M to mid-L dwarfs and for which the locus spans a narrow range of bluer colors in the same luminosity range. This allowed for identifying HIP 64892B and HIP 65426b as promising candidates before they could be confirmed as co-moving objects. We could also identify blindly known companions (e.g., HIP 78530B, $\eta$ Tel B, PZ Tel B, AB Pic b, CD-35 2722B, HIP 73990B) as promising candidates. 

Most of the remaining ambiguous candidates were observed at a second epoch to determine whether they are co-moving companions. Two crowded fields were observed with the J23 filter set. As shown in \cite{Bonnefoy18}, the locus of contaminants falls at a distinct location from the substellar objects in these CMDs and  allowed to classify most of these candidates as contaminants.  The locus in the K1K2 diagrams is more dispersed and the error bars on the photometry tend to be wider owing to the higher thermal background at these wavelengths. Therefore, detections in K1K2 requires follow-up observations to clarify the nature of the detections. 

The final status of the candidates is given in Table \ref{tab:detections_all}. The companions with a confirmed common-proper motion are flagged as 'C'. The contaminants either discriminated using the CMD or the comon-proper motion test are flagged as 'B'.Ambiguous cases are flagged as 'A'.  We report in addition several not clear detections ('NC') in that table which correspond to the identification of candidates with unreliable astrometry and/or photometry and for which our standard classification scheme could not be applied.

\section{Disks detected in SHINE}
\label{sec:disks}

\begin{figure*}
    \centering
\begin{tabular}{ccc}
    \includegraphics[width=0.66\columnwidth]{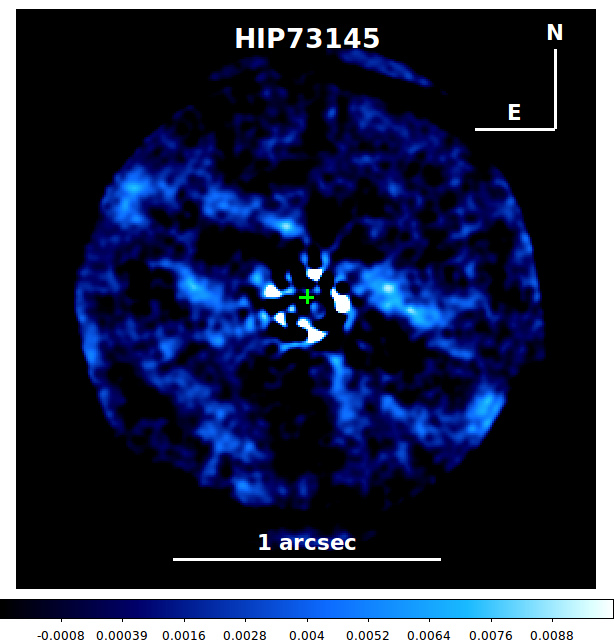} &
    \includegraphics[width=0.66\columnwidth]{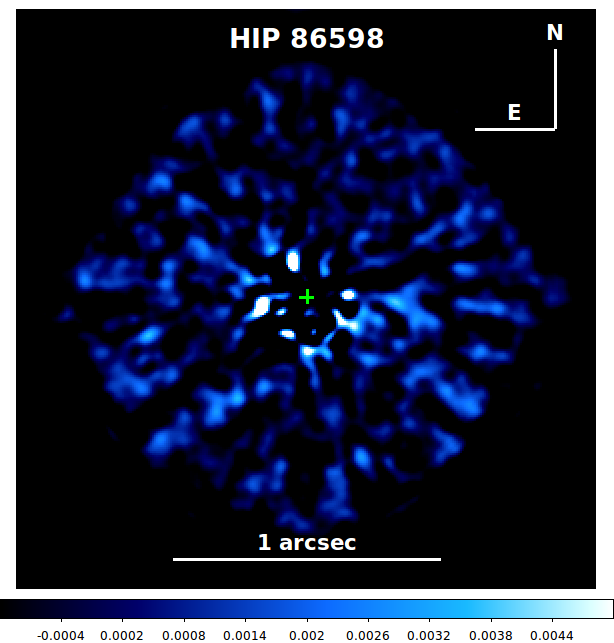} &
    \includegraphics[width=0.66\columnwidth]{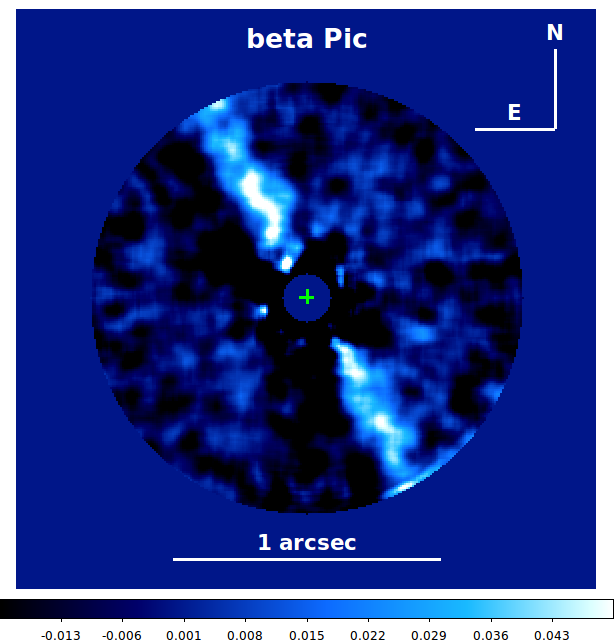} \\
    \includegraphics[width=0.66\columnwidth]{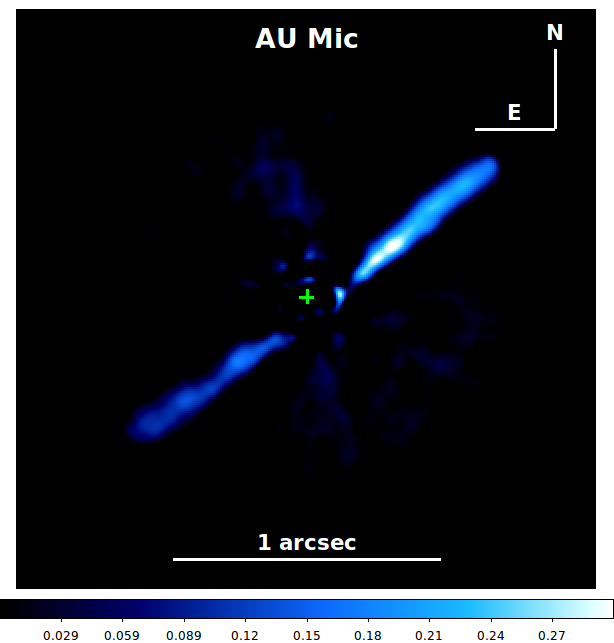} &
    \includegraphics[width=0.66\columnwidth]{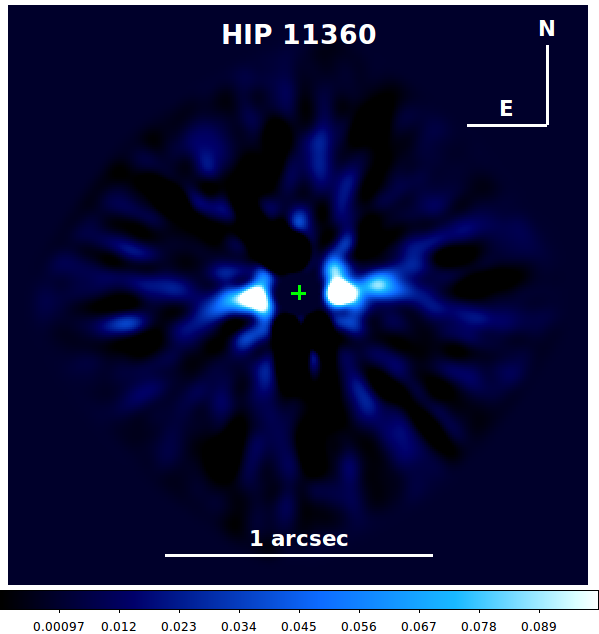} &
    \includegraphics[width=0.66\columnwidth]{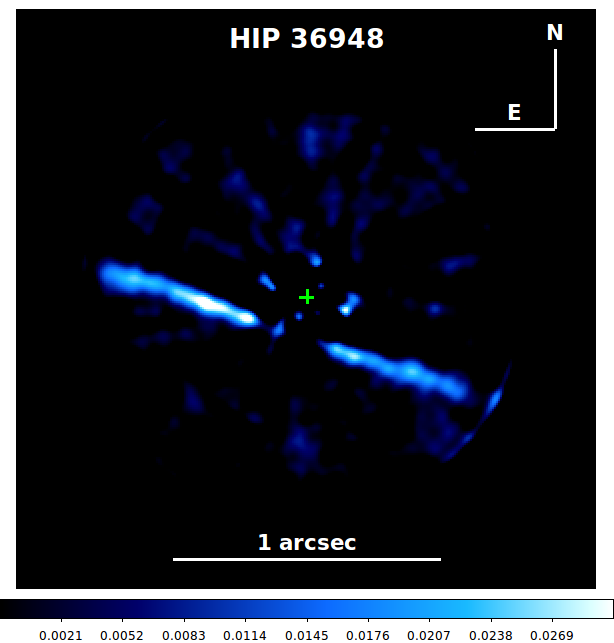} \\
    \includegraphics[width=0.66\columnwidth]{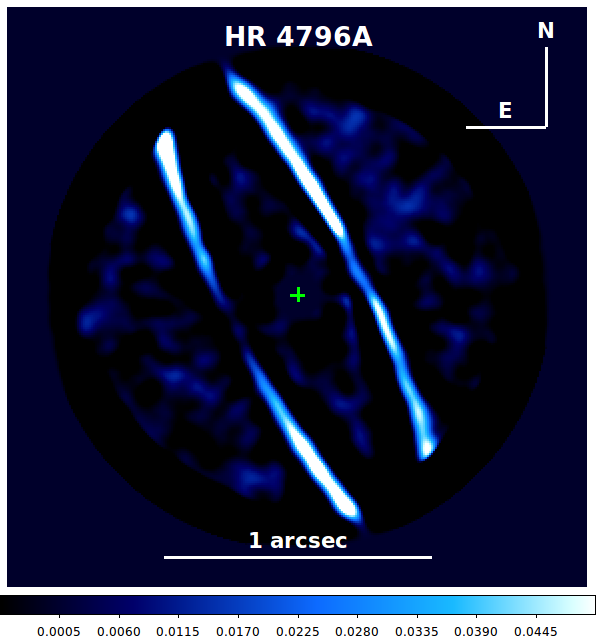} &
    \includegraphics[width=0.66\columnwidth]{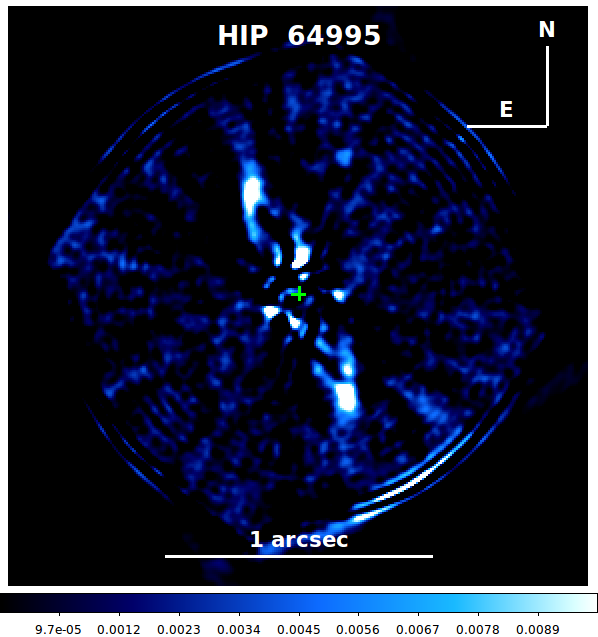} &
    \includegraphics[width=0.66\columnwidth]{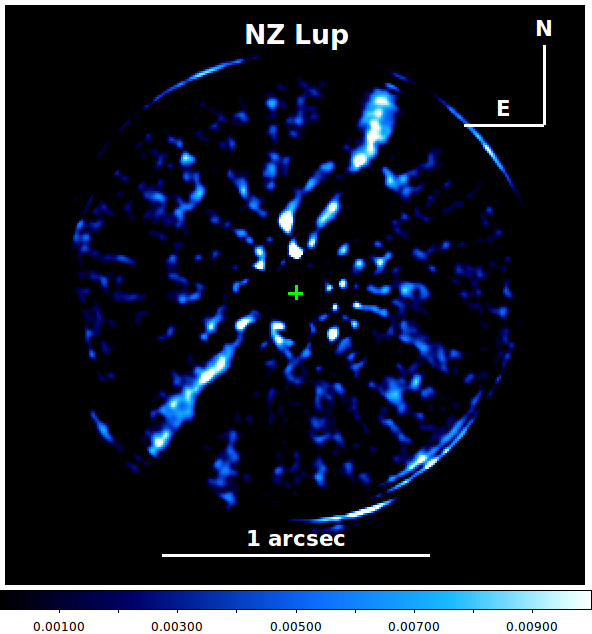}
    \\
    \end{tabular}
    \caption{Gallery of images of debris disks in the F100 sample detected with IFS. Some of detection is very marginal  (e.g. HIP86598) . The disk is not detectable for TWA7 and HIP682 (it was however detected with SPHERE using other approaches). Whenever possible, we combined several images to have a cleaner detection. This is the reason the planet is not visible in the beta Pic image. Generally, we used ASDI PCA with 25 modes, that is a quite conservative method, and smoothed the images to have a better view. In each panel, the green cross mark the star position}
    \label{fig:disks}
\end{figure*}

While not specifically optimized for this purpose, our observations allowed detection of circumstellar disks, thanks to the exquisite sensitivity and large FoV of IRDIS, twelve of SHINE targets, out of which three were new detections. By construction of our survey, they are debris disk - in fact, we did not consider stars with gas-rich disks to reduce the concern related to the impact of disks in planet detection. Table~\ref{tab:disk_detections} collects the main data for the disks we detected; a gallery of their appearance from the IFS data is presented in Figure~\ref{fig:disks}. Detailed analysis of each one of these disks is subject of specific studies, most of them already published (see Table~\ref{tab:disk_detections}). 

Not surprisingly, stars with detected debris disks are typically young; the median age is 24 Myr (data from Paper I) and there is only one target with an age older than 50 Myr (HIP682=HD377). This result is similar to that obtained by the survey by \citep{Esposito2020}. We notice that most of the detected disks are seen at high inclination $i$, with a median value of $i=82$~degree. The only disk that has a low inclination in our survey (TWA 7) was detected only in polarimetry \citep{olofsson2018}. The strong dependence on inclination is because of the combination of a deeper optical depth (debris disks being optically thin), of higher efficiency of forward scattering, and of the use of angular differential imaging to reach deeper contrast (see \citealt{Milli2012, sissa2018}). This selection effect would be less severe using polarimetric observations, as often done when studying disks (see the case of TWA 7; see also \citealt{Esposito2020}). The strong selection bias towards edge-on disks implies that debris disks at the typical age of our targets should be much more common than indicated by eleven detections over a sample of 150 stars. If we indeed consider only the 52 stars with age $<$30~ Myr in our sample, we detected disk around 8 of them. In five cases the inclination is $i>80$ degrees, an occurrence that has 17\% of the probability to happen. This suggests that debris disks similar to the ones we detected surround a large fraction of young stars (see \citealt{sissa2018} for a similar argument about the M-stars in our survey or \citealt{Meshkat2017}). 

Debris disks have a limited impact on detection limits of point sources. Low inclination debris disks have negligible effects because they are optically thin and well below detection limit. For what concerns high inclination disks, the effect is more relevant (see \citealt{Milli2012} for a discussion). They typically cover a small fraction of the sky area; however, planet orbit is possibly coplanar with the disk, so that the sky region covered by the disk is where most likely the planets are. As discussed by \citet{Milli2012}, the flux loss depends on the technique used, and may be large for some of them.  On the other hand, in addition to flux losses, \citet{Milli2012} showed that disk features may lead to false positive point sources detections.
Furthermore, the multi-wavelength approach allowed by simultaneously using both IFS and IRDIS provides more diagnostics for an appropriate classification of detected features and more particullarly point sources. The good sensitivity of ASDI-PCA to structures within disks is clearly shown by the case of AU Mic, where this technique allows to detect very faint point-like sources with contrasts of $\sim 14.5$~mag at SNR$>10$ along the edge-on disk \citep{Boccaletti2018}. The SNR of these detections agree with the expectation based on our estimate of the contrast limit at this separation based on the whole images. Note that these structures are however not planets: the spectrum is flat, inconsistent with that of young planets, and the objects are by far too bright to be reflecting planets. We carefully considered a similar possibility for all our targets, and we are confident that there is no false positive among our detections.

Concerning the detection of the disk structure themselves, the standard ADI methods used in this paper lead to several known causes of artifacts altering both the morphology and the photometry of the disks. In particular they suffer from positive / negative replicas and from the  self-subtraction phenomenon which can only be alleviated by using a physical model of instrumental effects as recently proposed by \citep{Pairet2020, Flasseur2021} but which are beyond the scope of this paper. 

\begin{table*}
    \caption[]{Debris disk detected from the SHINE survey (F150 sample)}
    \label{tab:disk_detections}
    \centering
    \begin{tabular}{lccccc}
    \hline \hline
Star & SpT & $M_G$ & Age & i & References\\
     &     & mag   & Myr & Degree &         \\
    \hline
    \multicolumn{6}{c}{New SHINE detections} \\
    \hline
HIP 73145             & A2IV   & 7.87 & 17 & 72.6 & \cite{feldt2017}\\
HIP 86598 / HD 160305 & F8/G0V & 8.18 & 24 & 82.0 & \cite{perrot2019}\\
HIP 59960 / HD 106906 & F5V&7.68&13&5&\cite{Lagrange2016}\\
    \hline
    \multicolumn{6}{c}{Previously known disks} \\
    \hline
$\beta$\,Pic b        & A6V    & 3.72 & 24 & $>$89 & \cite{Lagrange2019}   \\
AU Mic                & M1V    & 7.84 & 24 & $>$89 & \cite{Boccaletti2018}\\
HIP 682/ HD 377       & G2V    & 7.44 & 150 & & Langlois et al. in prep \\ 
HIP 11360 / HD 15115  & F4IV   & 6.68 & 45 & 85.8 & \cite{engler2019}\\ 
HIP 36948 / HD 61005 & G8V    & 8.00 & 50 & 84.5 & \cite{olofsson2016} \\ 
TWA 7                 & M2V    & 10.62& 10 & 13 & \cite{olofsson2018}\\
TWA 11 / HR4796       & A0V    & 5.77 & 10 & 76.4 & \cite{Milli2017_disk} \\
HIP 64995 / HD 115600 & F2IV/V & 8.14 & 16 & 80 & \cite{gibbs2019} \\
NZ Lup                & G2     & 7.78 & 16 & 86 &\cite{boccaletti2019} \\
    \hline 
    \end{tabular} 
\end{table*}

\begin{figure*}
	\centering
	\begin{tabular}{ccc}
		\includegraphics[width=0.66\columnwidth]{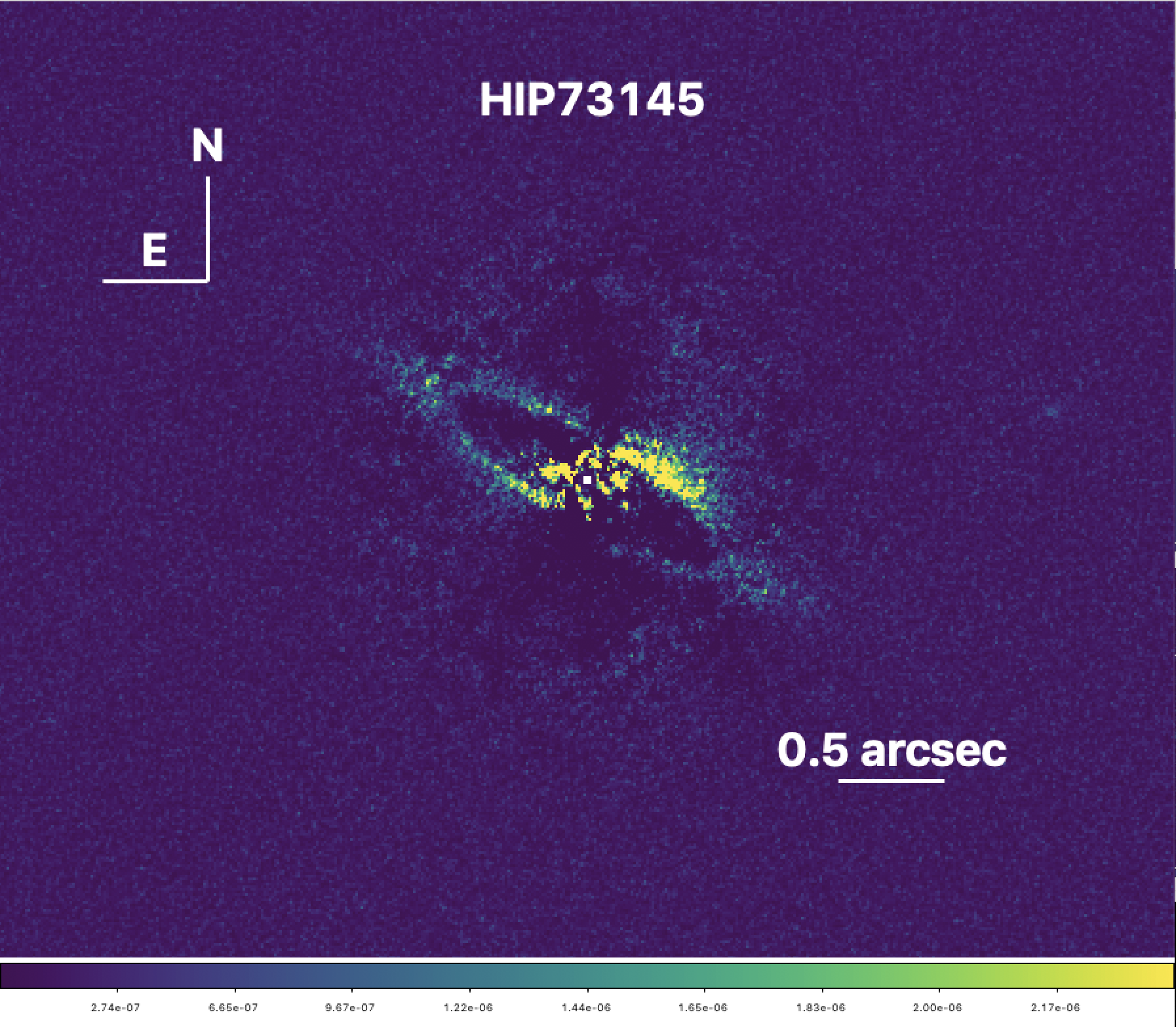} &
		\includegraphics[width=0.66\columnwidth]{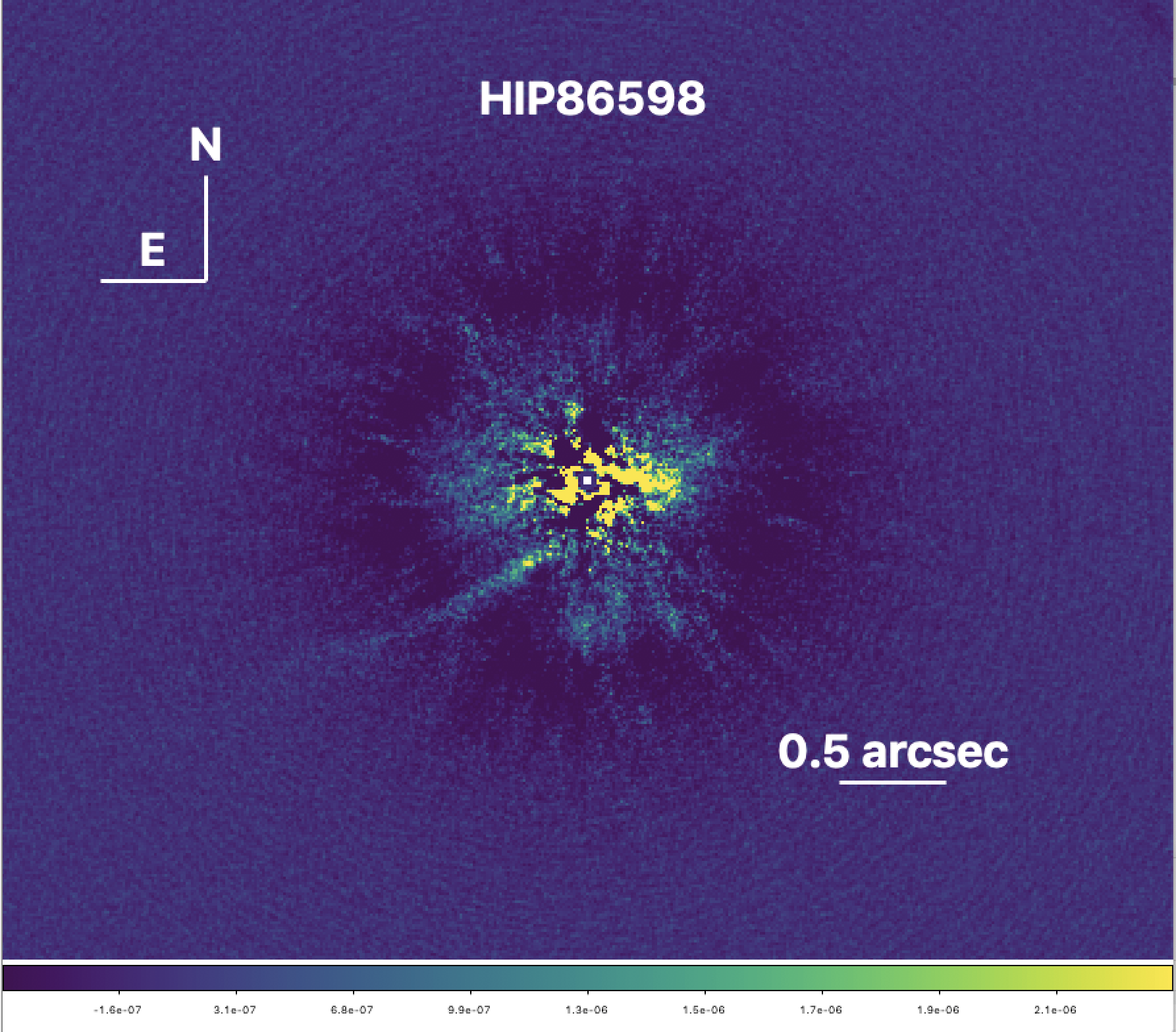} &
		\includegraphics[width=0.66\columnwidth]{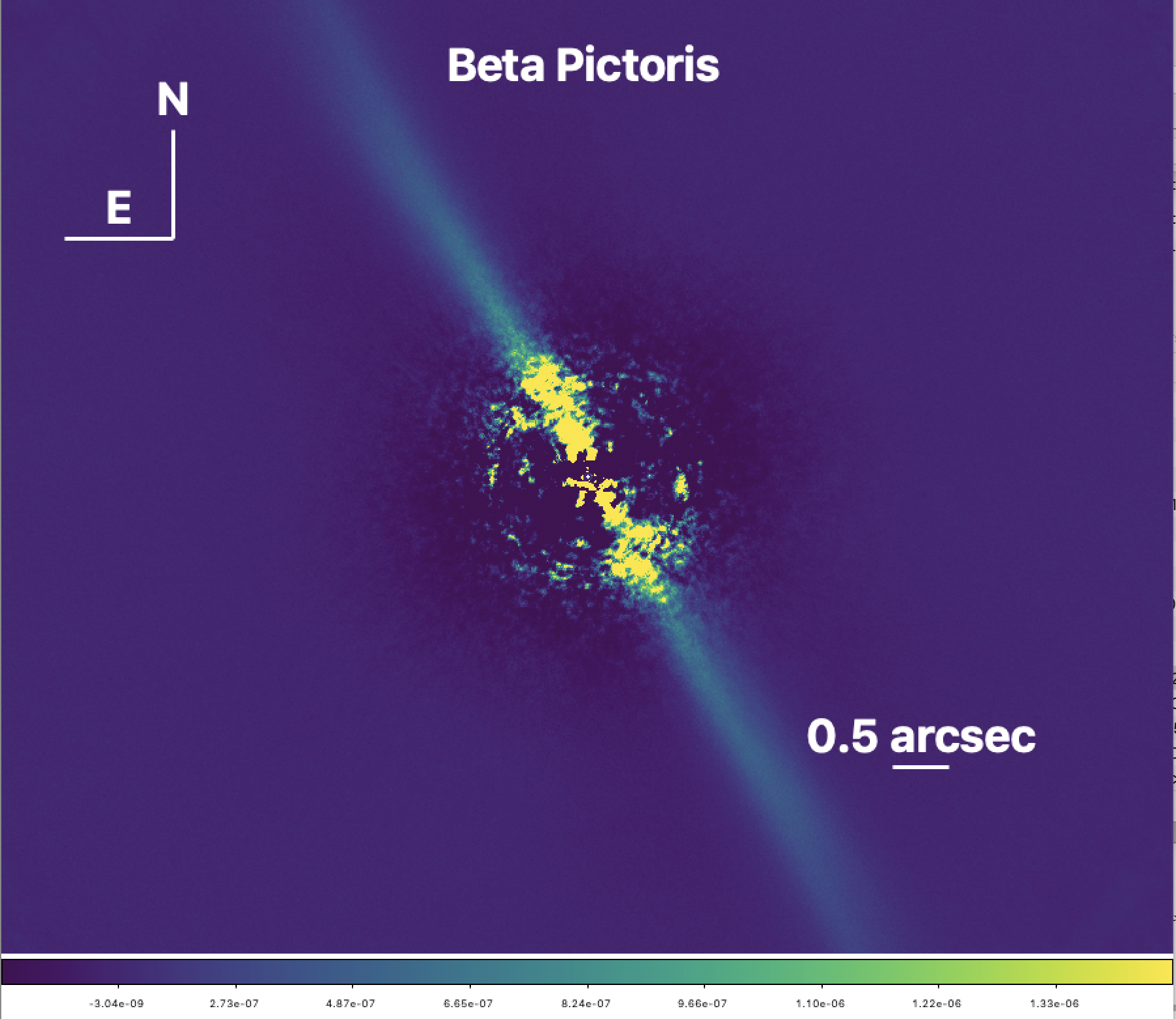} \\
		\includegraphics[width=0.66\columnwidth]{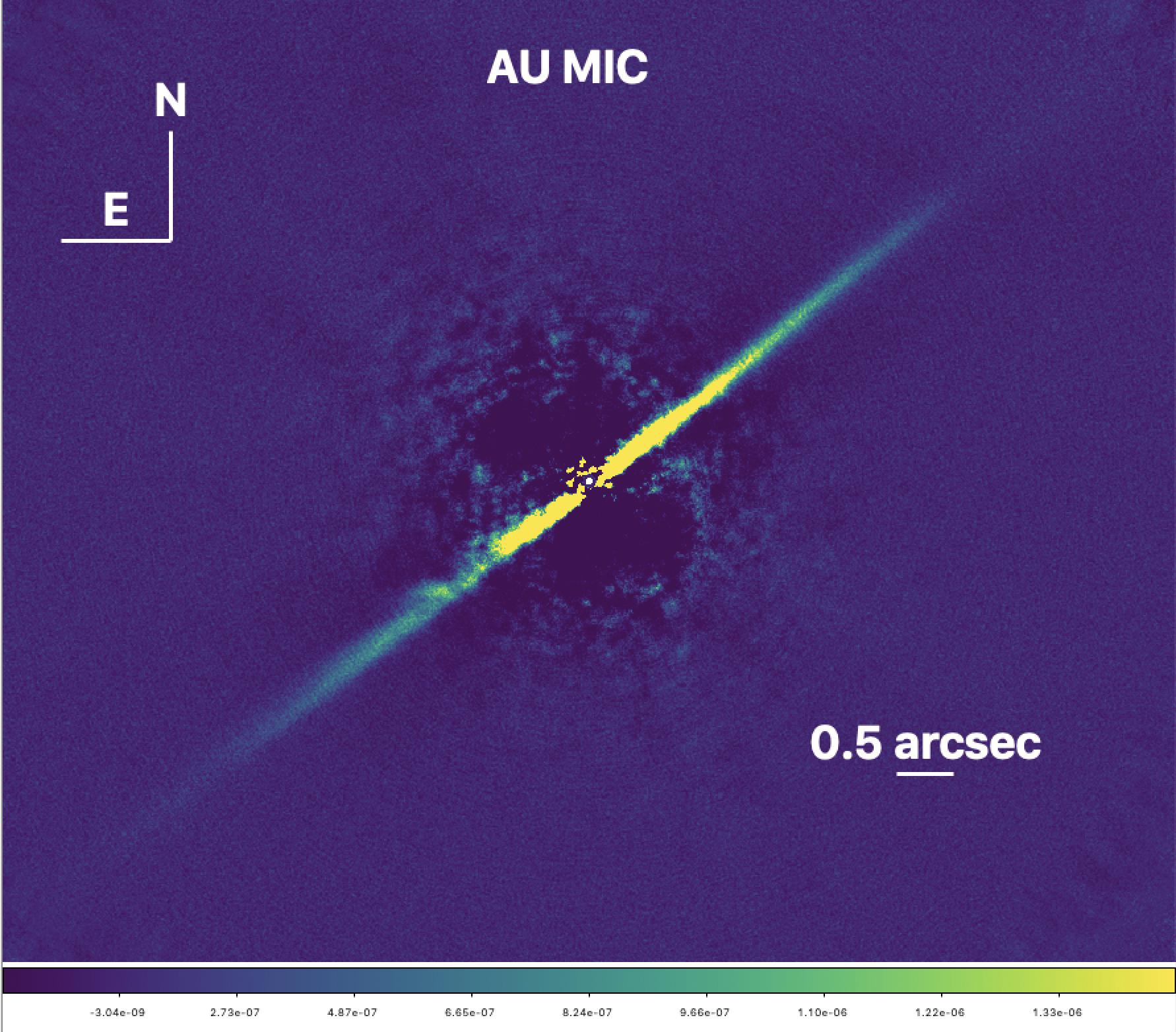} &
		\includegraphics[width=0.66\columnwidth]{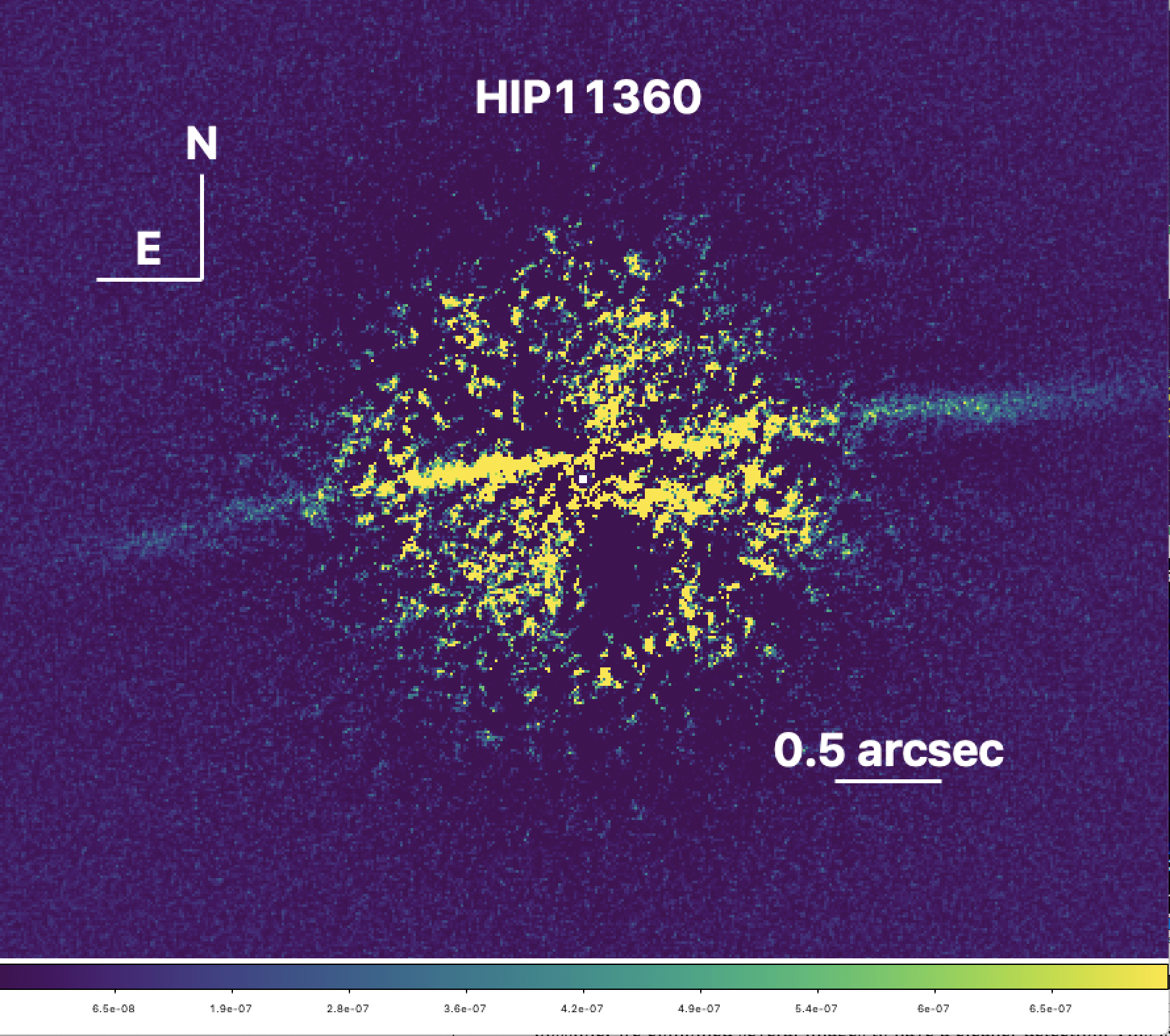} &
		\includegraphics[width=0.66\columnwidth]{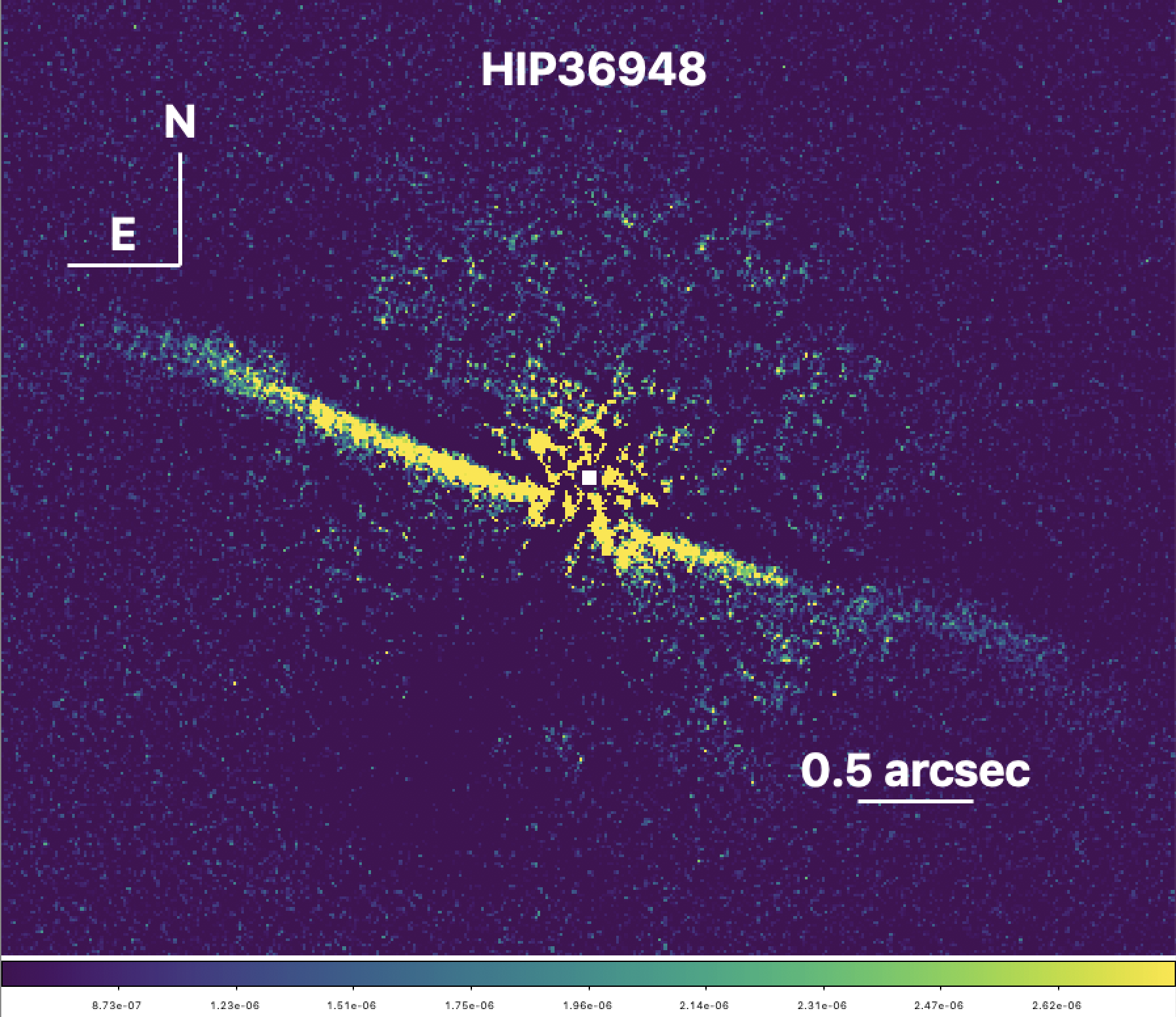} \\
		\includegraphics[width=0.66\columnwidth]{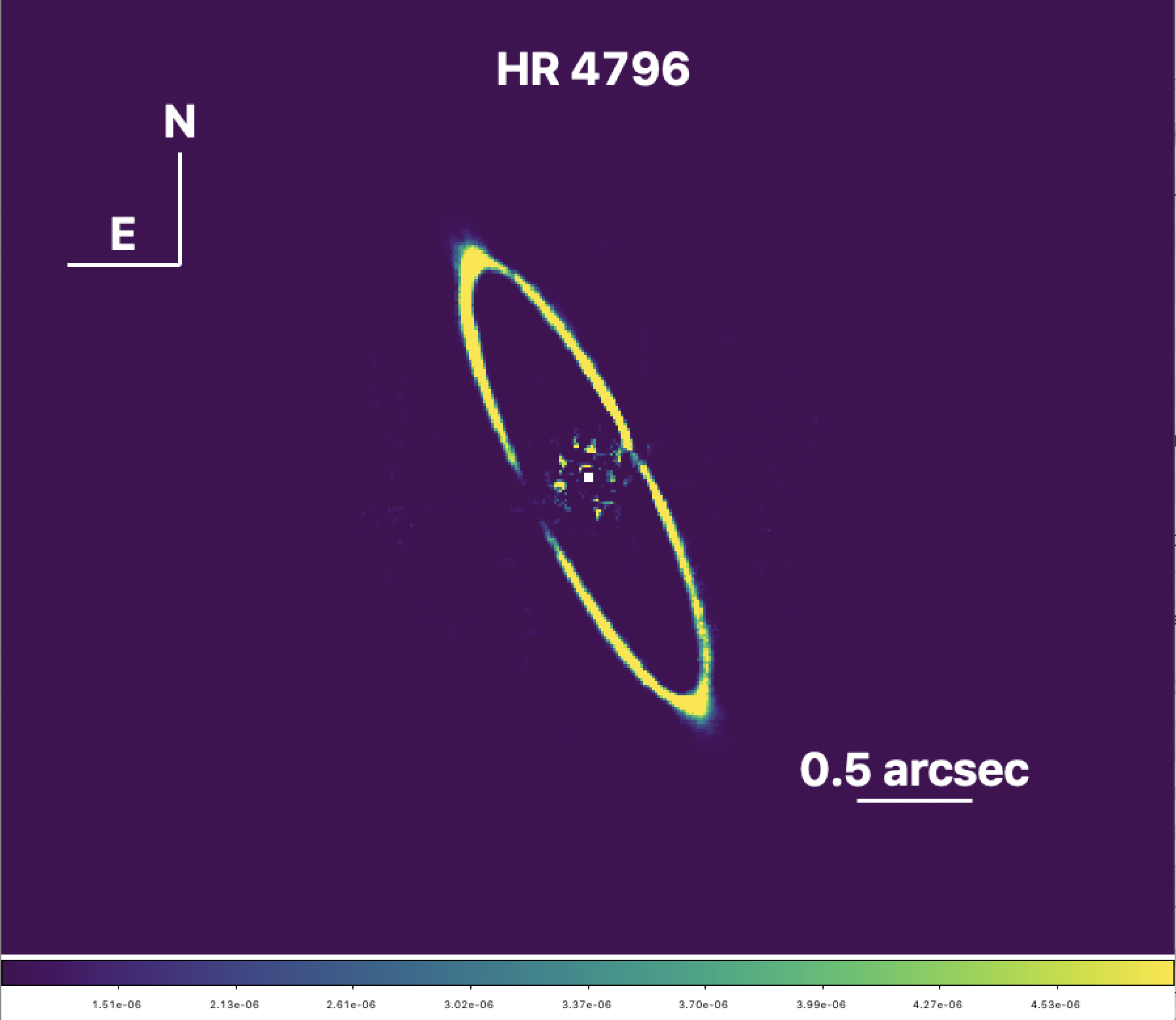} &
		\includegraphics[width=0.66\columnwidth]{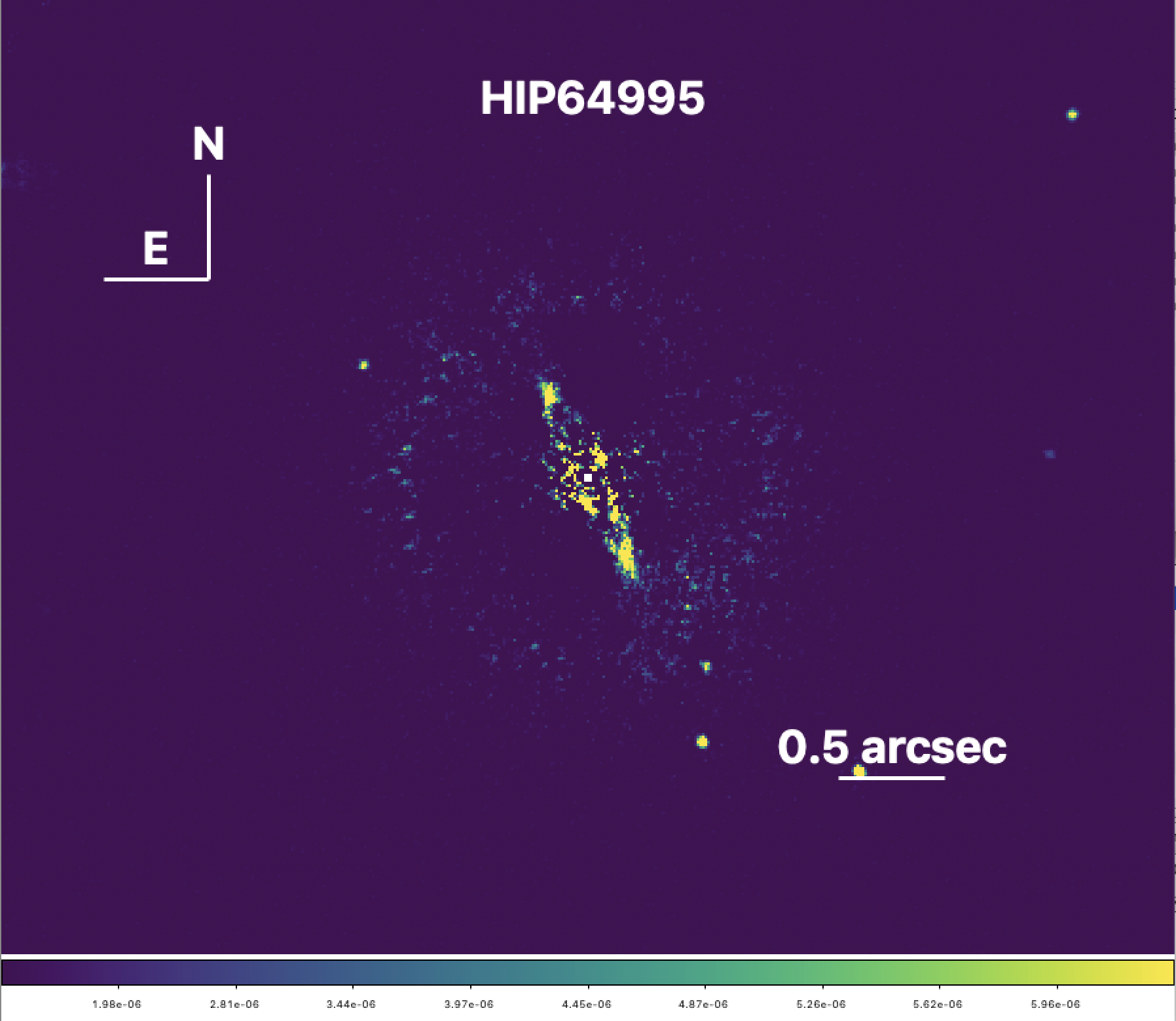} &
		\includegraphics[width=0.66\columnwidth]{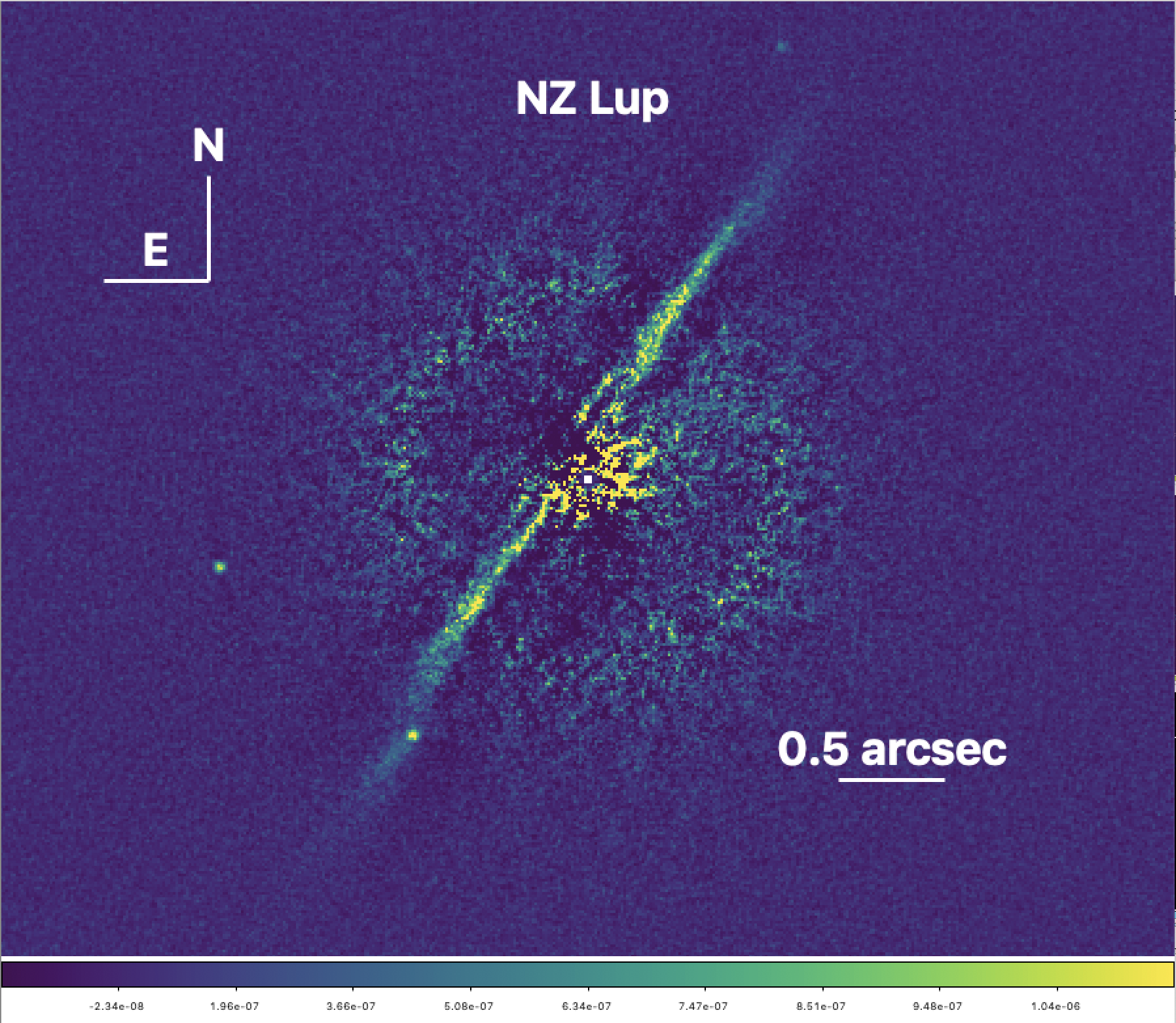}
		\\
	\end{tabular}
	\caption{Gallery of images of debris disks in the F100 sample detected with IRDIS obatained using PCA with 5 modes in H$_2$. Some of detections are marginal they are however better detected with  other signal extraction approaches.}
	\label{fig:Ird_disks}
\end{figure*}

\section{Conclusions and prospects}
\label{sec:conclusions}

We process in a uniform manner more than 300 datasets from the SPHERE/SHINE Survey obtained at the VLT/ESO in visitor mode and assess SHINE survey typical sensitivity as a function of the host star and observing conditions. From these first 150 stars observed out of the planned 400-star survey, we reach typical contrasts of $10^6$  at 1’’  separation and reach up to $10^7$ contrast at 2’’ separation for the best observations. SPHERE/SHINE delivers one of the few largest and deepest direct imaging surveys for exoplanets conducted to date. Compared to GPIES, SHINE has been operating in priority visitor mode, the observing conditions for SHINE in regular visitor mode are highly variable and on average not as good leading to a wider spreading of the contrast performances.

Sixteen substellar companions around twelve host stars have been detected within the first half of the SHINE survey, eight brown dwarfs and eight planetary-mass companions. The paper summarizes the measured position, contrast, and derived mass for each companion as well as it produces astro-photometric data for all the candidates detections. The two new discoveries in the first half of the SHINE survey are the planetary mass companion to HIP 65426 b \citep{Chauvin2017} and the brown dwarf companion to HIP 64892 B \citep{Cheetham2018}. 
All other substellar companions characterized during SHINE were discovered in previous direct imaging campaigns.

The SHINE survey is due to be completed in 2021, but will certainly extend over a few more years to become complete in terms of follow-up for all candidates within at least a 300 au and possibly even farther away. The final sample will include over 500 stars, which will make SHINE the largest high contrast imaging survey to date, covering from B to M stars in the solar neighborhood. The reanalysis of the complete SHINE data with advanced processing techniques \citep{Cantalloube2015,Ruffio2017,Flasseur2018,Flasseur2020b, Flasseur2020,Berdeu2020} will be performed in order to provide improved detection limits and reach the full power of this survey using the full sample.

There are a limited number of bright, young, nearby stars available to high-contrast direct imagers like SPHERE and GPI. As a result, at the current achievable contrasts we cannot expect a significant number of new detections of imaged giant planets following the completion of both GPIES and SPHERE/SHINE (\cite{Desidera2020}) campaigns. An improvement in performance from instrument upgrades (\cite{Boccaletti2020}, \cite{Chilcote2018} ), however, could unlock mass/separation phase space around these target stars that are currently inaccessible, which are in addition likely to host more planets \cite{Wagner2019}. Additional planets discovered with upgraded GPI and SPHERE would allow us to test the robustness of the trends with stellar mass and planetary mass uncovered in this paper. The Gaia mission will identify astrometric signatures of planets that may be confirmed with direct imaging, which can also boost the number of companions in this separation range. Finally, cross checks with results from other indirect techniques, e.g. transits or high precision radial velocity, will allow a much better picture of the architecture of planetary systems.

\begin{acknowledgements}
SPHERE is an instrument designed and built by a consortium consisting of IPAG (Grenoble, France), MPIA (Heidelberg, Germany), LAM (Marseille, France), LESIA (Paris, France), Laboratoire Lagrange (Nice, France), INAF - Osservatorio di Padova (Italy), Observatoire de Gen\`eve (Switzerland), ETH Z\"urich (Switzerland), NOVA (Netherlands), ONERA (France) and ASTRON (Netherlands) in collaboration with ESO. SPHERE was funded by ESO, with additional contributions from CNRS (France), MPIA (Germany), INAF (Italy), FINES (Switzerland) and NOVA (Netherlands). SPHERE also received funding from the European Commission Sixth and Seventh Framework Programmes as part of the Optical Infrared Coordination Network for Astronomy (OPTICON) under grant number RII3-Ct-2004-001566 for FP6 (2004-2008), grant number 226604 for FP7 (2009-2012) and grant number 312430 for FP7 (2013-2016). This work has made use of the SPHERE Data Centre, jointly operated by OSUG/IPAG (Grenoble), PYTHEAS/LAM/CeSAM (Marseille), OCA/Lagrange (Nice), Observatoire de Paris/LESIA (Paris), and Observatoire de Lyon (OSUL/CRAL).This work is supported by the French National Research Agency in the framework of the Investissements d’Avenir program (ANR-15-IDEX-02), through the funding of the "Origin of Life" project of the Univ. Grenoble-Alpes. This work is jointly supported by the French National Programms (PNP and PNPS) and by the Action Spécifique Haute Résolution Angulaire (ASHRA) of CNRS/INSU co-funded by CNES. We also thank the anonymous referee for her/his careful reading of the manuscript as well as her/his insightful comments and suggestions. AV acknowledges funding from the European Research Council (ERC) under the European Union's Horizon 2020 research and innovation programme (grant agreement No. 757561). A-M Lagrange acknowledges funding from French National Research Agency (GIPSE project). C. P. acknowledge financial support from Fondecyt (grant 3190691) and financial support from the ICM (Iniciativa Cient\'ifica Milenio) via the N\'ucleo Milenio  de  Formaci\'on Planetaria grant, from the Universidad de Valpara\'iso. T.H. acknowledges support from the European Research Council under the Horizon 2020 Framework Program via the ERC Advanced Grant Origins 83 24 28.
\end{acknowledgements}

\bibliographystyle{aa}
\bibliography{paper}

\begin{appendix}

\section{Detection Astrophotometric Table}
\label{sec:astrophot}

\tabcolsep=0.081cm

\onecolumn 

\begin{landscape}
	\tabcolsep=0.051cm
	\begin{longtable}{llllllllllllllllllllll}
		\caption{Detected point sources parameters (see electronic version for the full table)}
		\tiny
		\label{tab:detections_all}\\
		Main\_ID & date & filt & ncc & id & sep(mas) & sep & pa(deg) & pa & dm0 & dm0 & dm1 & dm1 & absm0 & absm0 & absm1 & absm1 & snr0 & snr1 & proba & color & status \\
		&  &  &  & &  & error & & error & & error & &error & &error& &error & & &  & & \\
		\endfirsthead
		%
		Main\_ID & date & filt & ncc & id & sep(mas) & sep & pa(deg) & pa & dm0 & dm0 & dm1 & dm1 & absm0 & absm0 & absm1 & absm1 & snr0 & snr1 & proba & color & status \\
		
		& & & & & & error & &error& & error & &error &  & error & & error & & &  & &  \\
		\endhead
		HIP53524 & 05/05/15 & K12 & 11 & 0.0 & 618.58 & 9.46 & 148.66 & 1.07 & 12.37 & 0.16 & 11.91 & 0.27 & 14.48 & 0.16 & 14.01 & 0.27 & 9.17 & 4.28 & 1.48 & 0.466 & C \\ \hline
		HIP53524 & 05/05/15 & K12 & 11 & 1.0 & 2212.95 & 17.41 & 35.14 & 0.41 & 13.02 & 0.12 & 13.13 & 0.21 & 15.13 & 0.13 & 15.24 & 0.21 & 17.28 & 5.81 & 22.24 & -0.111 & B \\ \hline
		HIP53524 & 05/05/15 & K12 & 11 & 2.0 & 2331.48 & 39.49 & 305.16 & 0.8 & 0 & 1.09 & 13.64 & 0.34 & 2.11 & 1.09 & 15.75 & 0.34 & 0 & 3.37 & 29.74 & -13.641 & B \\ \hline
		HIP53524 & 05/05/15 & K12 & 11 & 3.0 & 4064.85 & 21.13 & 140.74 & 0.1 & 11.87 & 0.11 & 11.79 & 0.11 & 13.98 & 0.11 & 13.89 & 0.12 & 40.09 & 17.63 & 39.61 & 0.086 & B \\ \hline
		HIP53524 & 05/05/15 & K12 & 11 & 4.0 & 4432.96 & 22.47 & 320.19 & 0.07 & 6.35 & 0.11 & 6.26 & 0.1 & 8.45 & 0.11 & 8.37 & 0.1 & 88.62 & 105.39 & 1.42 & 0.084 & B \\ \hline
		HIP53524 & 05/05/15 & K12 & 11 & 5.0 & 4669.81 & 23.82 & 246.71 & 0.07 & 11.21 & 0.11 & 11.06 & 0.1 & 13.31 & 0.11 & 13.17 & 0.11 & 43.92 & 29.37 & 37.46 & 0.143 & B \\ \hline
		HIP53524 & 05/05/15 & K12 & 11 & 6.0 & 5024.38 & 28.76 & 244.75 & 0.13 & 13.17 & 0.12 & 13.16 & 0.32 & 15.27 & 0.12 & 15.27 & 0.32 & 18.23 & 3.62 & 74.52 & 0.005 & B \\ \hline
		HIP53524 & 05/05/15 & K12 & 11 & 7.0 & 5204.32 & 27.25 & 238.69 & 0.09 & 13.62 & 0.13 & 0 & 1.09 & 15.73 & 0.13 & 2.11 & 1.09 & 16.12 & 0 & 81.89 & 13.619 & B \\ \hline
		HIP53524 & 05/05/15 & K12 & 11 & 8.0 & 5585.85 & 28.78 & 322.51 & 0.09 & 12.02 & 0.11 & 11.97 & 0.15 & 14.12 & 0.11 & 14.08 & 0.15 & 31.09 & 9.21 & 63.66 & 0.048 & B \\ \hline
		HIP53524 & 05/05/15 & K12 & 11 & 9.0 & 5817.68 & 29.73 & 324 & 0.08 & 11.63 & 0.11 & 11.51 & 0.11 & 13.73 & 0.11 & 13.62 & 0.11 & 35.02 & 18.32 & 59.45 & 0.114 & B \\ \hline
		HIP53524 & 05/05/15 & K12 & 11 & 10.0 & 6192.48 & 31.54 & 105.04 & 0.07 & 11.29 & 0.11 & 11.24 & 0.11 & 13.39 & 0.11 & 13.34 & 0.11 & 33.59 & 22.64 & 57.65 & 0.047 & B \\ \hline
		HIP53524 & 28/03/18 & K12 & 11 & 0.0 & 631.9 & 5.65 & 145.52 & 0.63 & 12.27 & 0.08 & 11.83 & 0.14 & 14.37 & 0.08 & 13.93 & 0.14 & 19.91 & 8.82 & 1.46 & 0.439 & C \\ \hline
		HIP53524 & 28/03/18 & K12 & 11 & 1.0 & 2248.47 & 8.96 & 38.04 & 0.26 & 13.12 & 0.09 & 13.03 & 0.24 & 15.23 & 0.1 & 15.14 & 0.24 & 14 & 4.76 & 23.68 & 0.095 & B \\ \hline
		HIP53524 & 28/03/18 & K12 & 11 & 2.0 & 4120.91 & 24.28 & 175.67 & 3.28 & 14.16 & 0.13 & 14.2 & 0.39 & 16.27 & 0.13 & 16.31 & 0.39 & 8.91 & 2.83 & 73.05 & -0.039 & B \\ \hline
		HIP53524 & 28/03/18 & K12 & 11 & 3.0 & 4157.41 & 6.8 & 140.03 & 0.09 & 12.07 & 0.06 & 11.94 & 0.09 & 14.17 & 0.07 & 14.05 & 0.09 & 30.94 & 14.86 & 43.89 & 0.126 & B \\ \hline
		HIP53524 & 28/03/18 & K12 & 11 & 4.0 & 4345.53 & 6.01 & 320.9 & 0.07 & 6.25 & 0.06 & 6.18 & 0.06 & 8.35 & 0.06 & 8.28 & 0.06 & 38.34 & 52.26 & 1.23 & 0.071 & B \\ \hline
		HIP53524 & 28/03/18 & K12 & 11 & 5.0 & 4608.14 &  & 245.48 &  & 0 & 1.09 & 0 & 1.09 & 2.11 & 1.09 & 2.11 & 1.09 & 0 & 0 & 0.01 & 0 & B \\ \hline
		HIP53524 & 28/03/18 & K12 & 11 & 6.0 & 4953.55 & 12.75 & 243.92 & 0.11 & 13.15 & 0.07 & 13.3 & 0.29 & 15.25 & 0.08 & 15.41 & 0.29 & 20.6 & 3.84 & 73.16 & -0.157 & B \\ \hline
		HIP53524 & 28/03/18 & K12 & 11 & 7.0 & 5135.98 & 15.49 & 237.8 & 0.14 & 13.5 & 0.09 & 13.77 & 0.56 & 15.61 & 0.1 & 15.87 & 0.56 & 13.91 & 1.96 & 79.82 & -0.266 & B \\ \hline
		HIP53524 & 28/03/18 & K12 & 11 & 8.0 & 5501.79 & 8.1 & 323.15 & 0.08 & 11.89 & 0.06 & 11.84 & 0.08 & 14 & 0.07 & 13.95 & 0.08 & 31.18 & 22 & 60.31 & 0.054 & B \\ \hline
		HIP53524 & 28/03/18 & K12 & 11 & 9.0 & 5774.7 &  & 325.03 &  & 0 & 1.09 & 0 & 1.09 & 2.11 & 1.09 & 2.11 & 1.09 & 0 & 0 & 0.02 & 0 & B \\ \hline
		HIP53524 & 28/03/18 & K12 & 11 & 10.0 & 6292.78 & 9.14 & 105.18 & 0.07 & 11.48 & 0.08 & 11.42 & 0.08 & 13.58 & 0.08 & 13.53 & 0.08 & 18.98 & 20.73 & 62.31 & 0.053 & B \\ \hline
		HIP53524 & 18/05/19 & K12 & 9 & 0.0 & 628.43 & 4.76 & 144.11 & 0.53 & 12.21 & 0.14 & 11.72 & 0.13 & 14.32 & 0.14 & 13.83 & 0.13 & 27.33 & 17.6 & 1.4 & 0.492 & C \\ \hline
		HIP53524 & 18/05/19 & K12 & 9 & 1.0 & 2268.61 & 9.16 & 39.36 & 0.26 & 13.08 & 0.14 & 13.29 & 0.23 & 15.19 & 0.14 & 15.4 & 0.23 & 22.04 & 5.25 & 23.72 & -0.21 & B \\ \hline
		HIP53524 & 18/05/19 & K12 & 9 & 2.0 & 4201.34 & 3.81 & 139.74 & 0.08 & 11.97 & 0.14 & 11.82 & 0.12 & 14.08 & 0.14 & 13.92 & 0.12 & 51.51 & 26.64 & 43.16 & 0.155 & B \\ \hline
		HIP53524 & 18/05/19 & K12 & 9 & 3.0 & 4305.8 & 2.67 & 321.2 & 0.07 & 6.36 & 0.14 & 6.28 & 0.11 & 8.47 & 0.14 & 8.38 & 0.11 & 51.96 & 80.05 & 1.36 & 0.088 & B \\ \hline
		HIP53524 & 18/05/19 & K12 & 9 & 4.0 & 4562.91 & 3.67 & 245.38 & 0.07 & 11.28 & 0.14 & 11.14 & 0.12 & 13.39 & 0.14 & 13.25 & 0.12 & 40.38 & 33.68 & 37.36 & 0.143 & B \\ \hline
		HIP53524 & 18/05/19 & K12 & 9 & 5.0 & 4922.52 & 10.86 & 243.49 & 0.12 & 13.24 & 0.15 & 13.27 & 0.2 & 15.34 & 0.15 & 15.38 & 0.21 & 19.19 & 6.33 & 74.05 & -0.036 & B \\ \hline
		HIP53524 & 18/05/19 & K12 & 9 & 6.0 & 5462.93 & 4.69 & 323.43 & 0.08 & 12.01 & 0.14 & 11.96 & 0.12 & 14.12 & 0.14 & 14.07 & 0.12 & 34.68 & 21.48 & 61.93 & 0.052 & B \\ \hline
		HIP53524 & 18/05/19 & K12 & 9 & 7.0 & 5697.63 & 4.18 & 324.9 & 0.08 & 11.61 & 0.14 & 11.53 & 0.12 & 13.71 & 0.14 & 13.63 & 0.12 & 35.06 & 28.29 & 57.6 & 0.08 & B \\ \hline
		HIP53524 & 18/05/19 & K12 & 9 & 8.0 & 6339.33 & 4.79 & 105.24 & 0.07 & 11.43 & 0.14 & 11.35 & 0.12 & 13.53 & 0.15 & 13.45 & 0.12 & 20.34 & 24.84 & 61.93 & 0.082 & B \\ \hline
	\end{longtable}
\end{landscape}
\twocolumn	
\section{Contrast PLots}
\begin{figure*}
	\begin{tabular}{cc}
		\centering
		IRDIS RAW CONTRASTS & IRDIS PROCESSED CONTRASTS\\
		\includegraphics[width=1.05\columnwidth]{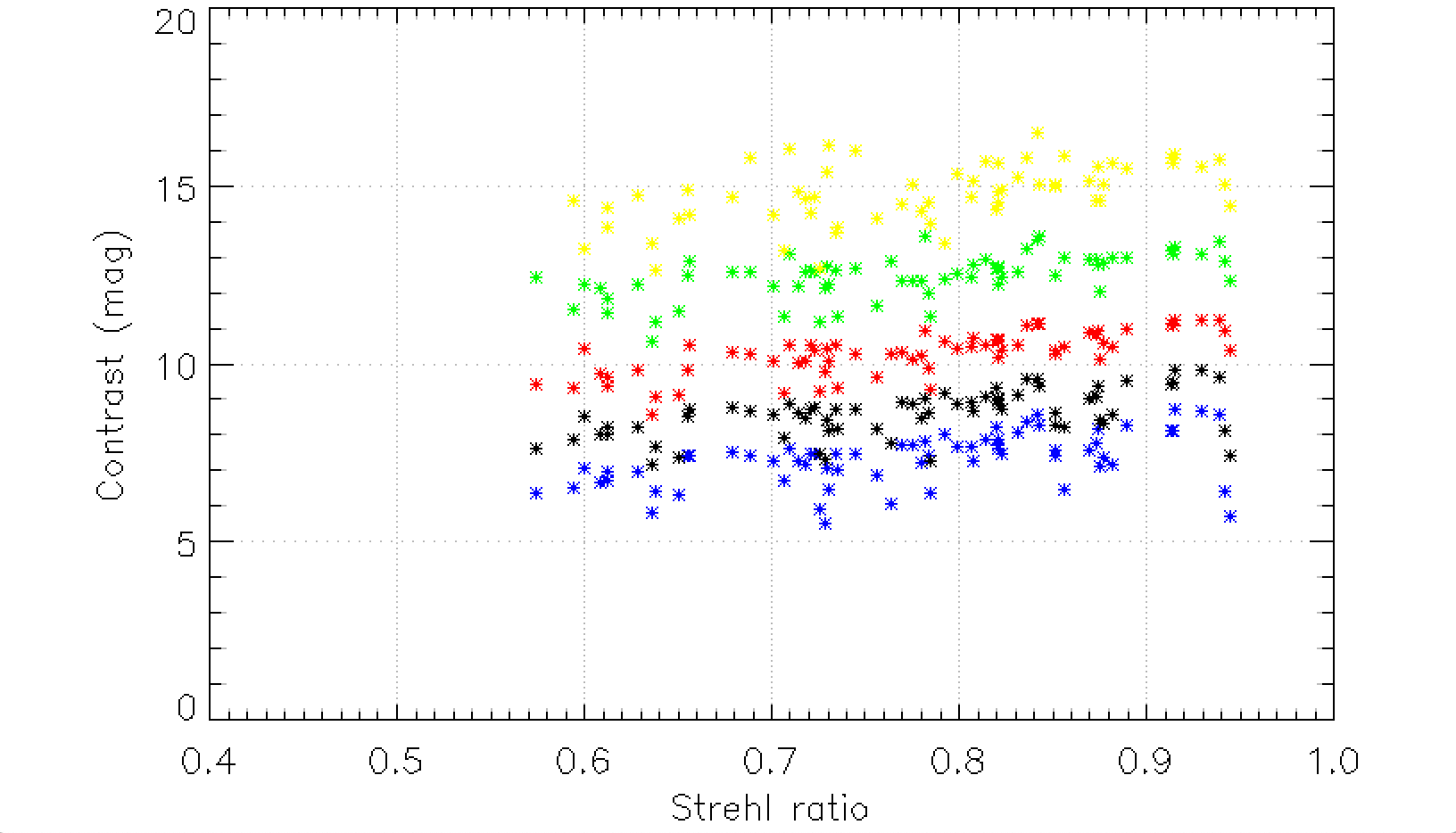}&
		\includegraphics[width=1.05\columnwidth]{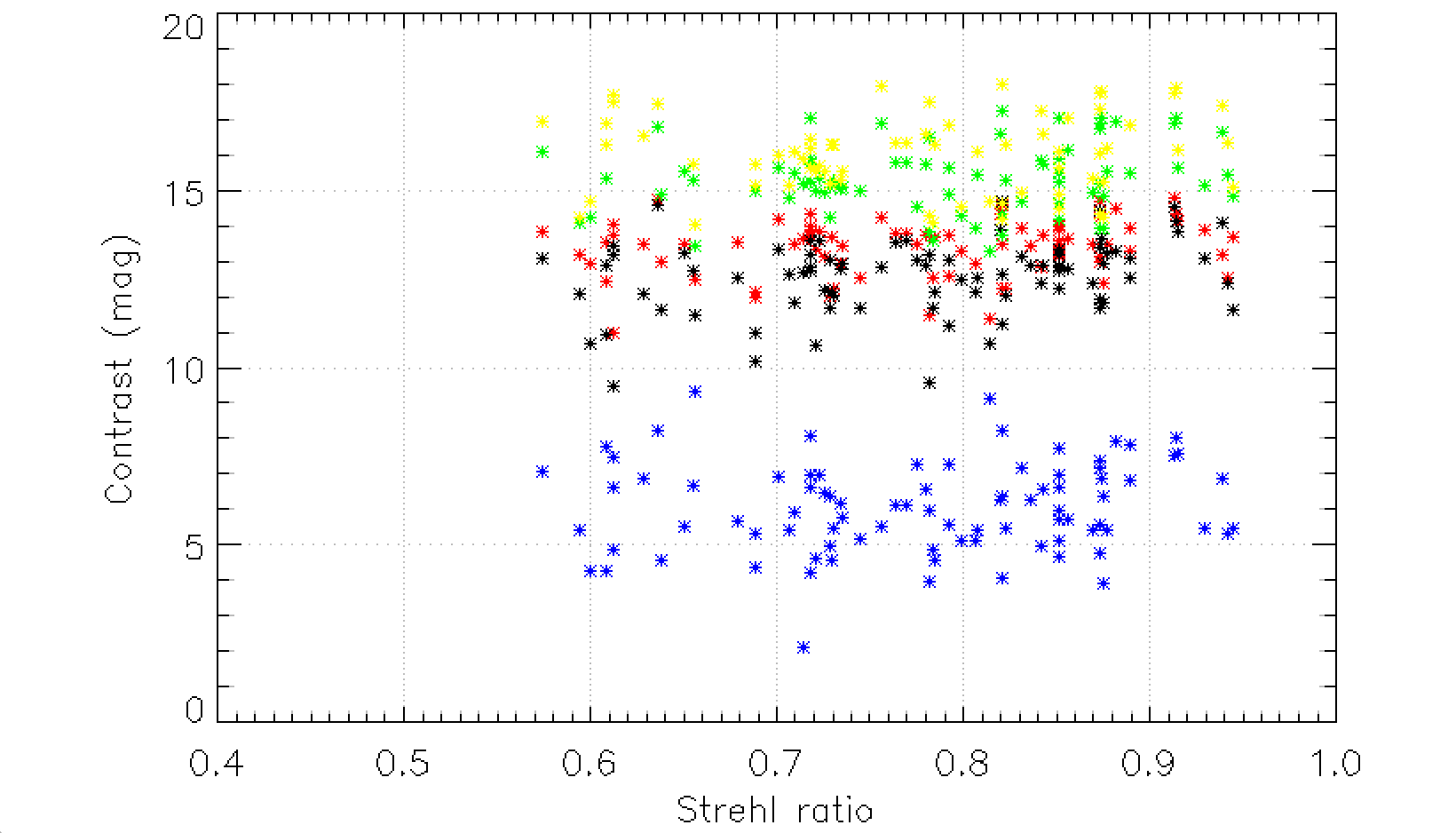}\\
		\includegraphics[width=1.05\columnwidth]{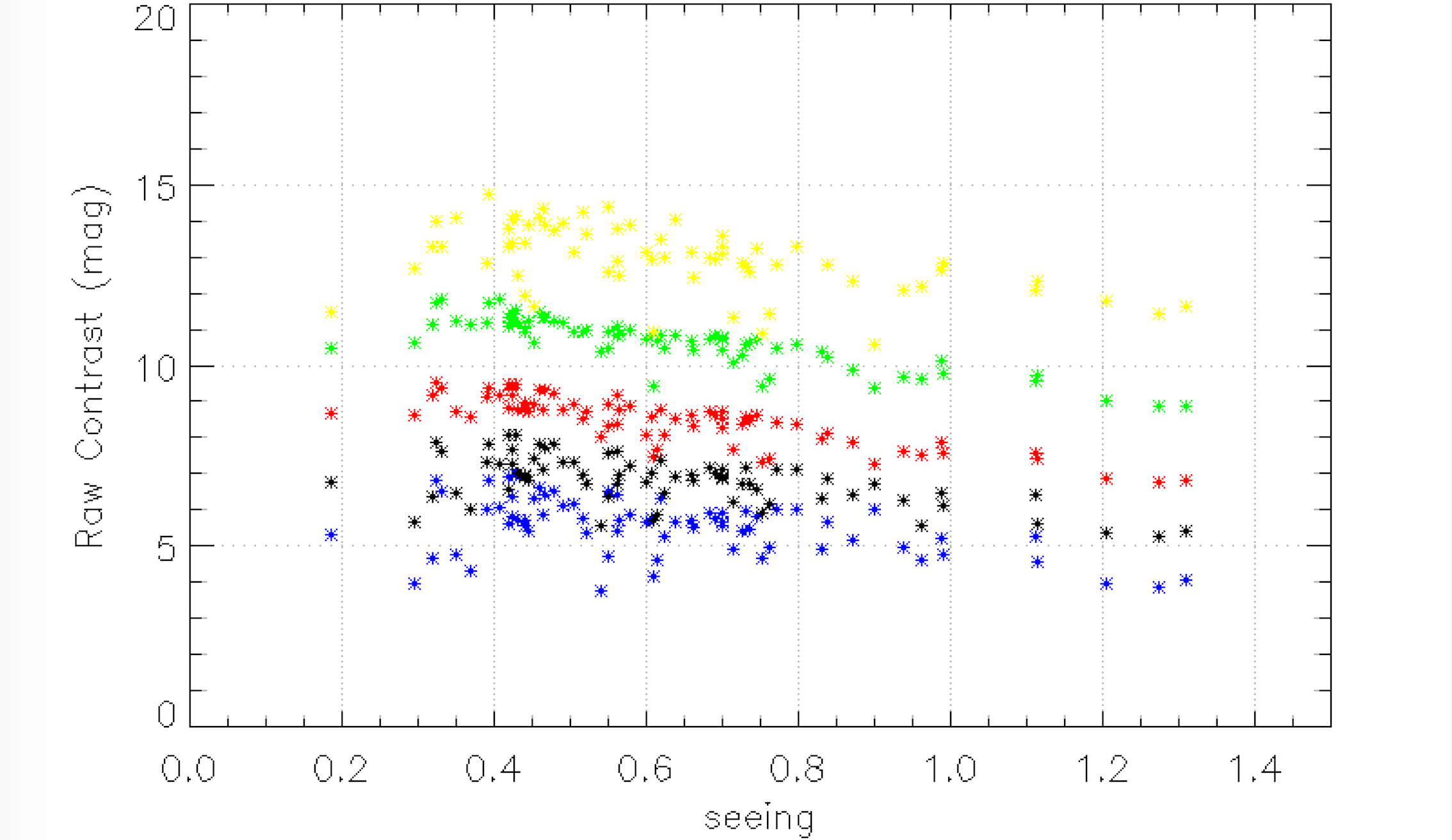}&
		\includegraphics[width=1.05\columnwidth]{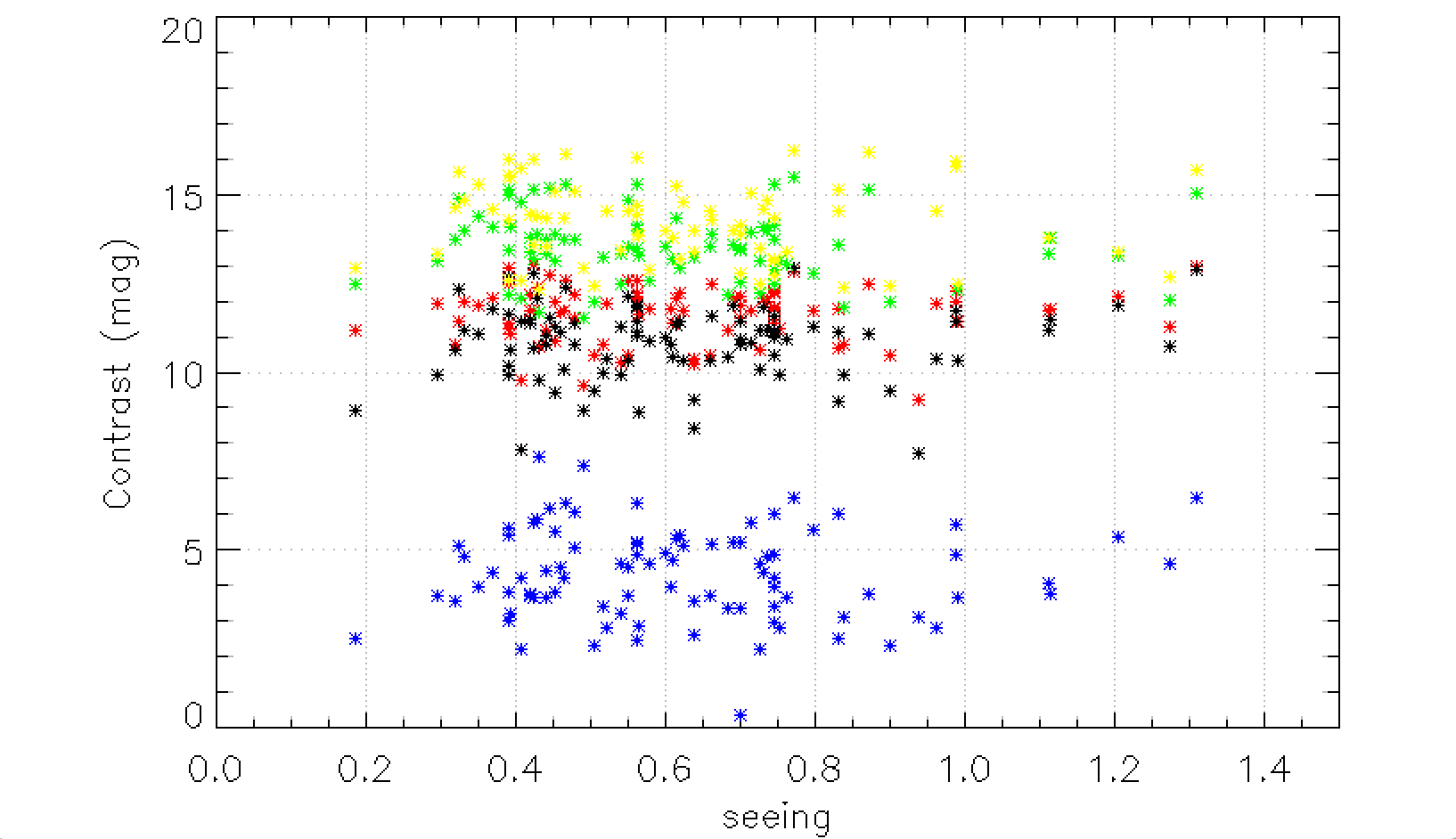}\\
		\includegraphics[width=1.05\columnwidth]{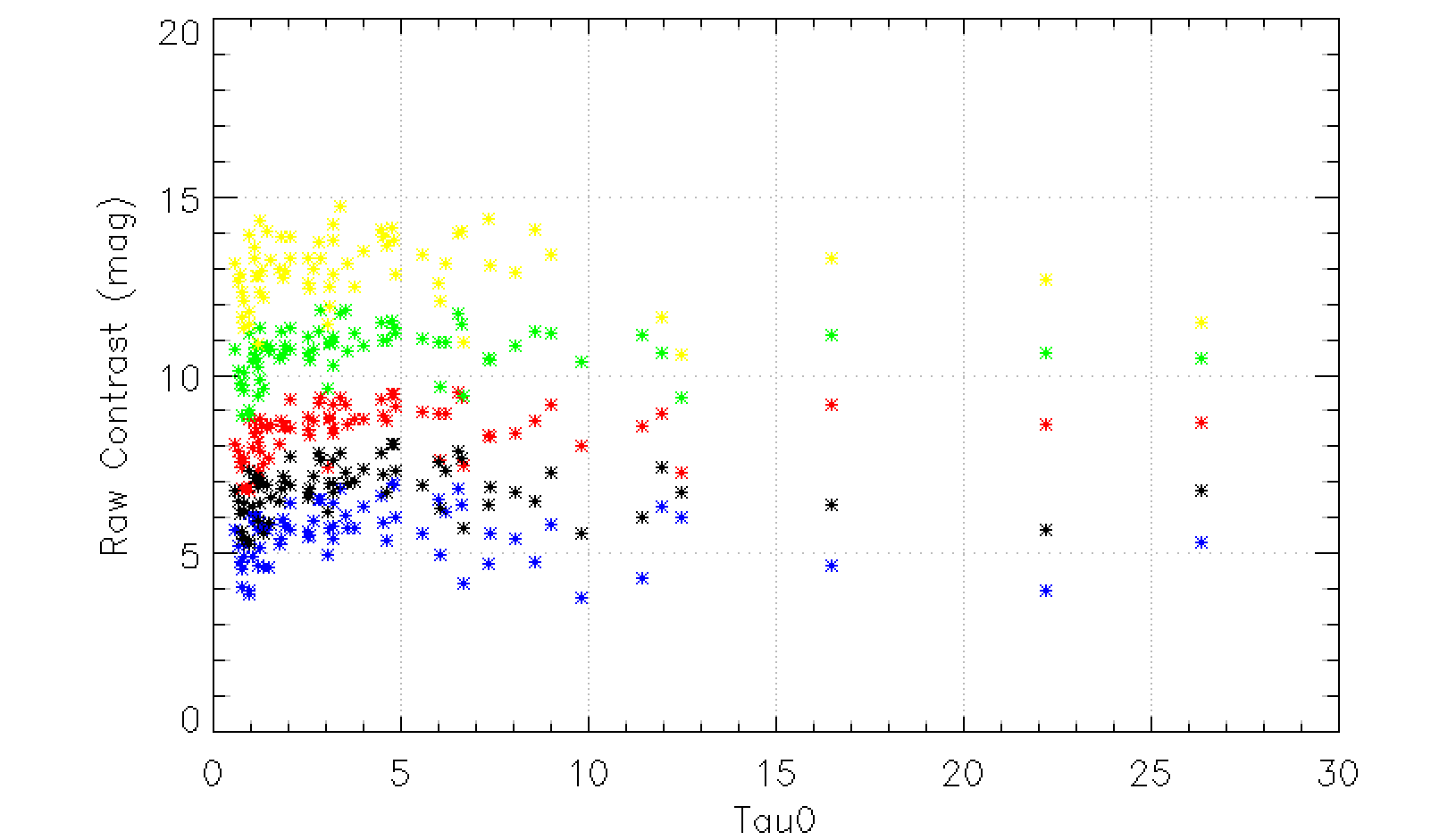}&
		\includegraphics[width=1.05\columnwidth]{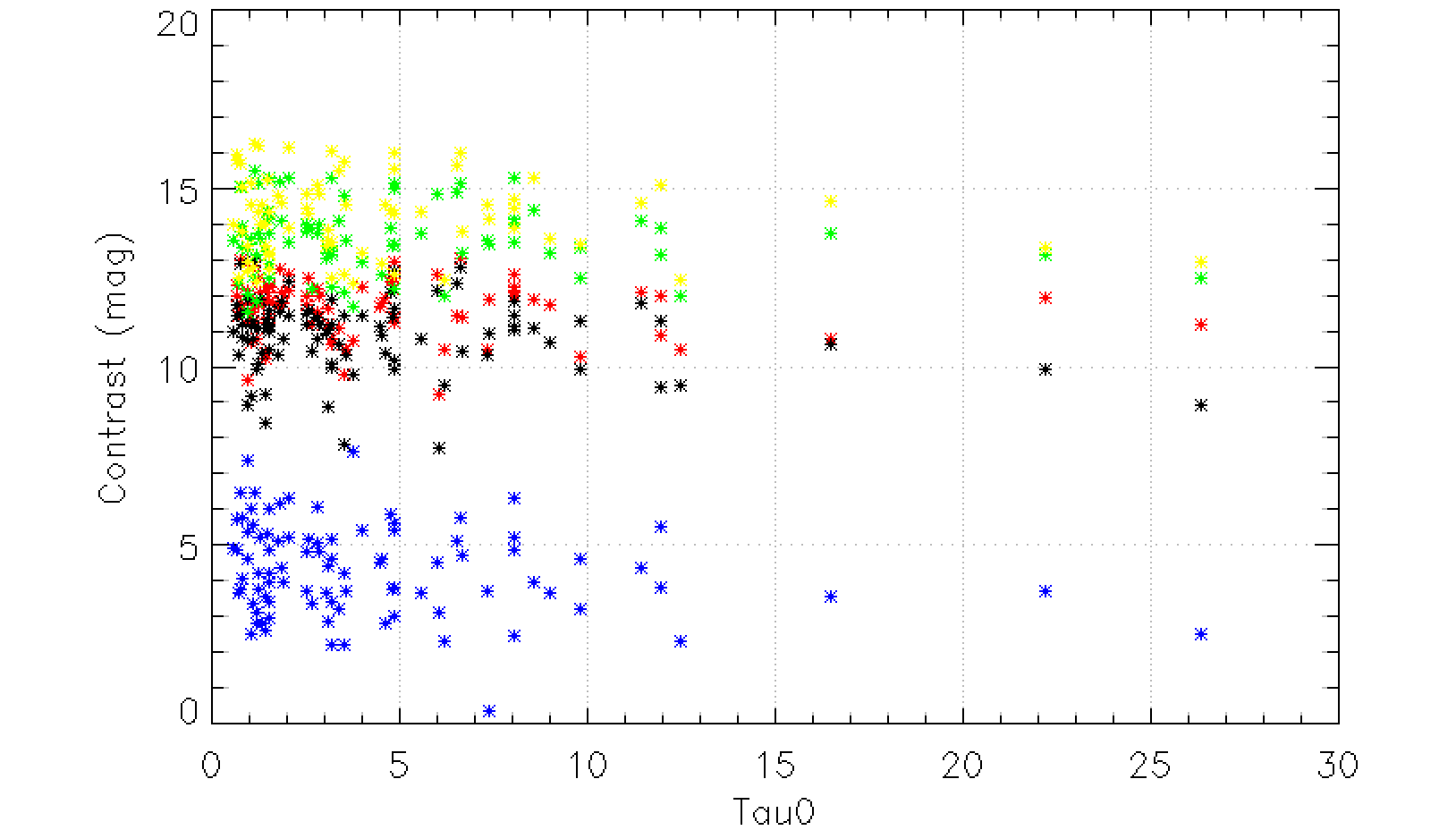}\\
	\end{tabular}
	\caption{IRDIS raw (Left) and processed (Right) contrasts in H2 band at various separations computed using TLOCI estimated as function of the observing conditions in irdifs mode. Blue is for 100 mas separation, black for 500 mas separation, red for 1000 mas separation, green for 2000 mas separation and yellow for 4000 mas separations. The seeing is expressed in arcseconds and the $\tau_0$ in milliseconds.}
	\label{fig:cont_sparta_IRDIS}
\end{figure*}
\begin{figure*}
	\begin{tabular}{cc}
		\centering
		\includegraphics[width=1.05\columnwidth]{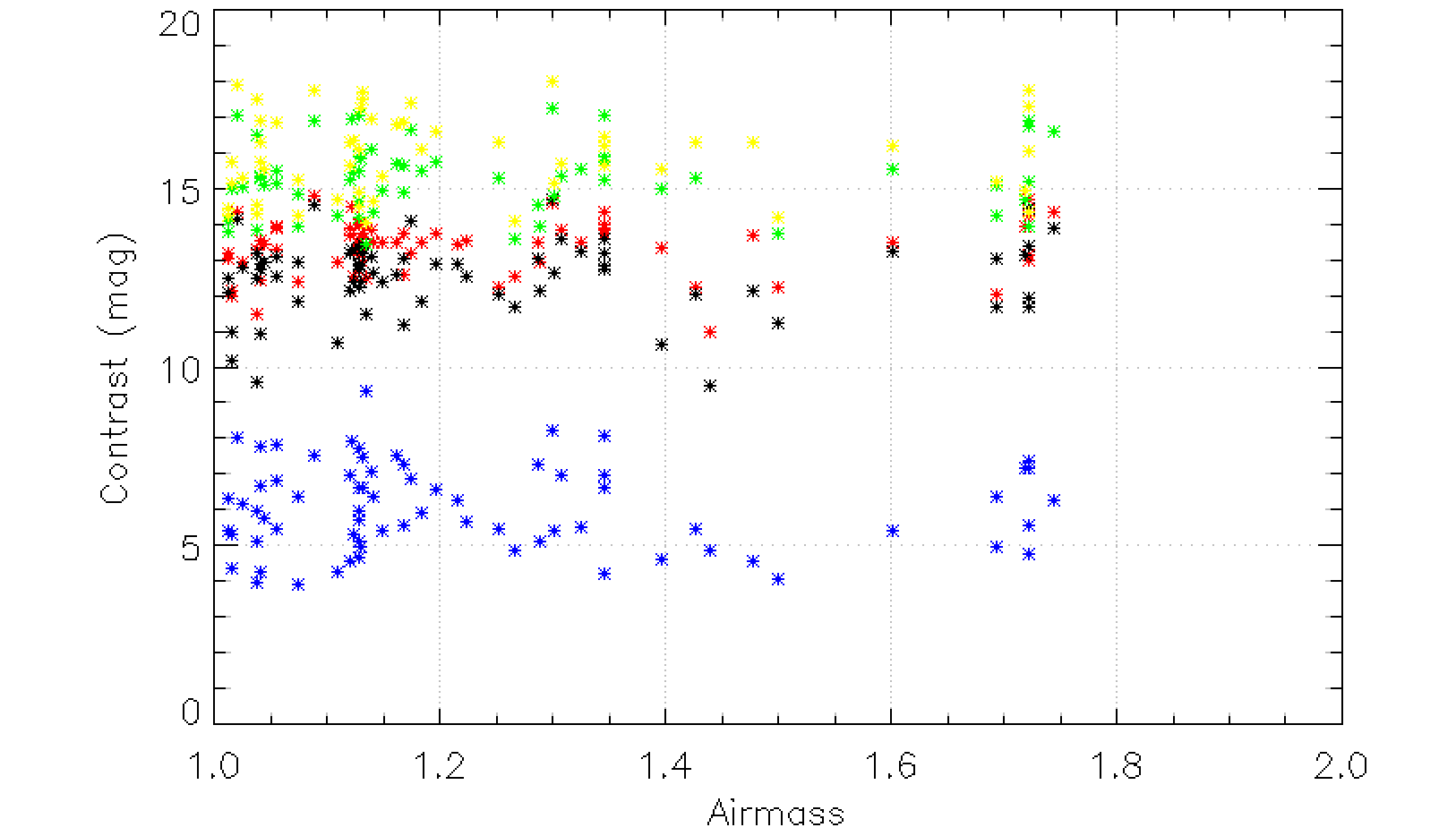} &
		\includegraphics[width=1.05\columnwidth]{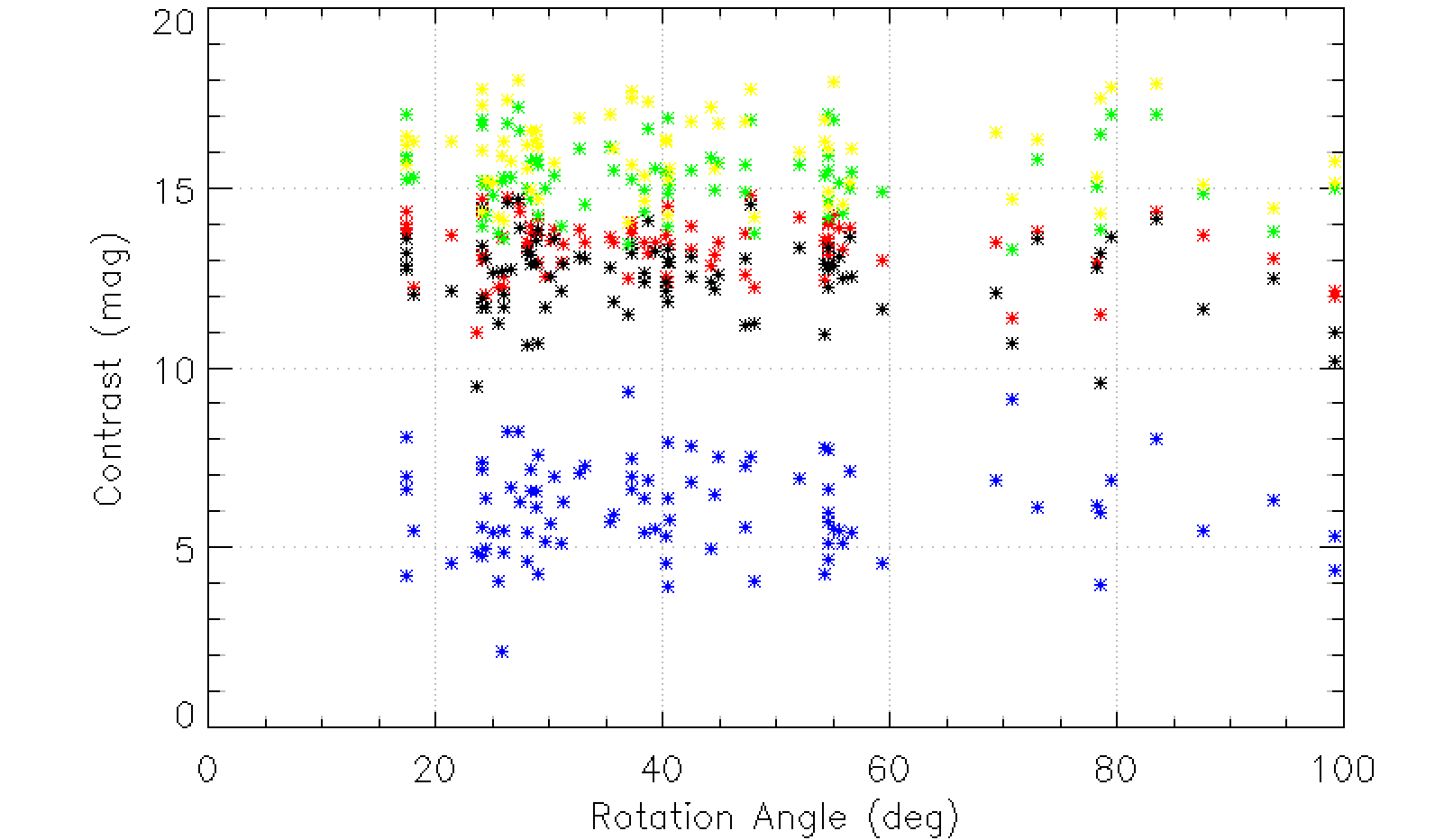}//
	\end{tabular}
	\caption{5sigma IRDIS processed contrasts at various separations computed using TLOCI estimated as function of the observing conditions (airmass and rotation angle) in irdifs mode. Blue is for 100 mas separation, black for 500 mas separation, red for 1000 mas separation , green for 2000 mas separation and yellow for 4000 mas separations.}
	\label{fig:cont_sparta_IRDISb}
\end{figure*}

\begin{figure*}
	\centering
	\begin{tabular}{ccc}
		\includegraphics[width=0.6\columnwidth]{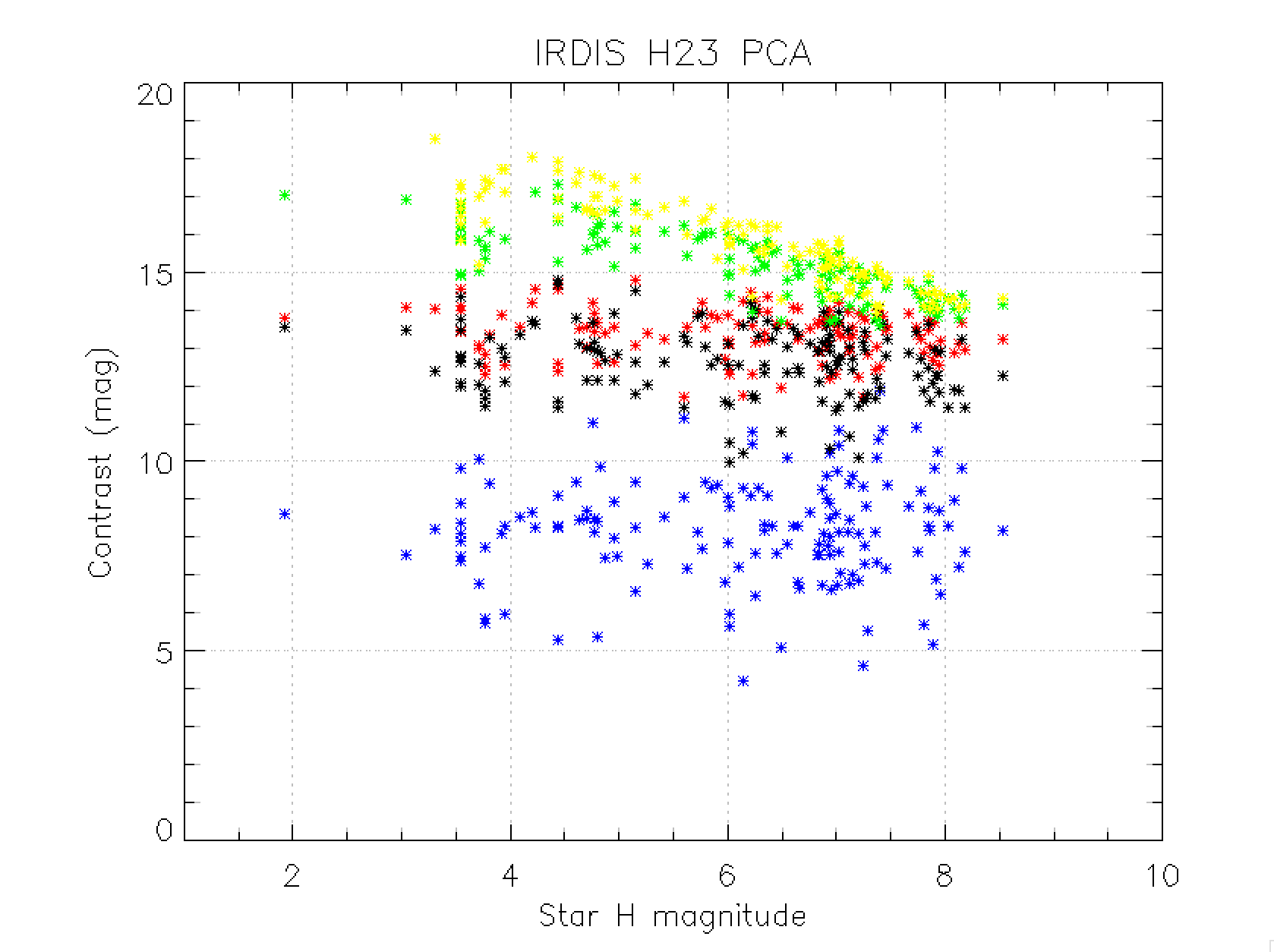} &
		\includegraphics[width=0.6\columnwidth]{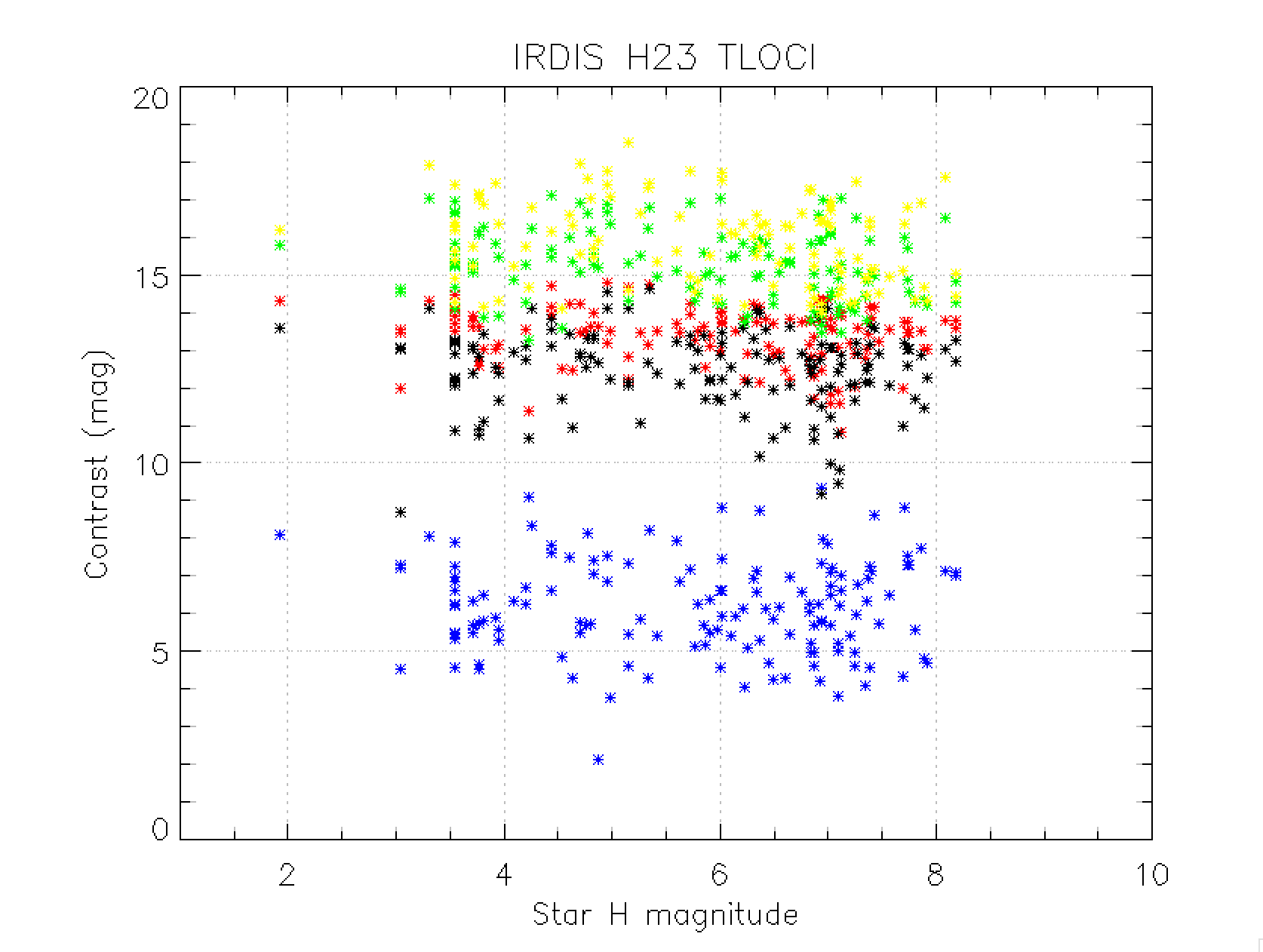} & \includegraphics[width=0.61\columnwidth]{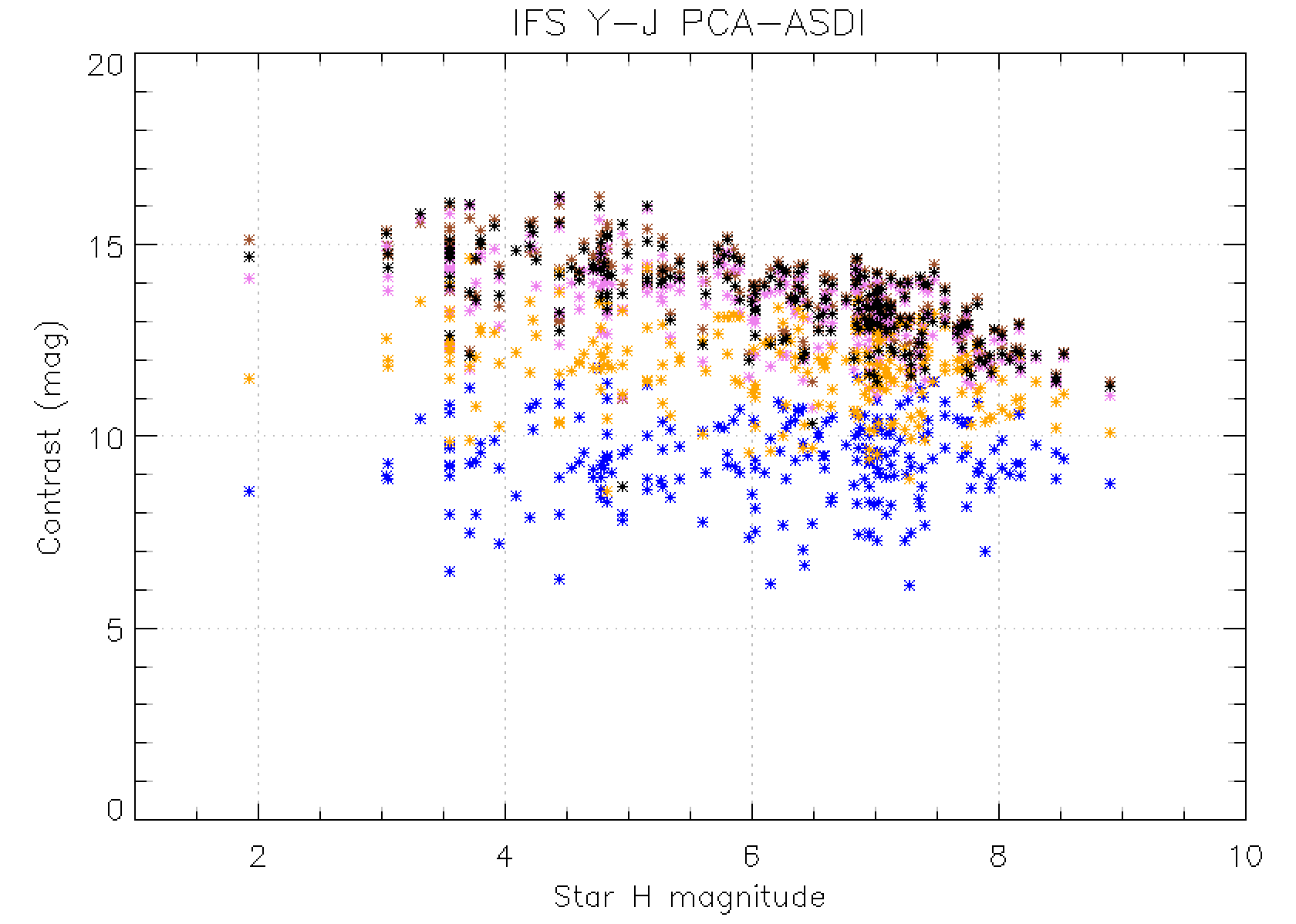}\\
		\includegraphics[width=0.6\columnwidth]{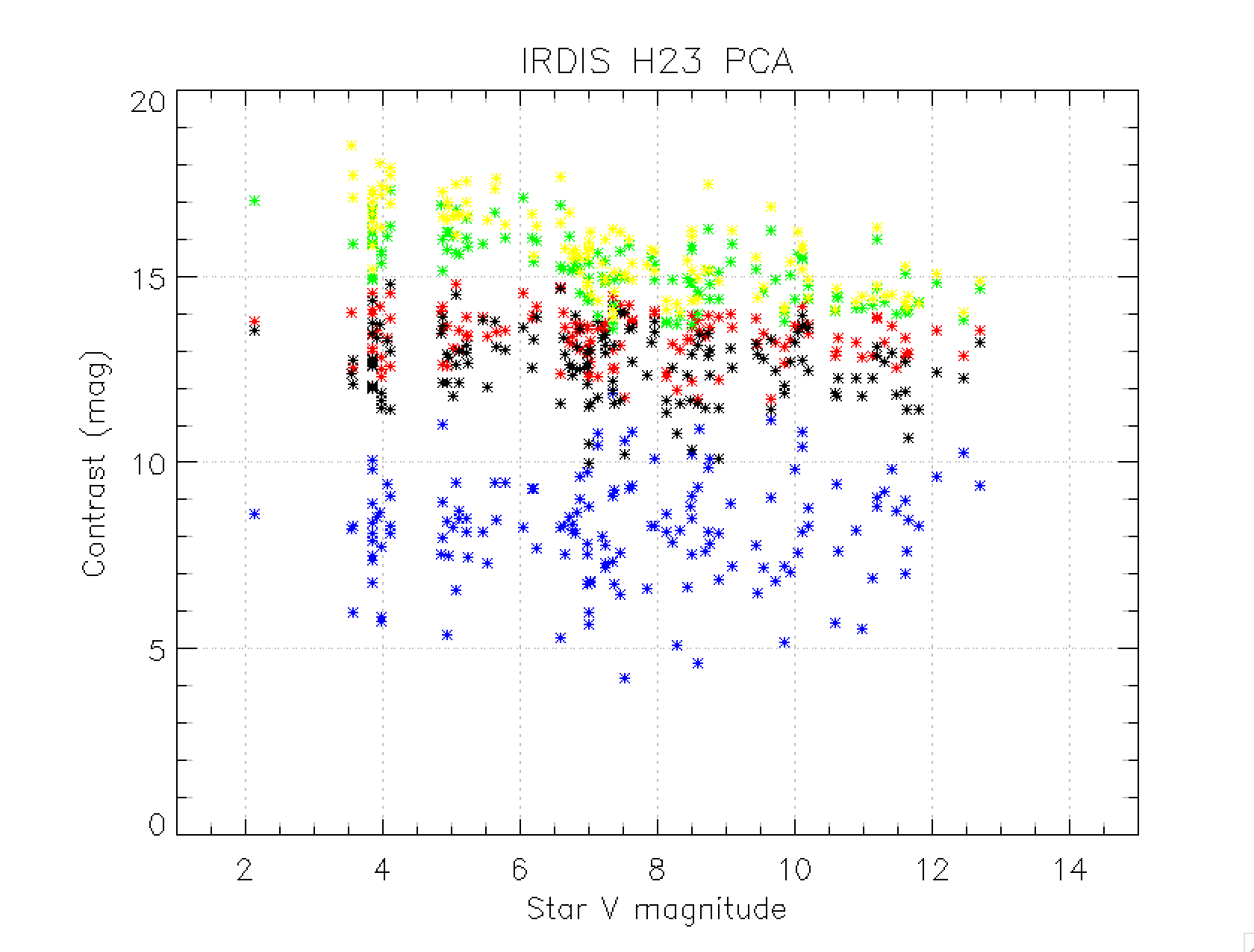} &
		\includegraphics[width=0.6\columnwidth]{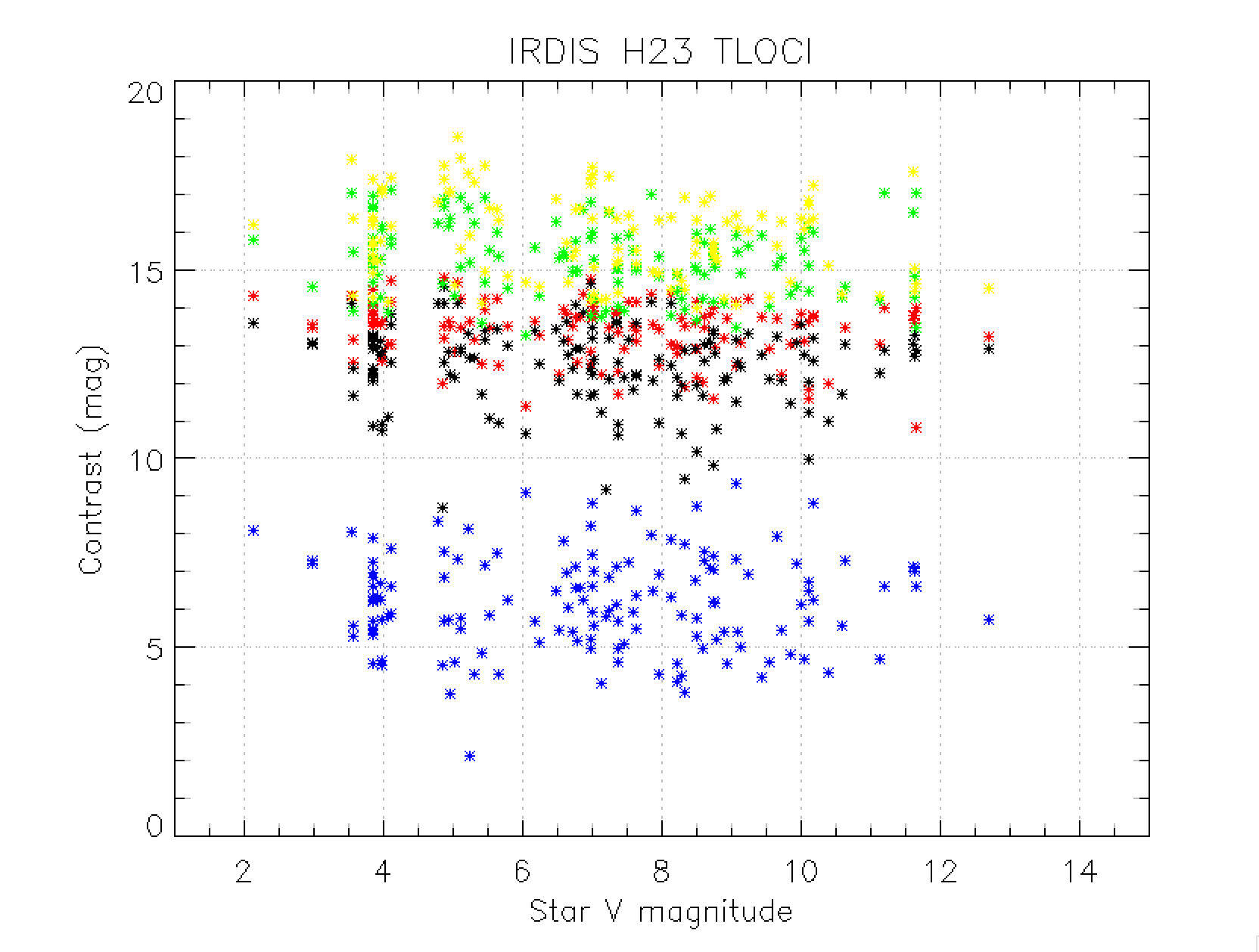} & \includegraphics[width=0.61\columnwidth]{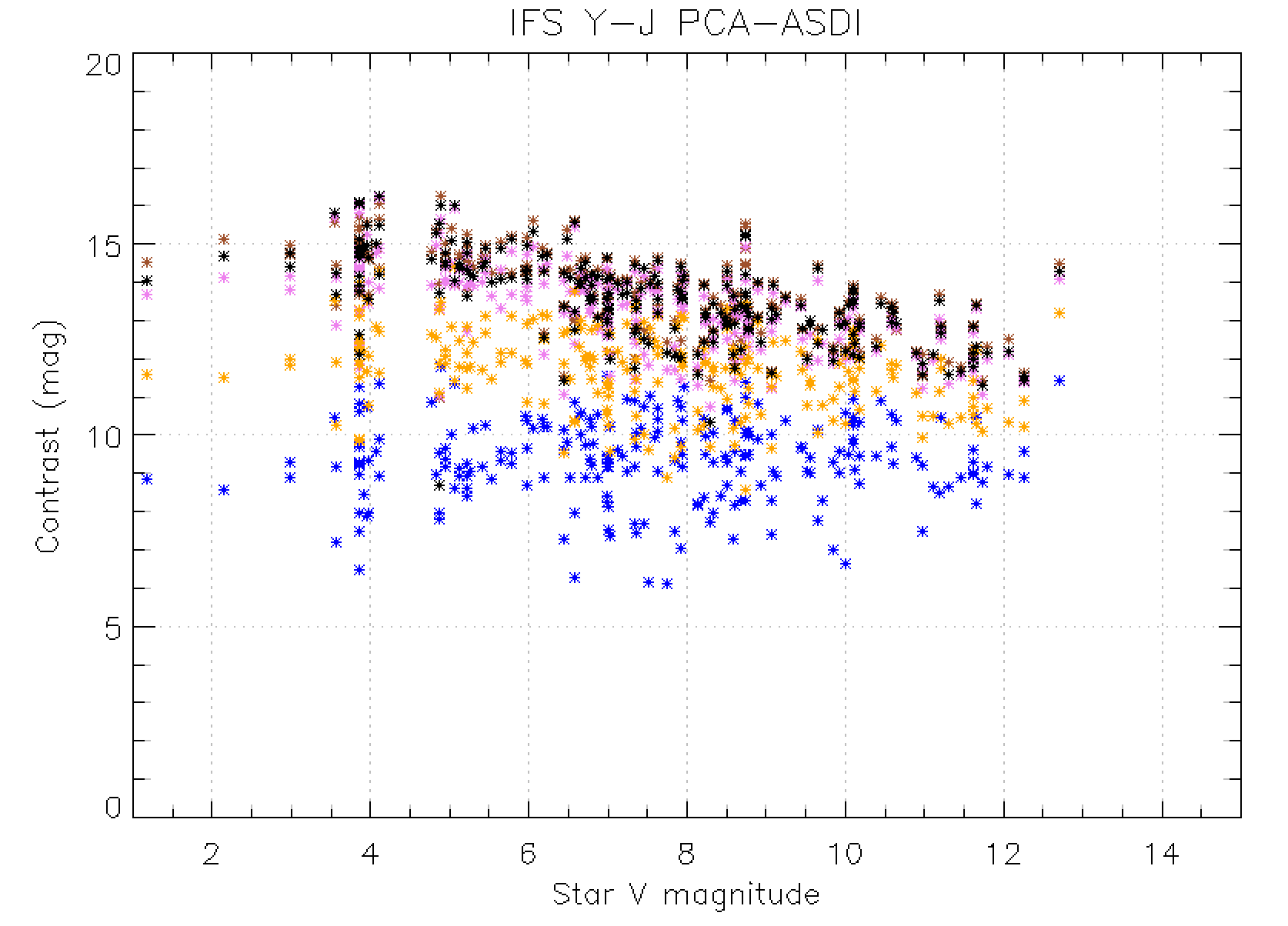}\\
		
	\end{tabular}
	\caption{IRDIFS Contrasts estimated at various separations as function of the sample stars V and H magnitude. Both IRDIS and IFS contrasts are shown separatly. IRDIS contrasts using PCA reduction and TLOCI (Left and middle figures).  The color code for IRDIS s as follow. Blue is for 100 mas separation, black for 500 mas separation, red for 1000 mas separation , green for 2000 mas separation and yellow for 4000 mas separations.IFS contrasts using PCA-ASDI reduction are shown on the right side figure. The color code for IFS is as follow. Blue is for 100 mas separation, orange is for 200 mas separation, pink is for 400 mas separation, black for 500 mas separation, brown is for 700 mas separations.}
	\label{fig:contvsmag}
\end{figure*}

\begin{figure*}
	\begin{tabular}{cc}
		\centering
		IFS RAW CONTRASTS & IFS PROCESSED CONTRASTS\\
		\includegraphics[width=1.05\columnwidth]{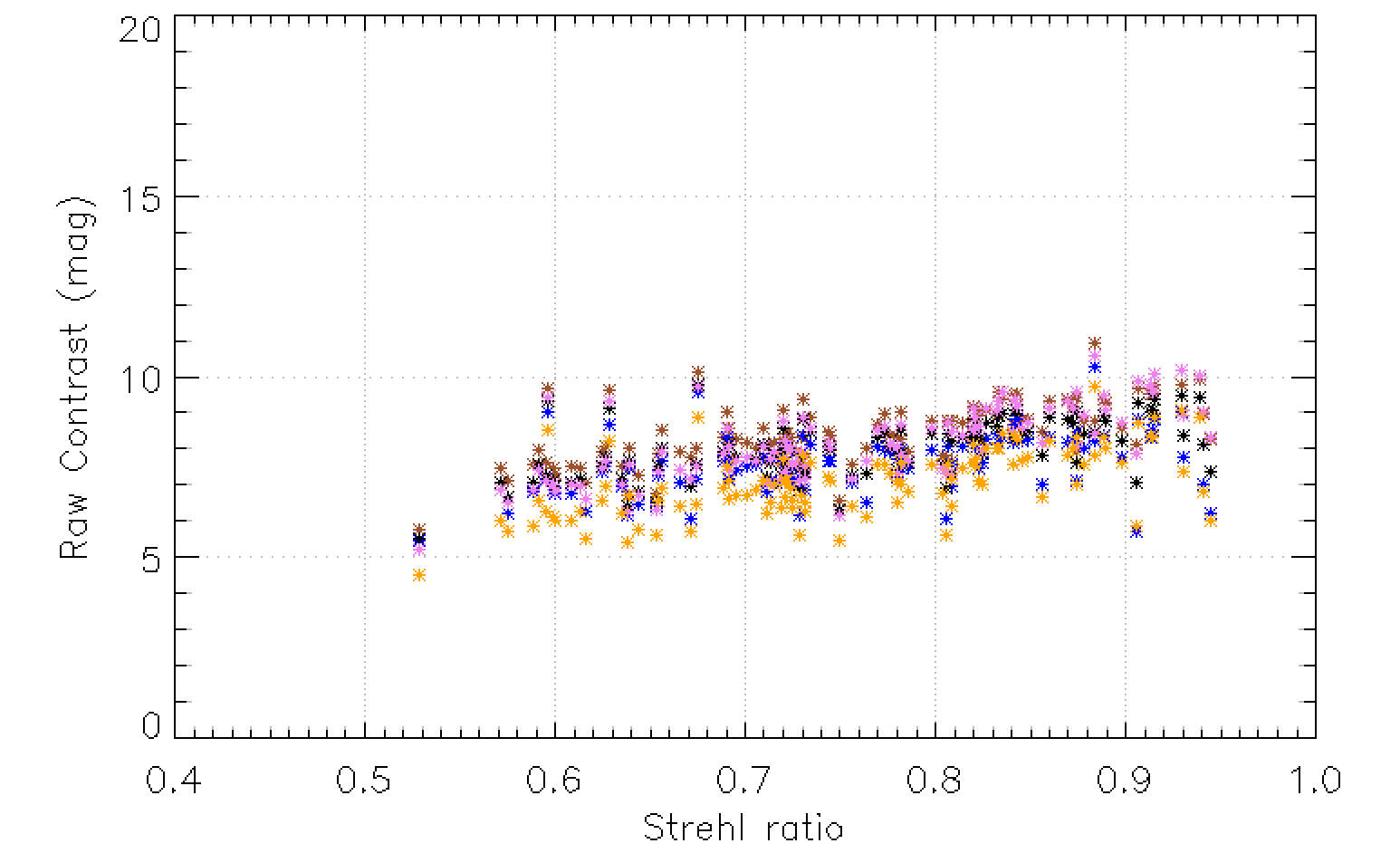}&
		\includegraphics[width=1.05\columnwidth]{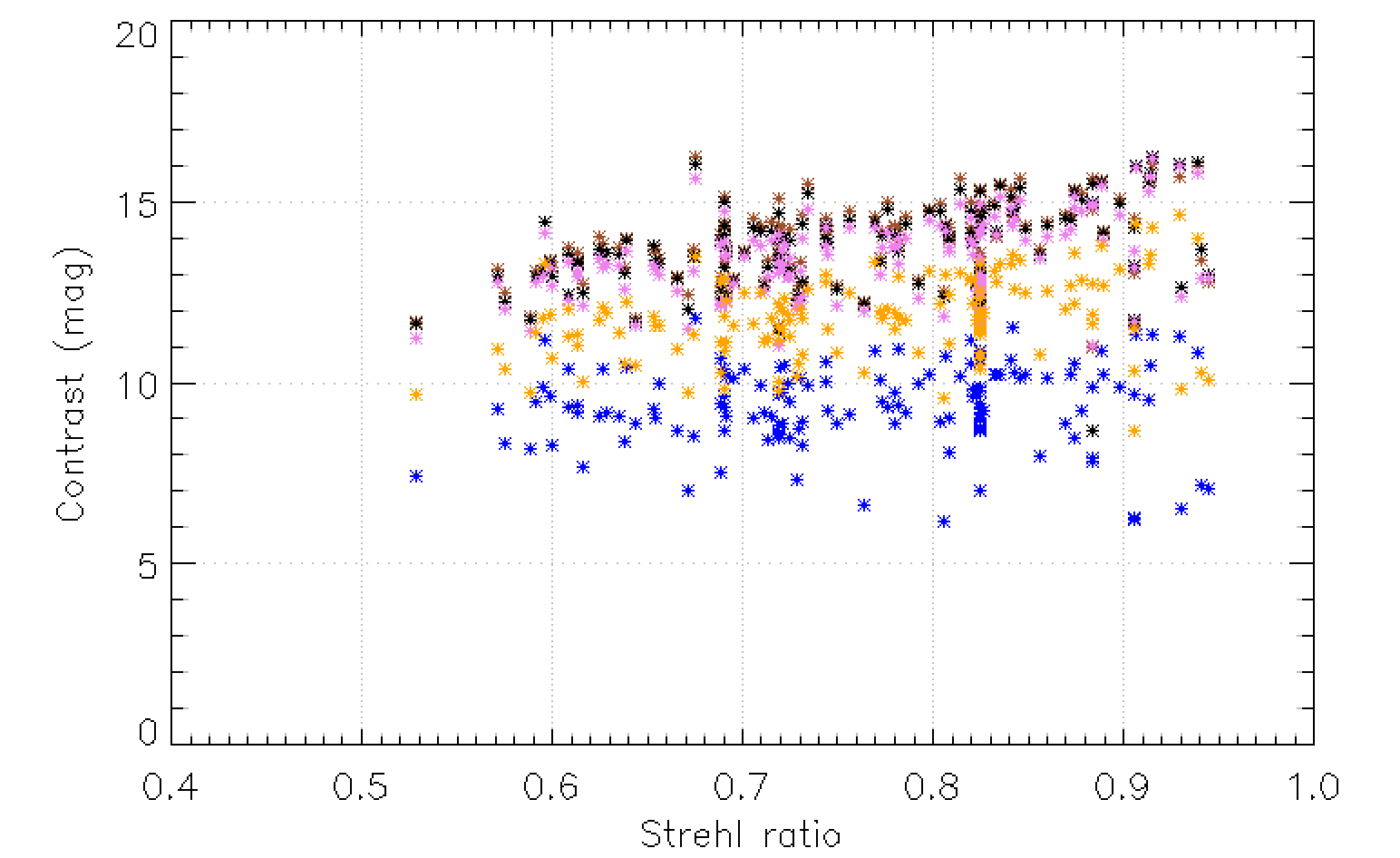}\\
		\includegraphics[width=1.05\columnwidth]{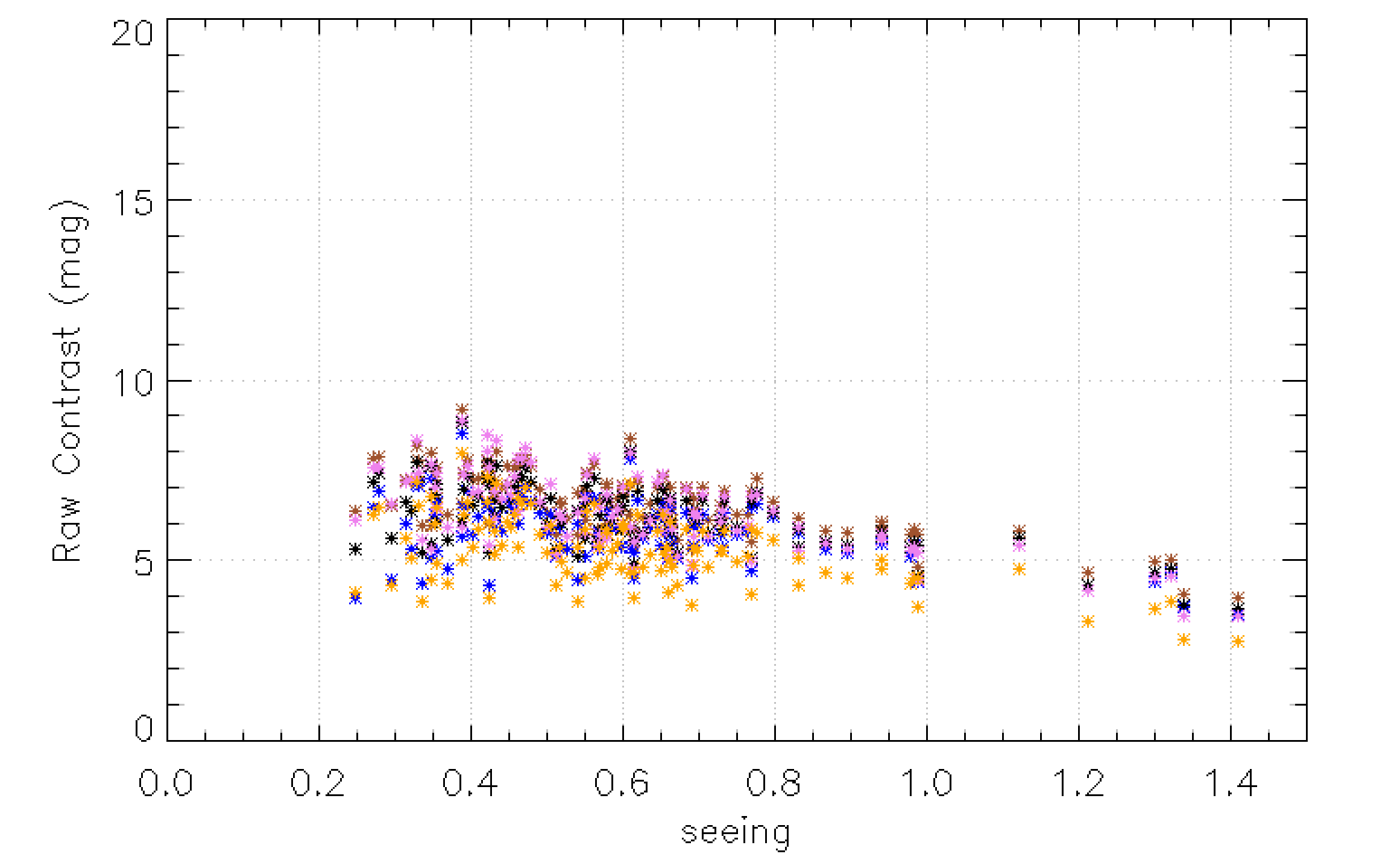}&
		\includegraphics[width=1.05\columnwidth]{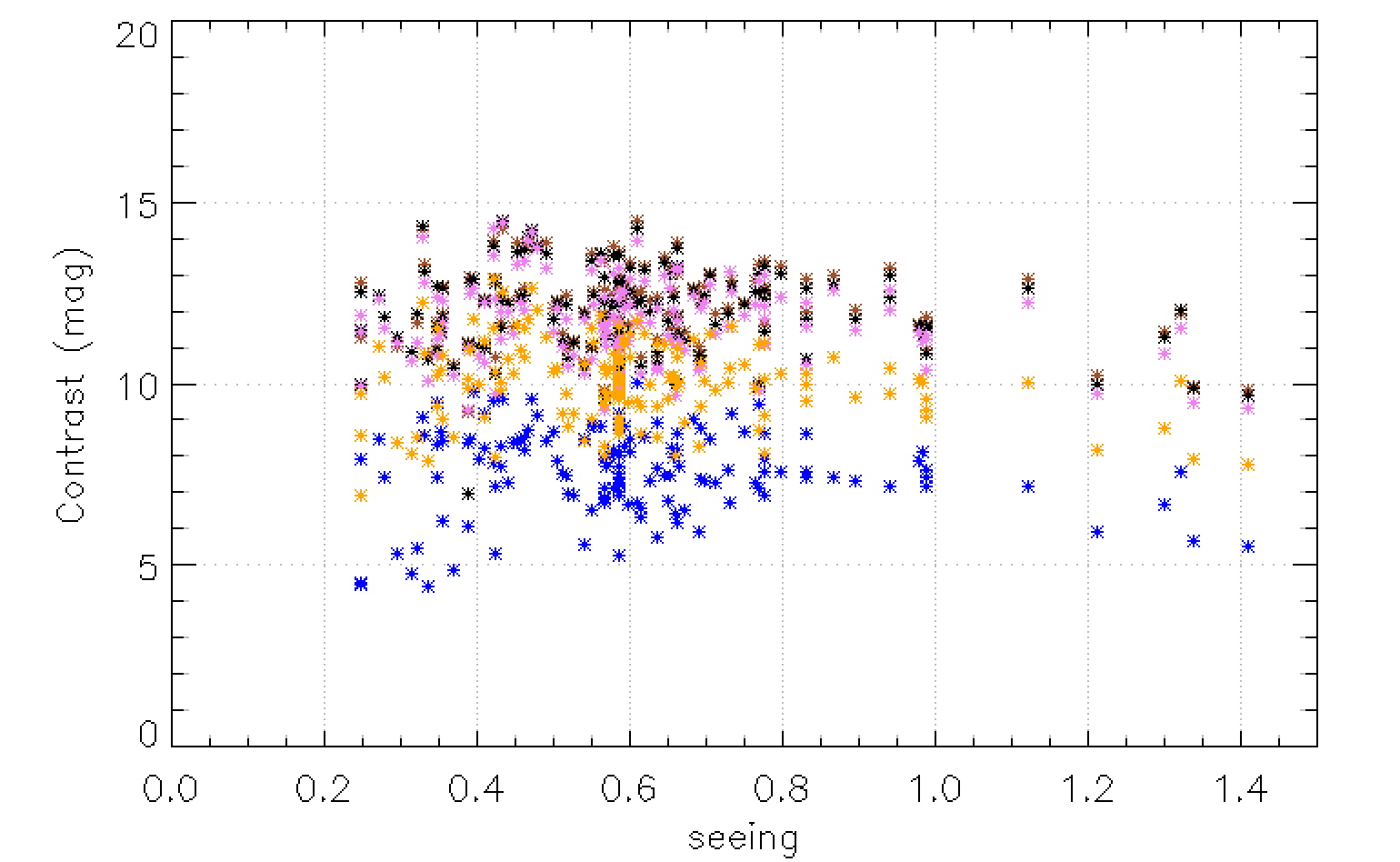}\\
		\includegraphics[width=1.05\columnwidth]{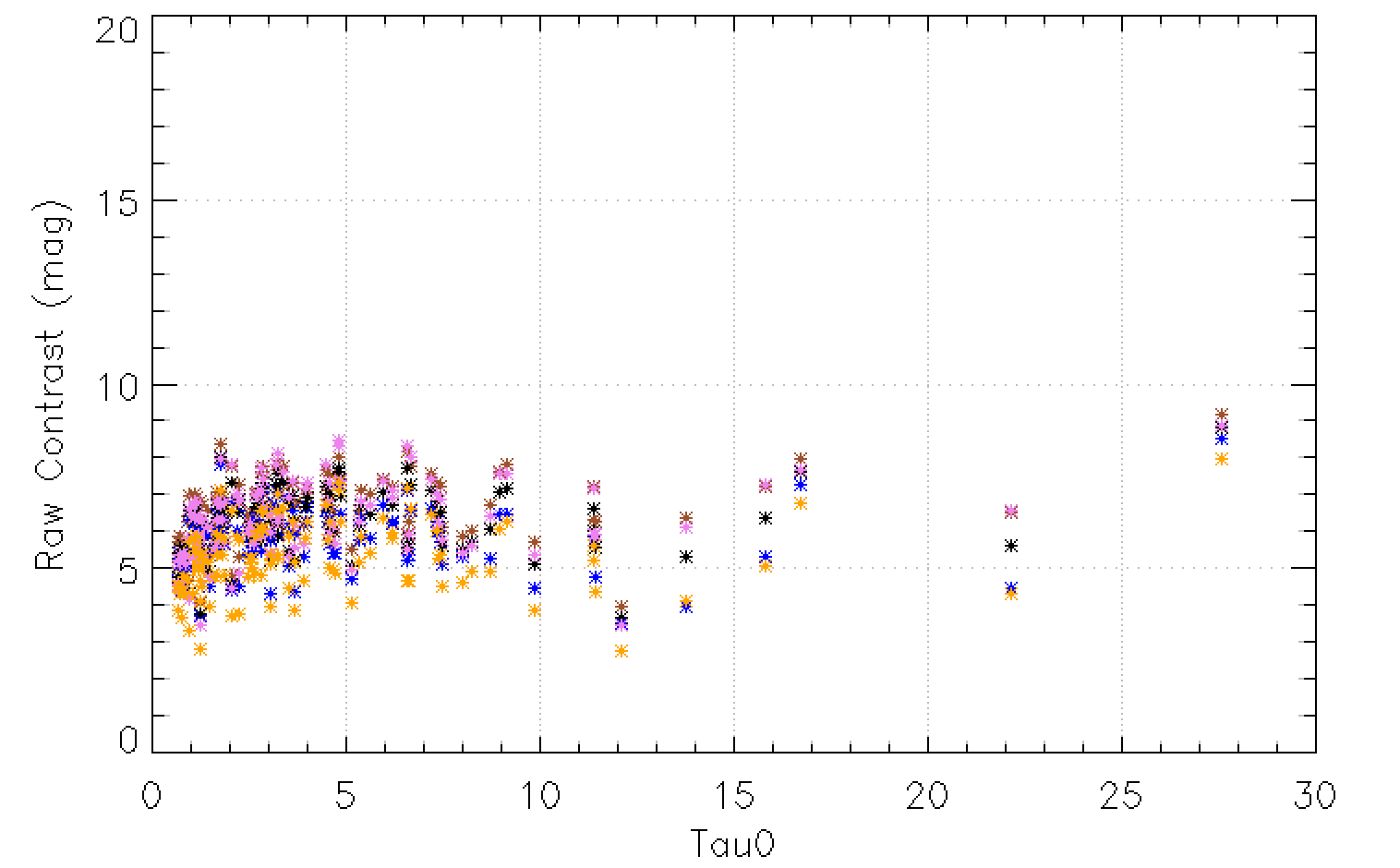}&
		\includegraphics[width=1.05\columnwidth]{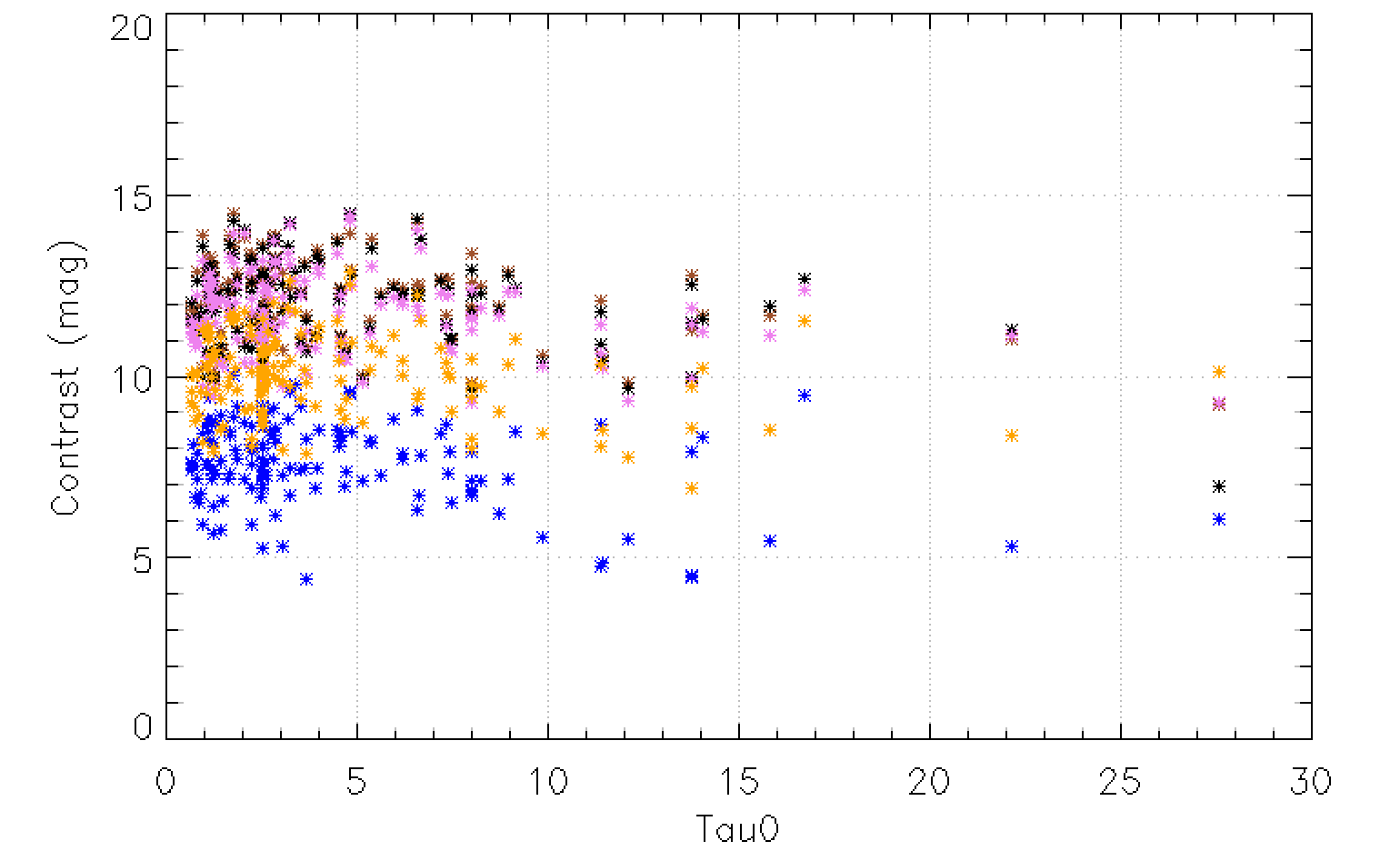}\\
	\end{tabular}
	\caption{IFS raw (Left) and processed (Right) contrasts in YJ bands at various separations computed using PCA ASDI estimated as function of the observing conditions in irdifs mode. Blue is for 100 mas separation, orange for 200 mas separation and pink for 400 mas separations, black for 500 mas separation, , brown for 700 mas separation.}
	\label{fig:cont_sparta_IFS}
\end{figure*}

\end{appendix}
\end{document}